\documentclass[a4paper,10pt,twoside]{cpc-hepnp}

\usepackage{multicol}
\usepackage{float}
\usepackage{fixltx2e}
\usepackage{graphicx}
\usepackage{booktabs}
\usepackage{amssymb,bm,mathrsfs,bbm,amscd}
\usepackage[tbtags]{amsmath}
\usepackage{lastpage}
\usepackage{siunitx}

\graphicspath{{figures/}}

\usepackage{multirow}
\usepackage{amsmath}
\usepackage{amssymb}
\usepackage{placeins}
\usepackage{soul} 
\usepackage{tabularx}
\usepackage{array}

\usepackage{epstopdf}
\usepackage{subfigure}
\usepackage{lineno}
\usepackage{longtable}
\usepackage{booktabs}

\newcommand{\abi}{{\rm ab}^{-1}}
\newcommand{\fbi}{{\rm fb}^{-1}}

\newcommand{\abih}{5.6\,{\rm ab}^{-1}}

\newcommand{\ZH}{e^+e^-\to ZH}
\newcommand{\ZX}{e^+e^-\to ZX}
\newcommand{\vvH}{e^+e^-\to \nu_e\bar{\nu}_e H}
\newcommand{\eeH}{e^+e^-\to e^+e^- H}

\newcommand{\sigmaBR}{\sigma\times\mathrm{BR}}
\newcommand{\mrecoil}{M_{\rm recoil}}

\newcommand{\BR}{{\rm BR}}
\newcommand{\bcg}{b\bar{b}/c\bar{c}/gg}
\newcommand{\CP}{C\!P}
\newcommand{\BRinvBSM}{{\rm BR_{inv}^{BSM}}}

\newcommand{\cmsi}{{\rm cm}^{-2}\, {\rm s}^{-1}}

\newcommand{\hcp}{\hspace*{0.2cm}}

\newcommand{\tsp}{\rule{0pt}{2.5ex}}

\newcommand{\ee}{e^+e^-}
\newcommand{\mm}{\mu^+\mu^-}
\newcommand{\vv}{\nu\bar{\nu}}
\newcommand{\qq}{q\bar{q}}
\newcommand{\bb}{b\bar{b}}
\newcommand{\cc}{c\bar{c}}

\newcommand{\Zee}{Z\to e^+e^-}
\newcommand{\Zmm}{Z\to \mu^+\mu^-}
\newcommand{\Zqq}{Z\to q\bar{q}}
\newcommand{\Zvv}{Z\to \nu\bar{\nu}}
\newcommand{\Zll}{Z\to \ell^+\ell^-}

\newcommand{\Ztt}{Z\to \tau^+\tau^-}

\newcommand{\Wqq}{W\to q\bar{q}}

\newcommand{\Hbb}{H\to b\bar{b}}
\newcommand{\Hcc}{H\to c\bar{c}}
\newcommand{\Hgg}{H\to gg}
\newcommand{\Hmm}{H\to \mu^+\mu^-}
\newcommand{\Htt}{H\to \tau^+\tau^-}
\newcommand{\Hyy}{H\to \gamma\gamma}
\newcommand{\HZy}{H\to Z\gamma}
\newcommand{\HWW}{H\to WW^*}
\newcommand{\HZZ}{H\to ZZ^*}
\newcommand{\Hinv}{H\to {\rm inv}}
\newcommand{\Hbcg}{H\to\bb/\cc/gg}

\def\met{\slash{\!\!\!\!\!\!E}_{\text{T}}}

\newcommand{\scidgts}[2]{\ensuremath{#1\mkern-4mu\times\mkern-4mu 10^{#2}}}

\usepackage{hyperref}
\hypersetup{
  colorlinks=true,linkcolor=blue,citecolor=blue,urlcolor=blue,
  pdftitle={CepC Higgs couplings prospects},
  pdfauthor={Yaquan et al}
  pdfpagemode={UseOutlines},
  bookmarksopen=true,bookmarksnumbered=true,pdfstartview={Fit}
}
\usepackage{feynmp}
\begin{document}
\fancyhead[c]{\small Chinese Physics C~~~Vol. 43, No. 4 (2019)
043002} \fancyfoot[C]{\small 043002-\thepage}

\footnotetext[0]{Received 9 November 2018, Revised 21 January 2019, Published Online}


\title{Precision Higgs Physics at the CEPC\thanks{Supported by the National Key Program for S\&T Researh and Development (2016YFA0400400); CAS Center for Excellence in Particle Physics; Yifang Wang's Science Studio of the Ten Thousand Talents Project; the CAS/SAFEA International Partnership Program for Creative Research Teams (H751S018S5); IHEP Innovation Grant (Y4545170Y2); Key Research Program of Frontier Sciences, CAS (XQYZDY-SSW-SLH002); Chinese Academy of Science Special Grant for Large Scientific Project (113111KYSB20170005); the National Natural Science Foundation of China(11675202); the Hundred Talent Programs of Chinese Academy of Science (Y3515540U1); the National 1000 Talents Program of China; Fermi Research Alliance, LLC (DE-AC02-07CH11359); the NSF under Grant No. PHY1620074 and by the Maryland Center for Fundamental Physics (MCFP); Tsinghua University Initiative Scientific Research Program; The Beijing Municipal Science and Technology Commission project(Z181100004218003).}}





\author{
Fenfen An$^{4,23}$%
\quad Yu Bai$^{9}$
\quad Chunhui Chen$^{23}$
\quad Xin Chen$^{5}$
\quad Zhenxing Chen$^{3}$ 
\quad Joao Guimaraes da Costa$^{4}$\\
\quad Zhenwei Cui$^{3}$
\quad Yaquan Fang$^{4,6,34}$
\quad Chengdong Fu$^{4}$
\quad Jun Gao$^{10}$
\quad Yanyan Gao$^{22}$
\quad Yuanning Gao$^{3}$\\
\quad Shao-Feng Ge$^{15,29}$
\quad Jiayin Gu$^{13}$
\quad Fangyi Guo$^{1,4}$
\quad Jun Guo$^{10}$
\quad Tao Han$^{5,31}$
\quad Shuang Han$^{4}$\\
\quad Hong-Jian He$^{11,10}$
\quad Xianke He$^{10}$
\quad Xiao-Gang He$^{11,10,20}$
\quad Jifeng Hu$^{10}$
\quad Shih-Chieh Hsu$^{32}$
\quad Shan Jin$^{8}$\\
\quad Maoqiang Jing$^{4,7}$
\quad Susmita Jyotishmati$^{33}$
\quad Ryuta Kiuchi$^{4}$
\quad Chia-Ming Kuo$^{21}$
\quad Pei-Zhu Lai$^{21}$
\quad Boyang Li$^{5}$\\
\quad Congqiao Li$^{3}$
\quad Gang Li$^{4,34}$
\quad Haifeng Li$^{12}$
\quad Liang Li$^{10}$
\quad Shu Li$^{11,10}$
\quad Tong Li$^{12}$\\
\quad Qiang Li$^{3}$
\quad Hao Liang$^{4,6}$
\quad Zhijun Liang$^{4,34}$
\quad Libo Liao$^{4}$
\quad Bo Liu$^{4,23}$
\quad Jianbei Liu$^{1}$\\
\quad Tao Liu$^{14}$
\quad Zhen Liu$^{26,30}$
\quad Xinchou Lou$^{4,6,33,34}$
\quad Lianliang Ma$^{12}$
\quad Bruce Mellado$^{17,18}$
\quad Xin Mo$^{4}$\\
\quad Mila Pandurovic$^{16}$
\quad Jianming Qian$^{24}$
\quad Zhuoni Qian$^{19}$
\quad Nikolaos Rompotis$^{22}$
\quad Manqi Ruan$^{4}$
\quad Alex Schuy$^{32}$\\
\quad Lian-You Shan$^{4}$
\quad Jingyuan Shi$^{9}$
\quad Xin Shi$^{4}$
\quad Shufang Su$^{25}$
\quad Dayong Wang$^{3}$
\quad Jin Wang$^{4}$\\
\quad Lian-Tao Wang$^{27}$
\quad Yifang Wang$^{4,6}$
\quad Yuqian Wei$^{4}$
\quad Yue Xu$^{5}$
\quad Haijun Yang$^{10,11}$
\quad Ying Yang$^{4}$\\
\quad Weiming Yao$^{28}$
\quad Dan Yu$^{4}$
\quad Kaili Zhang$^{4,6}$
\quad Zhaoru Zhang$^{4}$
\quad Mingrui Zhao$^{2}$
\quad Xianghu Zhao$^{4}$
\quad Ning Zhou$^{10}$}
\maketitle
\address{%
$^1$ Department of Modern Physics, University of Science and Technology of China, Anhui 230026, China\\
$^2$  China Institute of Atomic Energy, Beijing 102413, China\\
$^3$  School of Physics, Peking University, Beijing 100871, China\\
$^4$  Institute of High Energy Physics, Beijing 100049, China\\
$^5$  Department of Engineering Physics, Physics Department, Tsinghua University, Beijing 100084, China\\
$^6$  University of Chinese Academy of Science (UCAS), Beijing 100049, China\\
$^7$  School of Nuclear Science and Technology, University of South China, Hengyang 421001, China\\
$^8$  Department of Physics, Nanjing University, Nanjing 210093, China\\
$^9$  Department of Physics, Southeast University, Nanjing 210096, China\\
$^{10}$  School of Physics and Astronomy, Shanghai Jiao Tong University, KLPPAC-MoE, SKLPPC, Shanghai 200240, China\\
$^{11}$  Tsung-Dao Lee Institute, Shanghai 200240, China\\
$^{12}$  Institute of Frontier and Interdisciplinary Science and Key Laboratory of Particle Physics and Particle Irradiation\
 (MOE), Shandong University, Qingdao 266237, China\\
$^{13}$  PRISMA Cluster of Excellence \& Mainz Institute of Theoretical Physics, Johannes Gutenberg-Universit$\ddot{a}$t Mainz, Mainz 55128, Germany\\
$^{14}$  Department of Physics, Hong Kong University of Science and Technology, Hong Kong\\
$^{15}$  Kavli IPMU (WPI), UTIAS, The University of Tokyo, Kashiwa, Chiba 277-8583, Japan\\
$^{16}$  Vinca Institute of Nuclear Sciences, University of Belgrade, Belgrade 11000, Serbia\\
$^{17}$  School of Physics and Institute for Collider Particle Physics, University of the Witwatersrand, Johannesburg 2050, South Africa\\
$^{18}$  iThemba LABS, National Research Foundation, PO Box 722, Somerset West 7129, South Africa\\
$^{19}$  Center for Theoretical Physics of the Universe, Institute of Basic Science, Daejeon 34126, South Korea\\
$^{20}$  Department of Physics, National Taiwan University, Taipei 10617, Taiwan\\
$^{21}$  Department of Physics and Center for High Energy and High Field Physics, National Central University, Taoyuan City 32001, Taiwan\\
$^{22}$  Department of Physics, University of Liverpool, Liverpool L69 7ZX, United Kingdom\\
$^{23}$  Department of Physics and Astronomy, Iowa State University, Ames 50011-3160, USA\\
$^{24}$  Department of Physics, University of Michigan, Ann Arbor, Michigan 48109, USA\\
$^{25}$  Department of Physics, University of Arizona, Arizona 85721, USA\\
$^{26}$ Theoretical Physics Department, Fermi National Accelerator Laboratory, Batavia 60510, USA\\
$^{27}$  Department of Physics, University of Chicago, Chicago 60637, USA\\
$^{28}$  Lawrence Berkeley National Laboratory, Berkeley, California 94720, USA\\
$^{29}$  Department of Physics, University of California, Berkeley, California 94720, USA\\
$^{30}$  Maryland Center for Fundamental Physics, Department of Physics, University of Maryland, College Park, Maryland 20742, USA\\
$^{31}$  Department of Physics \& Astronomy, University of Pittsburgh, Pittsburgh 15260, USA\\
$^{32}$  Department of Physics, University of Washington, Seattle 98195-1560, USA\\
$^{33}$  Department of Physics, University of Texas at Dallas, Texas 75080-3021, USA\\
$^{34}$  Physical Science Laboratory, Huairou National Comprehensive Science Center, Beijing, 101400, China\\
}

\begin{abstract}
The discovery of the Higgs boson with its mass around 125 GeV by the ATLAS and CMS Collaborations marked the beginning of a new era in high energy physics. The Higgs boson will be the subject of extensive studies of the ongoing LHC program. At the same time, lepton collider based  Higgs factories have been proposed as a possible next step beyond the LHC, with its main goal to precisely measure  the properties of the Higgs boson and probe potential new physics associated with the Higgs boson. The Circular Electron Positron Collider~(CEPC) is one of such proposed Higgs factories. The CEPC is an $e^+e^-$ circular collider proposed by and to be hosted in China.  Located in a tunnel of approximately 100~km in circumference, it will operate at a center-of-mass energy of 240~GeV as the Higgs factory. In this paper, we present the first estimates on the precision of the Higgs boson property measurements achievable at the CEPC and discuss implications of these measurements.
\end{abstract}
%
\begin{keyword}
CEPC, Higgs boson, Higgs boson properties, Higgs boson couplings, Higgs factory, Effective Field Theory, EFT 
\end{keyword}

\begin{pacs}
1--3 PACS(Physics and Astronomy Classification Scheme, http://www.aip.org/pacs/pacs.html/)
\end{pacs}

\footnotetext[0]{\hspace*{-3mm}\raisebox{0.3ex}{$\scriptstyle\copyright$}2013
Chinese Physical Society and the Institute of High Energy Physics
of the Chinese Academy of Sciences and the Institute
of Modern Physics of the Chinese Academy of Sciences and IOP Publishing Ltd}%





\begin{multicols}{2}
\section{Introduction}
\label{sec:introduction}

The historic discovery of a Higgs boson in 2012 by the ATLAS and CMS collaborations~\cite{atlas:2012obs,cms:2012obs}  at the Large Hadron Collider (LHC)  has opened a new era in particle physics. Subsequent measurements of the properties of the new particle have indicated compatibility with the Standard Model (SM) Higgs boson~\cite{Aad:2013wqa,Aad:2013xqa,Chatrchyan:2013lba,Chatrchyan:2014vua,Khachatryan:2014kca,Khachatryan:2016vau,Aad:2015zhl}.   While the SM  has been remarkably successful in describing experimental phenomena, it is important to recognize that it is not a complete theory.    In particular, it does not {\it predict} the parameters in the Higgs potential, such as the Higgs boson mass. The vast difference between the Planck scale and the weak scale remains a major mystery. There is not a complete understanding of the  nature of electroweak phase transition.  The discovery of a spin zero Higgs boson,  the first elementary particle of its kind, only sharpens these questions. It is clear that any attempt of addressing these questions will involve new physics beyond the SM~(BSM).
Therefore, the Higgs boson discovery marks the beginning of a new era of theoretical and experimental explorations.

A physics program of the precision measurements of the Higgs boson properties will be a critical component of any road map for  high energy physics in the coming decades.
Potential new physics beyond the SM could lead to observable deviations  in the Higgs boson couplings from the SM expectations. Typically, such deviations can be parametrized as 
\begin{equation}
\delta = c \frac{v^2}{ M_{\rm NP}^2},
\label{eq:scale}
\end{equation}
where $v$ and $M_{\rm NP}$ are the vacuum expectation value of the Higgs field and the typical mass scale of new physics, respectively. The size of the proportionality constant $c$ depends on the model, but it should not be much larger than ${\mathcal{O}}(1)$. The high-luminosity LHC~(HL-LHC) will measure the Higgs boson couplings to about $5 \% $~\cite{ATL-PHYS-PUB-2014-016,CMS:2013xfa}. At the same time, the LHC will directly search for new physics from a few hundreds of  GeV to  at least one TeV. Eq.~\ref{eq:scale} implies that probing new physics significantly beyond the LHC reach would require the measurements of the Higgs boson couplings at least at percent level accuracy.  
To achieve such precision will need new facilities, a lepton collider operating as a Higgs factory is a natural next step. 

The Circular Electron-Positron Collider (CEPC)~\cite{CEPCStudyGroup:2018rmc}, proposed by the Chinese particle physics community, is one of such possible facilities. The CEPC will be placed in a tunnel with a circumference of approximately 100 km  and will operate at a center-of-mass energy of $\sqrt{s}\sim 240$~GeV, near the maximum of the Higgs boson production cross section through the $e^+e^-\to ZH$ process. At the CEPC, in contrast to the LHC, Higgs boson candidates can be identified through a technique known as the recoil mass method without tagging its decays. Therefore,  the Higgs boson production can be disentangled from its decay in a model independent way.  Moreover, the cleaner environment at a lepton collider allows much better exclusive measurements of Higgs boson decay channels.  All of these give the CEPC an impressive reach in probing Higgs boson properties. With the expected integrated luminosity of $\abih$, over one million Higgs bosons will be produced. With this sample, the CEPC will be able to measure the Higgs boson coupling to the $Z$ boson with an accuracy of $0.25 \% $, more than a factor of 10 better than the HL-LHC~\cite{ATL-PHYS-PUB-2014-016,CMS:2013xfa}. Such a precise measurement gives the CEPC unprecedented reach into interesting new physics scenarios which are difficult to probe at the LHC. The CEPC also has strong capability in detecting Higgs boson invisible decay. It is sensitive to the invisible decay branching ratio down to $0.3 \%$. In addition,  it is expected to have good sensitivities to exotic decay channels which are swamped by backgrounds at the LHC.    It is also important to stress that an $e^+ e^-$ Higgs factory can perform {\it model independent} measurement of the Higgs boson width. This unique feature in turn allows for the model independent determination of the Higgs boson couplings. 

This paper documents the first studies of a precision Higgs boson physics program at the CEPC. It is organized as follows: Section~\ref{sec:detector} briefly summarizes the collider and detector performance parameters assumed for the studies. Section~\ref{sec:samples} gives an overview of relevant $\ee$ collision processes and Monte Carlo simulations.  Sections~\ref{sec:massXS} and~\ref{sec:decays} describe inclusive and exclusive Higgs boson measurements. Section~\ref{sec:combinations} discusses the combined analysis to extract Higgs boson production and decay properties. Section~\ref{sec:kappaEFT} interprets the results in the coupling and effective theory frameworks. Section~\ref{sec:CPandExotic} estimates the reaches in the test of Higgs boson spin/$\CP$ properties and in constraining the exotic decays of the Higgs boson based on previously published phenomenological studies. Finally the implications of all these measurements are discussed in Section~\ref{sec:implications}.


\section{CEPC Detector Concept}
\label{sec:detector}
\subsection{The CEPC operating scenarios}

The CEPC is designed to operate as a Higgs factory at $\sqrt{s} = 240$~GeV and as a $Z$ factory at $\sqrt{s}=91.2$~GeV. It will also perform $WW$ threshold scans around $\sqrt{s}=160$~GeV. Table~\ref{tab:Luminosity} shows potential CEPC operating scenarios and the expected numbers of $H$, $W$ and $Z$ bosons produced in these scenarios.

\begin{table*}
\begin{center}
\caption{\label{tab:Luminosity}\small CEPC operating scenarios and the numbers of Higgs, $W$ and $Z$ bosons produced. The integrated luminosity and the event yields assume two interaction points. The ranges of luminosities and the $Z$ yield of the $Z$ factory operation correspond to detector solenoid field of 3 and 2~Tesla.}
\begin{tabular}{lccc}\hline\hline
Operation mode & $Z$ factory & $WW$ threshold & Higgs factory \\ \hline
$\sqrt{s}$ (GeV)                            &   91.2   &  160  &     240   \\
Run time (year)                             &    2     &    1  &       7   \\
Instantaneous luminosity ($10^{34}\,\cmsi$) &  16--32  &   10  &       3   \\
Integrated luminosity ($\abi$)              &  8--16    &  2.6  &     5.6   \\ \hline \\[-3mm]
Higgs boson yield                           &   --         &   --    &   $10^6$ \\
$W$ boson yield                             &   --         & $10^7$  &  $10^8$ \\
$Z$ boson yield                             & $10^{11}$--$10^{12}$ & $10^8$  &  $10^8$ \\
\hline\hline
\end{tabular}
\end{center}
\end{table*}

The CEPC operation as a Higgs factory will run for 7 years and produce a total of 1 million Higgs bosons with two interaction points. Meanwhile, approximately 100 million $W$ bosons and 1 billion $Z$ bosons will also be produced in this operation. These large samples of $W$ and $Z$ bosons will allow for in-situ detector characterization as well as for the precise measurements of electroweak parameters. 

Running at the $WW$ threshold around $\sqrt{s}=160$~GeV,  $10^{7}$ $W$ bosons will be produced in one year. Similarly running at the $Z$ pole around $\sqrt{s}=91.2$~GeV (the $Z$ factory), CEPC will produce $10^{11}$--$10^{12}$ $Z$ bosons. These large samples will enable high precision measurements of the electroweak observables such as $A_{FB}^{b}$, $R_b$, the $Z$ boson line-shape parameters, the mass and width of the $W$ boson. An order of magnitude or more improvement in the precision of these observables are foreseen.

\subsection{Conceptual detector design}

The primary physics objective of the CEPC is the precise determination of the Higgs boson properties. Therefore CEPC detectors must be able to reconstruct and identify all key physics objects that the Higgs bosons are produced with or decay into with high efficiency, purity and accuracy. 
These objects include charged leptons, photons, jets, missing energy and missing momentum. 
Moreover, the flavor tagging of jets, such as those from $b$, $c$ and light quarks or gluons, are crucial for identifying the hadronic decays of the Higgs bosons. The detector requirements for the electroweak and flavor physics are similar. One notable additional requirement is the identification of charged particles such as $\pi^\pm$ and $K^\pm$ for the flavor physics program.

\begin{center}
	\includegraphics[width=0.45\textwidth]{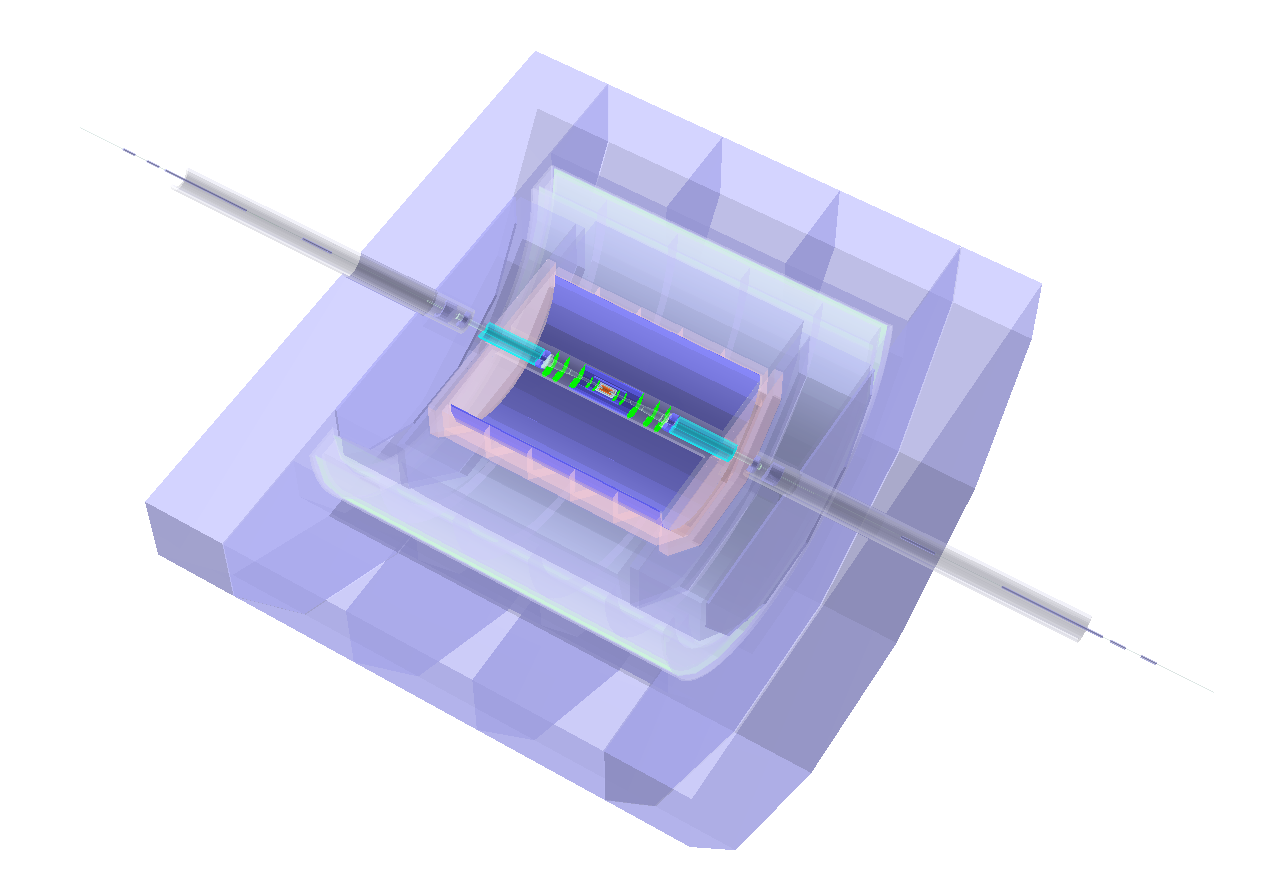}
	\figcaption{\label{fig:CEPCDetector}\small Conceptual CEPC detector, CEPC-v1, implemented in {\sc Mokka}~\cite{MoradeFreitas:2002kj} and {\sc Geant} 4~\cite{geant4}. It is comprised of a silicon vertexing and tracking system of both pixel and strips geometry, a Time-Projection-Chamber tracker, a high granularity calorimeter system, a solenoid of 3.5~Tesla magnetic field, and a muon detector embedded in a magnetic field return yoke.}
\end{center}

Using the International Large Detector (ILD)~\cite{Abe:2010aa,Behnke:2013lya} as a reference, a particle flow oriented conceptual detector design, CEPC-v1 (see Fig.~\ref{fig:CEPCDetector}), has been developed for the CEPC. 
A detailed description of the CEPC-v1 detector can be found in Ref.~\cite{CEPC-SPPCStudyGroup:2015csa}.
Originally developed for the LEP experiments~\cite{PatrickALEPHEnergyFlow, ALEPHEnergyFlow}, particle flow is a proven concept for event reconstruction~\cite{PandoraPFA, ArborPFA, CMS:2009nxa, Beaudette:2014cea}, based on the principle of reconstructing all visible final-state particles in the most sensitive detector subsystem. 
Specifically, a particle-flow algorithm reconstructs charged particles in the tracking system, measures photons in the electromagnetic calorimeter and neutral hadrons in both electromagnetic and hadronic calorimeters. 
Physics objects are then identified or reconstructed from this list of final state particles. Particle flow reconstruction provides a coherent interpretation of an entire physics event and, therefore, is particularly well suited 
for the identification of composite physics objects such as the $\tau$ leptons and jets.

The particle-flow algorithm requires good spatial separations of calorimeter showers induced by different final state particles for their reconstruction. It is imperative to minimize the amount of material before the calorimeter to reduce the uncertainty induced by the nuclear interactions and Bremsstrahlung radiations. Therefore, a high granularity calorimeter system and low material tracking system are implemented in the CEPC-v1 detector concept. 
The tracking system consists of silicon vertexing and tracking detectors as well as a Time Projection Chamber (TPC).
The calorimetry system is based on the sampling technology with absorber/active-medium combination of Tungsten-Silicon for the electromagnetic calorimeter (ECAL) and Iron-Resistive Plate Chamber~(RPC) for the hadronic calorimeter~(HCAL). The calorimeters are segmented at about 1~channel/cm$^{3}$, three orders of magnitude finer than those of the LHC detectors. Both the tracking and the calorimeter system are housed inside a solenoid of 3.5 Tesla magnetic field. 
The CEPC-v1 detector has a sophisticated machine-detector interface with an 1.5-meter L* (the distance between the interaction point and the final focusing quadrupole magnet) to accommodate the high design luminosity.
Table~\ref{tab:GeoPerform} shows the geometric parameters and the benchmark detector subsystem performance of the CEPC-v1 detector. A schematic of the detector is shown in  Fig.~\ref{fig:CEPCDetQuad}.

\begin{center}
\includegraphics[width=0.45\textwidth]{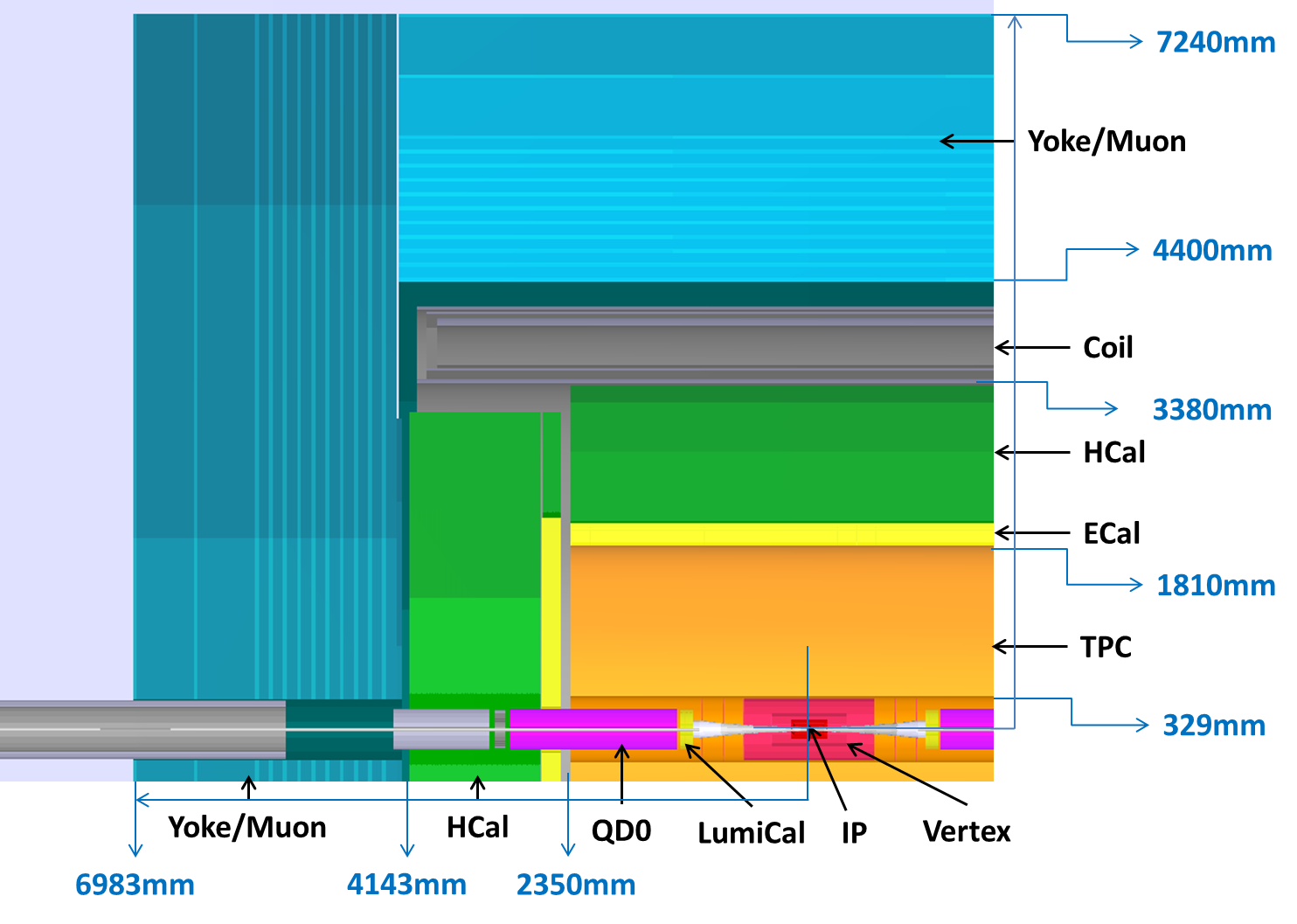}
\figcaption{\label{fig:CEPCDetQuad}\small The layout of one quarter of the CEPC-v1 detector concept.}
\end{center}
\begin{table*}[t!]
\begin{center}
\caption{\label{tab:GeoPerform}\small Basic parameters and performance of the CEPC-v1 detector. The radiation length ($X_0$) and the nuclear interaction length ($\lambda$) are measured at the normal incidence. The cell sizes are for transverse readout sensors and the layer numbers are for longitudinal active readouts. The $\theta$ is the track polar angle.}

\begin{tabular}{lll} \hline\hline
\multicolumn{2}{l}{Tracking system} \\ 
 & Vertex detector            & 6 pixel layers \\
 & Silicon tracker            & 3 barrel layers, 6 forward disks on each side \\
 & Time projection chamber    & 220 radial readouts \\ \\[-3.5mm]

\multicolumn{2}{l}{Calorimetry} \\
 & ECAL          & W/Si, $24X_0$,  $5\!\times\! 5$ mm$^{2}$ cell with 30 layers \\
 & HCAL          & Fe/RPC, $6\lambda$, $10\!\times\! 10$ mm$^{2}$ cell with 40 layers \\ \\[-3.5mm]

\multicolumn{2}{l}{Performance} \\ 
  & Track momentum resolution       &  $\Delta(1/p_T)\sim 2\times 10^{-5}$  (1/GeV)  \\
  & Impact parameter resolution     &  $5\,{\rm \mu m} \oplus 10\,{\rm \mu m} /[(p/{\rm GeV})\, (\sin\theta)^{3/2}]$       \\
  & ECAL energy resolution          &  $\Delta E / E \sim 16\% / \sqrt{E/{\rm GeV}} \oplus 1\%$  \\
  & HCAL energy resolution          &  $\Delta E / E \sim 60\% / \sqrt{E/{\rm GeV}} \oplus 1\%$  \\ \hline\hline
\end{tabular}
\end{center}
\end{table*}

\subsection{Object reconstruction and identification}
A dedicated particle flow reconstruction toolkit, {\sc Arbor}~\cite{ArborPFA}, has been developed for the CEPC-v1 detector. Inspired by the tree structure of particle showers, {\sc Arbor} attempts to reconstruct every visible final state particle. Figure~\ref{fig:EvtDisplay} illustrates a simulated $\ee \to ZH\to \qq\,\bb$ event as reconstructed by the {\sc Arbor} algorithm. The algorithm's performance for leptons, photons and jets are briefly summarized here. More details can be found in Refs.~\cite{Ruan:2018yrh,Zhao:2018jiq}.



\vspace*{0.3cm}
\subsubsection{Leptons and Photons}

Leptons ($\ell$)\footnote{Unless otherwise noted, leptons refer to electrons and muons or their antiparticles thereafter, i.e. $\ell=e,\,\mu$.} are fundamental for the measurements of the Higgs boson properties at the CEPC. About 7\% of the Higgs bosons are produced in association with a pair of leptons through the $\ee\to ZH\to \ell^+\ell^-\, H$ process. These events allow for the identifications of Higgs bosons using the recoil mass information and therefore enable the measurement of the $ZH$ production cross section and the Higgs boson mass. Moreover, a significant fraction of Higgs bosons decay into final states with leptons indirectly through the leptonic decays of the $W$ or $Z$ bosons as well as the $\tau$ leptons. These leptons serve as signatures for identifying different Higgs boson decay modes.

\begin{center}
	\includegraphics[width=.45\textwidth]{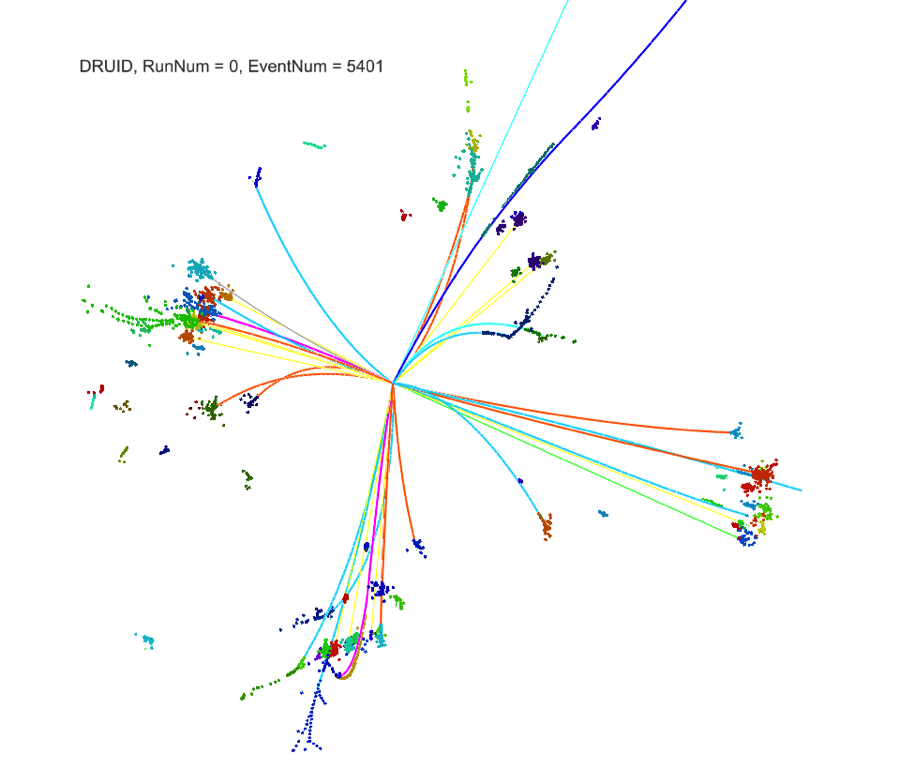}
	\figcaption{\label{fig:EvtDisplay}\small A simulated $e^+e^-\to ZH\to q\bar{q}\ b\bar{b}$ event reconstructed with the {\sc Arbor} algorithm. Different types of reconstructed final state particles are represented in different colors. }
\end{center}

\begin{figure*}
\begin{center}
\subfigure[]{\includegraphics[width=0.40\textwidth]{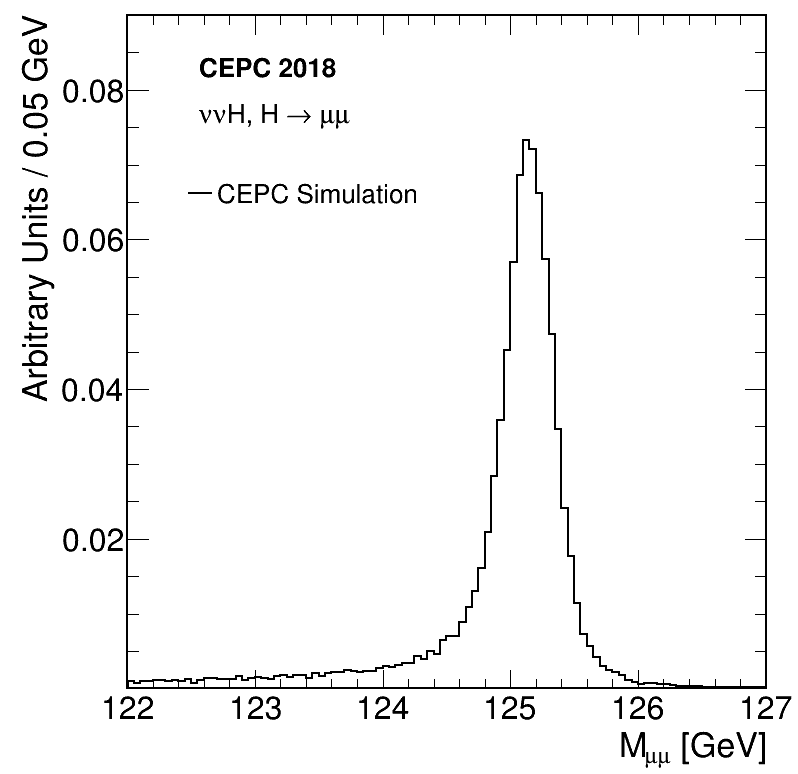}}
\subfigure[]{\includegraphics[width=0.40\textwidth]{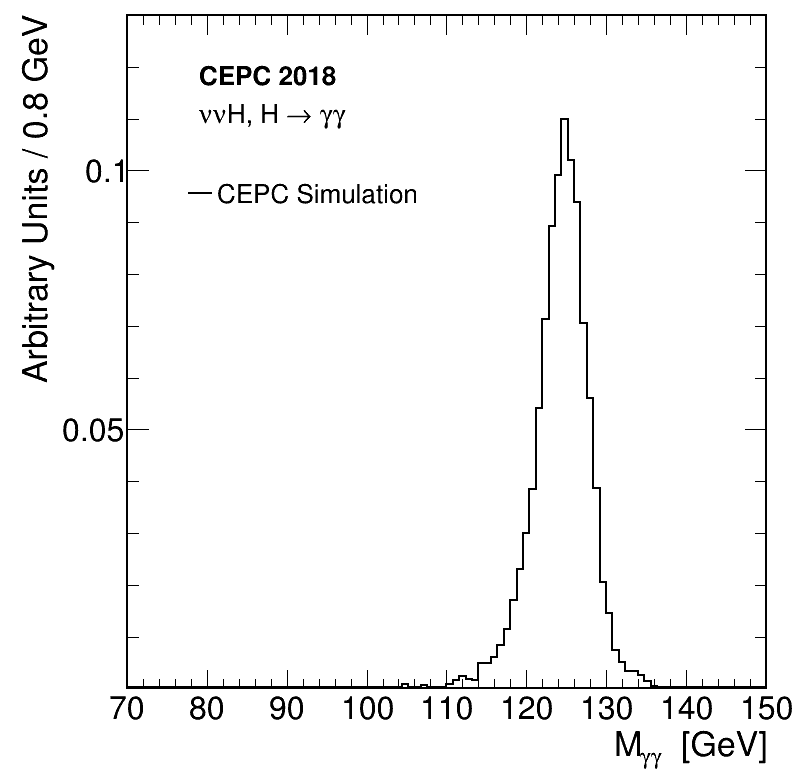}}
	\figcaption{\small Simulated invariant mass distributions of (a) muon pairs from $H\to\mm$ and (b) photon pairs from $H\to\gamma\gamma$, both from the $\ee\to ZH$ process with the $\Zvv$ decay. The $M_{\mm}$ distribution is fit with a Gaussian core plus a small low-mass tail from the Bremsstrahlung radiation. The Gaussian has a width of 0.2~GeV, corresponding to a relative mass resolution of 0.16\%.  The $M_{\gamma\gamma}$ distribution is described well by a Crystal Ball function with a width of 3.1~GeV, corresponding to a relative mass resolution of  2.5\%. }
\label{fig:HmmHyy}
\end{center}
\end{figure*}

A lepton identification algorithm, LICH~\cite{Yu:2017mpx}, has been developed and integrated into {\sc Arbor}. Efficiencies close to 99.9\% for identifying electrons and muons with energies above 2~GeV have been achieved while the mis-identification probabilities from hadrons are limited to be less than 1\%. 
The CEPC-v1 tracking system provides an excellent momentum resolution that is about ten times better than those of the LEP and LHC detectors. The good resolution is illustrated in the narrow invariant mass distribution of the muon pairs from the $H\to\mm$ decays as shown in Fig.~\ref{fig:HmmHyy}(a). 

Photons are essential for the studies of $H\to\gamma\gamma$ and $H\to Z\gamma$ decays. They are also important for the reconstruction and measurements of $\tau$ leptons and jets. The $H\to\gamma\gamma$ decay is an ideal process to characterize the photon performance of the CEPC-v1. Figure~\ref{fig:HmmHyy}(b) shows the invariant mass distribution of the photon pairs from the $H\to\gamma\gamma$ decays.

\vspace*{0.3cm}
\subsubsection{Jets}

Approximately 70\% of Higgs bosons decay directly into jets ($\bb, \cc, gg$) and an additional 22\% decay indirectly into final states with jets through the $H\to WW^*, ZZ^*$ cascades.  Therefore, efficient jet reconstruction and precise measurements of their momenta are pre-requisite for a precision Higgs physics program. In {\sc Arbor}, jets are reconstructed using the Durham algorithm~\cite{Catani:1991hj}. As a demonstration of the CEPC-v1 jet performance, Fig.~\ref{fig:BMR} shows the reconstructed dijet invariant mass distributions of the $W\to \qq$, $Z\to \qq$ and $H\to\bb/\cc/ gg$ decays from the $ZZ\to \vv\, \qq$, $WW\to\ell\nu\,\qq$ and $ZH\to\vv (\bcg)$ processes, respectively. Compared with $\Wqq$, the $\Zqq$ and $\Hbcg$ distributions have long low-mass tails, resulting from the heavy-flavor jets in these decays. The jet energy resolution is expected to be between 3--5\% for the jet energy range relevant at the CEPC. This resolution is approximately 2--4 times better than those of the LHC experiments~\cite{Aad:2012ag,Khachatryan:2016kdb}. The dijet mass resolution for the $W$ and $Z$ bosons is approximately 4.4\%, which allows for an average separation of $2\sigma$ or better of the the hadronically decaying $W$ and $Z$ bosons.
 
\begin{center}
  \includegraphics[width=0.45\textwidth]{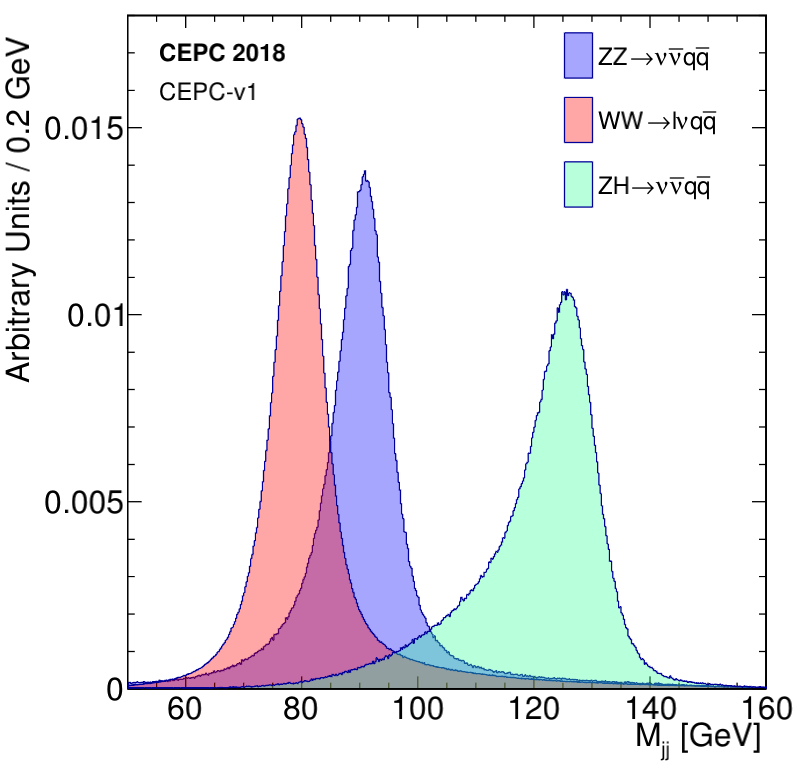}
  \figcaption{\label{fig:BMR}\small Distributions of the reconstructed dijet invariant mass for the $W\to \qq$, $Z\to \qq$ and $\Hbcg$ decays from, respectively, the $WW\to \ell\nu \qq$, $ZZ\to\vv \qq$ and $ZH\to\vv(\bcg)$ processes. All distributions are normalized to unit area. }
\end{center}
\begin{center}
	\includegraphics[width=0.45\textwidth]{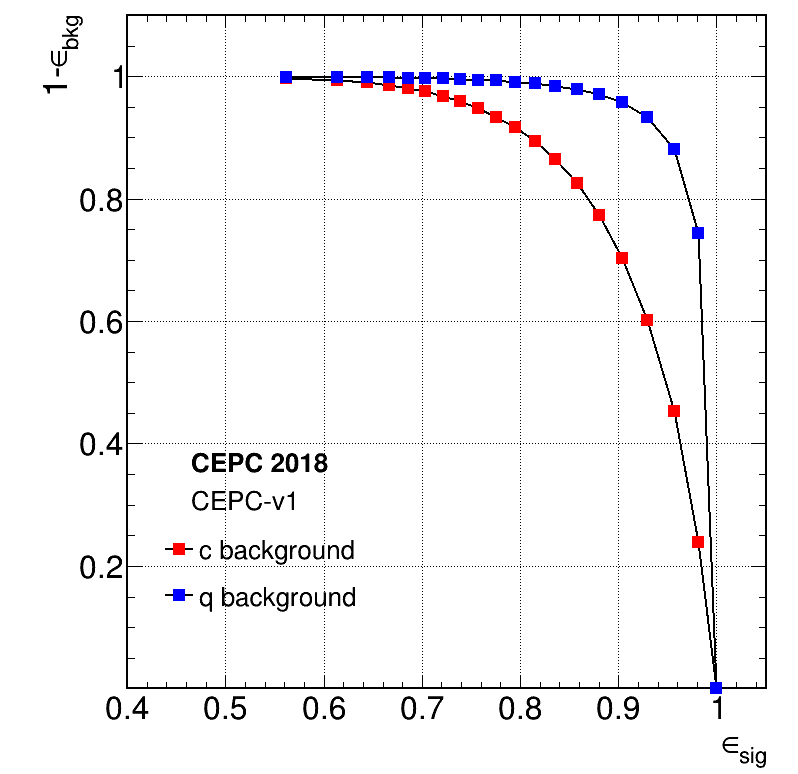}
	\figcaption{\label{fig:jetft}\small Efficiency for tagging $b$-jets vs rejection for light-jet background (blue) and $c$-jet background (red), determined from an inclusive $\Zqq$ sample from the $Z$ factory operation.  }
\end{center}

Jets originating from heavy flavors ($b$- or $c$-quarks) are tagged using the LCFIPlus algorithm~\cite{LCFIPlus}. The algorithm combines information from the secondary vertex, jet mass, number of leptons etc to construct $b$-jet and $c$-jet discriminating variables. The tagging performance characterized using the $Z\to \qq$ decays from the $Z$ factory operation is shown in Fig.~\ref{fig:jetft}.  For an inclusive $Z\to\qq$ sample, $b$-jets can be tagged with an efficiency of 80\% and a purity of 90\% while the corresponding efficiency and purity for tagging $c$-jets are 60\% and 60\%, respectively.



\subsection{Ongoing optimization}
\label{sec:detop}
The CEPC-v1 detector design is used as the reference detector for the studies summarized in this paper. A series of optimizations have been performed meanwhile, aiming to reduce power consumption and  construction cost and to improve the machine-detector interface while minimizing the impact on Higgs boson physics. An updated detector concept, CEPC-v4, has thus been developed. The CEPC-v4 has a smaller solenoidal field of 3~Tesla\footnote{For the $Z$ factory operation, the magnetic field may be reduced further to 2~Tesla to reduce the beam-field coupling and therefore achieve higher instantaneous luminosity.} and a reduced calorimeter dimensions along with fewer readout channels. In particular, the ECAL readout senor size is changed from $5\times 5$~mm$^2$ to $10\times 10$~mm$^2$. A new Time-of-Flight measurement capability is added to improve the  flavor physics potential.

The weaker magnetic field degrades momentum resolution for charged particles by 14\%, which translates directly into a degraded muon momentum resolution. The impact on other physics objects such as electrons, photons and jets are estimated to be small as the track momentum resolution is not a dominant factor for their performance. In parallel with the detector optimization, the accelerator design has chosen 240~GeV as the nominal center-of-mass energy for the Higgs factory. However, the simulation of CEPC-v1 assumes $\sqrt{s}=250$~GeV. The estimated precision of Higgs boson property measurements for CEPC-v1 operating at 250~GeV are therefore extrapolated to obtain those for CEPC-v4 at $\sqrt{s}=240$~GeV, as discussed Section~\ref{sec:v1tov4}.

\section{Theory and Monte Carlo Samples}
\label{sec:samples}

\subsection{Higgs boson production and decay}

\begin{figure*}[ht!]
\begin{center}
	\subfigure[]{\includegraphics[width=0.25\textwidth]{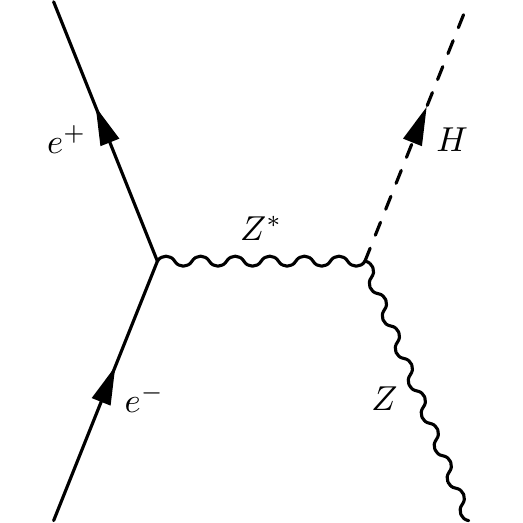}}
	\subfigure[]{\includegraphics[width=0.25\textwidth]{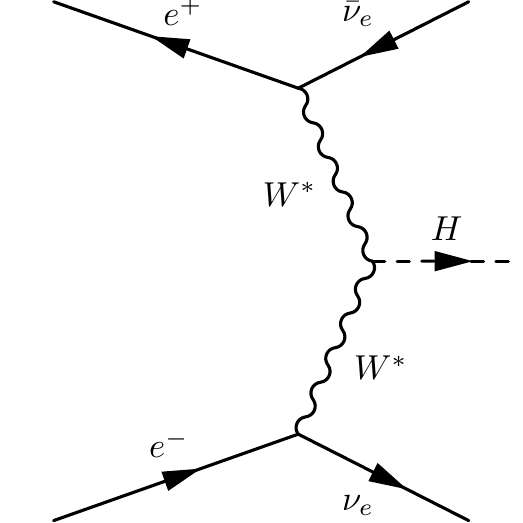}}
	\subfigure[]{\includegraphics[width=0.25\textwidth]{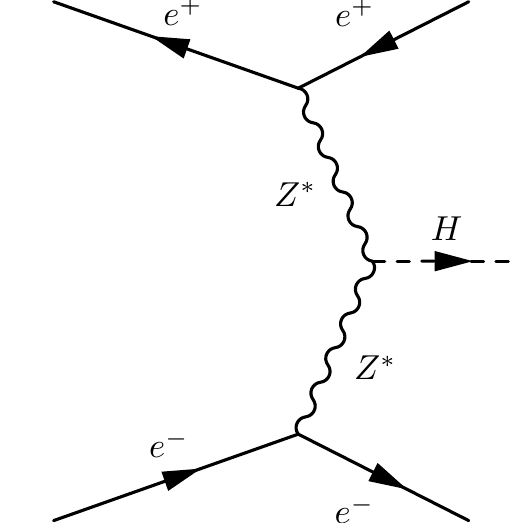}}
\figcaption{\label{fig:FeynmanDiagram}\small Feynman diagrams of the Higgs boson production processes at the CEPC: (a) $\ZH$, (b) $\vvH$ and (c) $\eeH$.}

\includegraphics[width=0.45\textwidth]{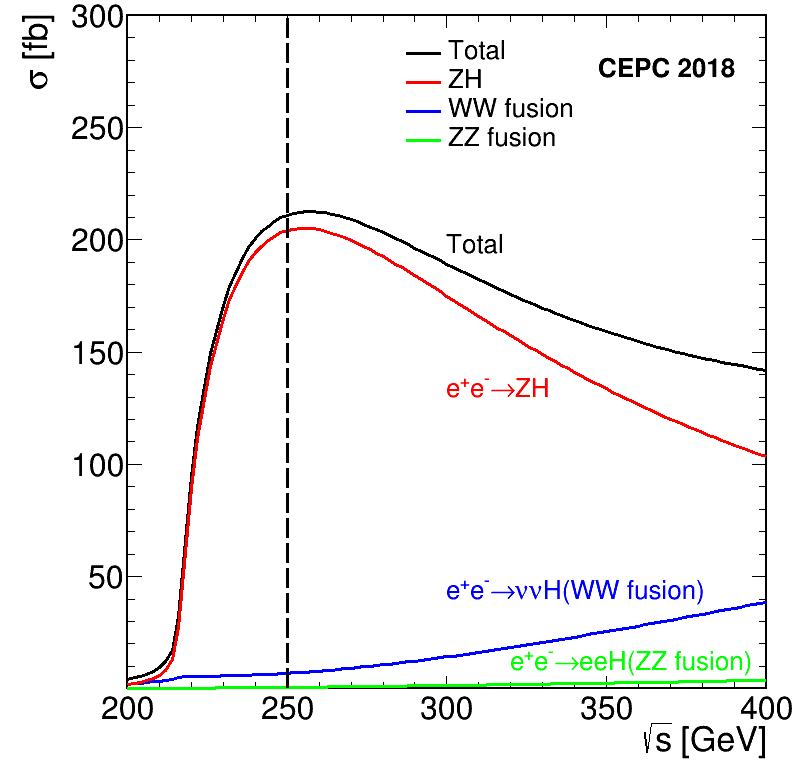}
 \figcaption{\small Production cross sections of $\ZH$ and $e^+e^-\to (\ee/\vv) H$ as functions of $\sqrt{s}$ for a 125 GeV SM Higgs boson. The vertical indicates $\sqrt{s}=250$~GeV, the energy assumed for most of the studies summarized in this paper.}  
\label{fig:HiggsXS}
\end{center}
\end{figure*}

Production processes for a 125~GeV SM Higgs boson at the CEPC operating at $\sqrt{s}\sim 240-250$~GeV are $\ZH$ ($ZH$ associate production or Higgsstrahlung), $\vvH$ ($W$ fusion) and $\eeH$ ($Z$ fusion) as illustrated in Fig.~\ref{fig:FeynmanDiagram}. In the following, the $W$ and $Z$ fusion processes are collectively referred to as vector-boson fusion (VBF) production.

The SM Higgs boson production cross sections as functions of center-of-mass energy are shown in Fig.~\ref{fig:HiggsXS}, assuming that the mass of the Higgs boson is 125~GeV. Similarly, the Higgs boson decay branching ratios and total width are shown in Table~\ref{tab:HiggsBR}.
As an $s$-channel process, the cross section of the $\ZH$ process reaches its maximum at $\sqrt{s}\sim 250$~GeV, and then decreases asymptotically as $1/{s}$.  The VBF process proceeds through $t-$channel exchange of vector bosons and its cross section increases logarithmically as $\ln^2(s/M_V^2)$. Because of the small neutral-current $Zee$ coupling, the VBF cross section is dominated by the $W$ fusion process. 

Numerical values of these cross sections at $\sqrt{s}=250$~GeV are listed in Table~\ref{tab:CrossSections}.  Because of the interference effects between $\ZH$ and $\vvH$ for the $Z\to\nu_e\bar{\nu}_e$ decay and between $\ZH$ and $\eeH$ for the $\Zee$ decay, the cross sections of these processes cannot be separated. The breakdowns in Fig.~\ref{fig:HiggsXS} and Table~\ref{tab:CrossSections} are for illustration only. The $\ZH$ cross section shown is from Fig.~\ref{fig:FeynmanDiagram}(a) only whereas the $\vvH$ and $\eeH$ cross sections include contributions from their interferences with the $\ZH$ process.

The CEPC as a Higgs boson factory is designed to deliver a combined integrated luminosity of $\abih$ to two detectors in 7 years. 
Over $10^6$ Higgs boson events will be produced during this period. 
The large statistics, well-defined event kinematics and clean collision environment will enable the CEPC to measure the Higgs boson production cross sections as well as its properties (mass, decay width and branching ratios, etc.) with precision far beyond those achievable at the LHC.
In contrast to hadron collisions, $e^+e^-$ collisions are unaffected by underlying events and pile-up effects. Theoretical calculations are less dependent on higher order QCD radiative corrections.  Therefore, more precise tests of theoretical predictions can be performed at the CEPC. 
The tagging of $\ZH$ events using the recoil mass method (see Section~\ref{sec:massXS}), independent of the Higgs boson decay, is unique to lepton colliders. It provides a powerful tool to perform model-independent measurements of the inclusive $\ee\to ZH$ production cross section, $\sigma(ZH)$, and of the Higgs boson decay branching ratios. Combinations of these measurements will allow for the determination of the total Higgs boson decay width and the extraction of the Higgs boson couplings to fermions and vector bosons. These measurements will provide sensitive probes to potential new physics beyond the SM. 

\begin{table*}[t!]
\begin{center}
\caption{\label{tab:HiggsBR}\small Standard model predictions of the decay branching ratios and total width of a 125~GeV Higgs boson~\cite{LHCCrossSectionPaper2011,LHCCrossSectionPaper2012,Heinemeyer:2013tqa}. The quoted uncertainties include contributions from both theoretical and parametric sources.} 
\begin{tabular}{clcrcrc}\hline\hline
& Decay mode   &\hspace*{0.5cm} & Branching ratio   &\hspace*{0.5cm} &  Relative uncertainty  & \\ \hline
& $\Hbb$         &&               57.7\%   &&  $+3.2\%,\, -3.3\%$   \\
& $\Hcc$         &&               2.91\%   &&  $+12\%,\, -12\%$     \\
& $\Htt$         &&               6.32\%   &&  $+5.7\%,\, -5.7\%$   \\
& $\Hmm$         &&  $2.19\times 10^{-4}$  &&  $+6.0\%,\, -5.9\%$   \\ \\[-3mm]

& $\HWW$         &&               21.5\%   &&  $+4.3\%,\, -4.2\%$   \\
& $\HZZ$         &&               2.64\%   &&  $+4.3\%,\, -4.2\%$   \\
& $\Hyy$         &&  $2.28\times 10^{-3}$  &&  $+5.0\%,\, -4.9\%$   \\
& $\HZy$         &&  $1.53\times 10^{-3}$  &&  $+9.0\%,\, -8.8\%$   \\
& $\Hgg$         &&               8.57\%   &&  $+10\%,\, -10\%$      \\ \\[-3mm]\hline

& $\Gamma_H$     &&    4.07 MeV            &&  $+4.0\%,\, -4.0\%$  \\ \hline\hline
\end{tabular}
\end{center}
\end{table*}

\begin{table*}[t!]
\begin{center}
\caption{\label{tab:CrossSections}\small Cross sections of the Higgs boson production and other SM processes at $\sqrt{s}=250$~GeV and numbers of events expected in 5.6~ab$^{-1}$. 
Note that there are interferences between the same final states from different processes after the $W$ or $Z$ boson decays, see text. 
With the exception of the Bhabha process, the cross sections are calculated using the Whizard program~\cite{Kilian:2007gr}. The Bhabha cross section is calculated using the BABAYAGA event generator~\cite{CarloniCalame:2003yt} requiring final-state particles to have $|\cos\theta|<0.99$. Photons, if any, are required to have $E_\gamma>0.1$~GeV and $|\cos\theta_{e^\pm\gamma}|<0.99$. } 
\begin{tabular}{clcc}\hline\hline
& Process  & \hcp Cross section \hcp  & \hcp Events in 5.6 ab$^{-1}$ \hcp \tsp \\ \hline
&\multicolumn{3}{c}{Higgs boson production, cross section in fb} \\ \hline
&$\ZH$     &       204.7           &   $1.15\times{10^{6}}   $               \\
&$\vvH$    &        6.85           &   $3.84\times{10^{4}}   $               \\
&$\eeH$    &        0.63           &   $3.53\times{10^{3}}   $               \\ \hline
&Total     &       212.1           &   $1.19\times{10^{6}}   $               \\[3mm]

&\multicolumn{3}{c}{Background processes, cross section in pb} \\ \hline
&$e^+e^-\to \ee\,(\gamma)$ (Bhabha)             & 850   &  $4.5\times{10^{9}}$  \\
&$e^+e^-\to \qq\,(\gamma)$                      & 50.2  &  $2.8\times{10^{8}}$  \\
&$e^+e^-\to \mm\,(\gamma)$ [or $\tau^+\tau^-\,(\gamma)$]&4.40  &  $2.5\times{10^{7}}$  \\
&$e^+e^-\to WW$                                 & 15.4  &  $8.6\times{10^{7}}$  \\ 
&$e^+e^-\to ZZ$                                 & 1.03  &  $5.8\times{10^{6}}$  \\ 
&$e^+e^-\to \ee Z$                              & 4.73  &  $2.7\times{10^{7}}$  \\
&$e^+e^-\to e^+\nu W^-/e^-\bar{\nu}W^+$         & 5.14  &  $2.9\times{10^{7}}$  \\
\hline\hline
\end{tabular}
\end{center}
\end{table*}

\subsection{Background processes}

\begin{figure*}[t!]
	\begin{center}
		\includegraphics[width=0.6\textwidth]{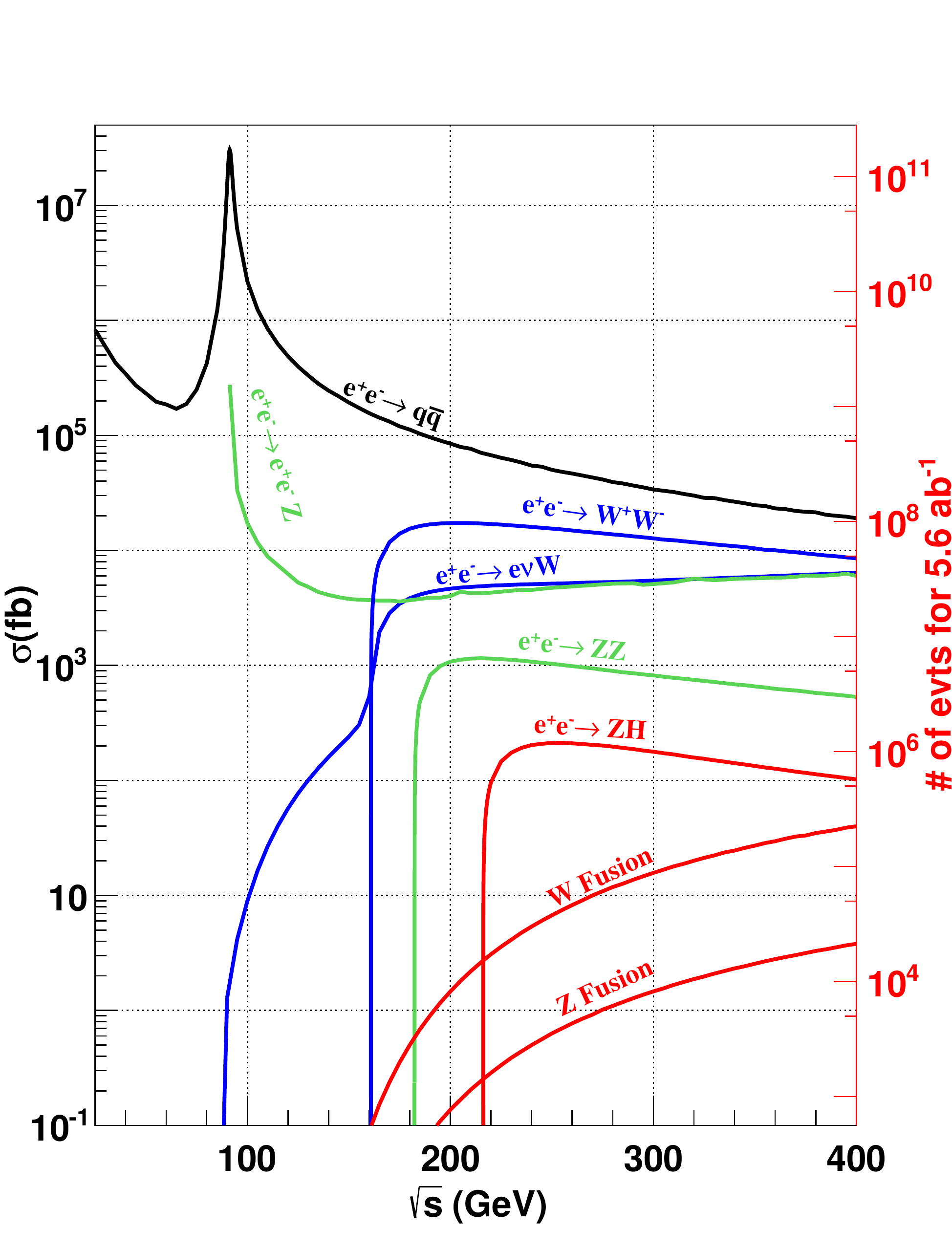}
		\figcaption{\label{fig:SMProcesses}\small Cross sections of main SM processes of $e^+e^-$ collisions  as functions of center-of-mass energy $\sqrt{s}$ obtained from the Whizard program~\cite{Kilian:2007gr}. The calculations include initial-state radiation (ISR). 
			The $W$ and $Z$ fusion processes refer to $\ee \to \vv H$ and $\ee\to \ee H$ production, respectively. Their numerical values at $\sqrt{s}=250$~GeV can be found in Table~\ref{tab:CrossSections}. }
	\end{center}
\end{figure*}

Apart from the Higgs boson production, other SM processes include $\ee\to\ee$ (Bhabha scattering), $\ee\to Z\gamma$ (initial-state radiation return), $\ee\to WW/ZZ$ (diboson) as well as the single boson production of  $\ee\to \ee Z$ and $e^+e^-\to e^+\nu W^-/e^-\bar{\nu}W^+$. Their cross sections and expected numbers of events for an integrated luminosity of $\abih$ at $\sqrt{s}=250$~GeV are shown in Table~\ref{tab:CrossSections} as well. The energy dependence of the cross sections for these and the Higgs boson production processes are shown in Fig.~\ref{fig:SMProcesses}. 
Note that many of these processes can lead to identical final states and thus can interfere. For example,  $e^+e^-\to e^+\nu_e W^-\to e^+\nu_e e^-\bar{\nu}_e$ and  $e^+e^-\to e^+e^- Z\to e^+e^-\nu_e\bar{\nu}_e$ have the same final state after the decays of the $W$ or $Z$ bosons. Unless otherwise noted, these processes are simulated together to take into account interference effects for the studies presented in this paper. The breakdowns shown in the table and figure assume stable $W$ and $Z$ bosons, and thus are, therefore, for illustration only.

Along with $1.2\times 10^{6}$ Higgs boson events, $5.8\times{10^6}$ $ZZ$, $8.6\times{10^7}$ $WW$ and $2.8\times{10^8}$ $q\bar{q}(\gamma)$ events will be produced. Though these events are  backgrounds to Higgs boson events, they are important for the calibration and characterization of the detector performance and for the measurements of electroweak parameters.

\subsection{Event generation and simulation}

The following software tools have been used to generate events, simulate detector responses and reconstruct simulated events. 
A full set of SM samples, including both the Higgs boson signal and SM background events, are generated with {\sc Whizard}~\cite{Kilian:2007gr}. 
The generated events are then processed with MokkaC~\cite{MoradeFreitas:2002kj}, the official CEPC simulation software based on the framework used for ILC studies~\cite{ilcsoft}. Limited by computing resources, background samples are often pre-selected with loose generator-level requirements or processed with fast simulation tools.

All Higgs boson signal samples and part of the leading background samples are processed with Geant4~\cite{geant4} based full detector simulation and reconstruction. The rest of backgrounds are simulated with a dedicated fast simulation tool, where the detector acceptance, efficiency, intrinsic resolution for different physics objects are parametrized. Samples simulated for ILC studies~\cite{Asner:2013psa} are used for cross checks of some studies.

\section{Higgs Boson Tagging using Recoil Mass}
\label{sec:massXS}

\begin{figure*}[t!]
\begin{center}
  \subfigure[]{\includegraphics[width=0.40\textwidth]{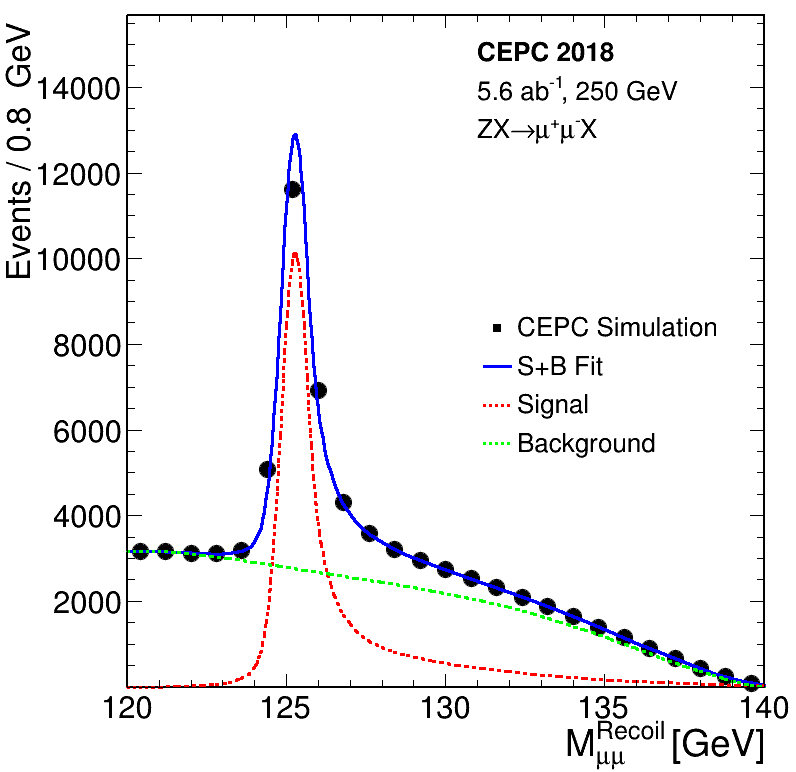}}
  \subfigure[]{\includegraphics[width=0.40\textwidth]{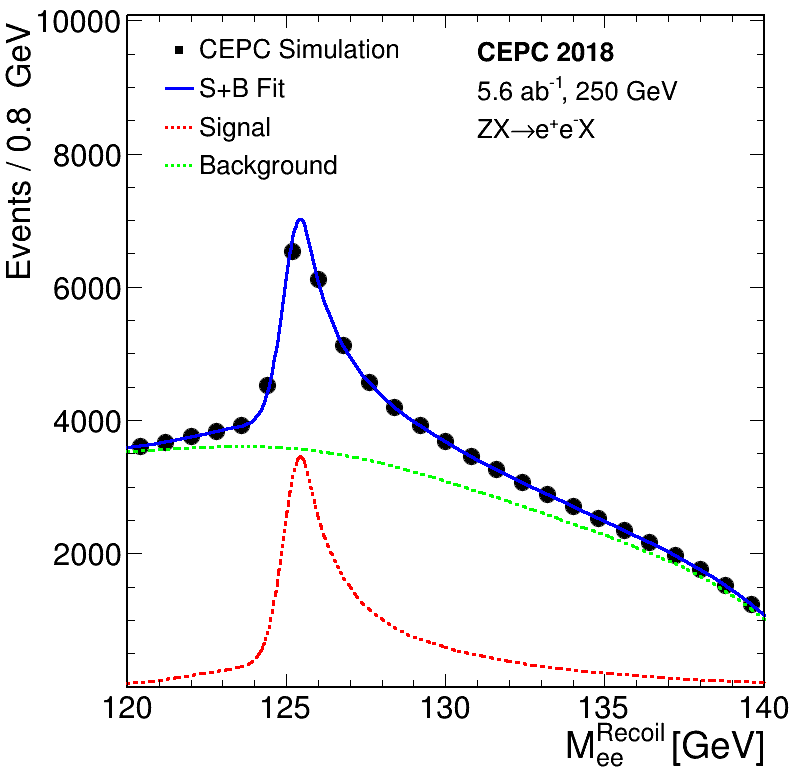}}
\figcaption{\label{fig:mrecoil_ll}\small The inclusive recoil mass spectra of $\ZX$ candidates for (a) $\Zmm$ and (b) $\Zee$. No attempt to identify $X$ is made. The markers and their uncertainties (too small to be visible) represent expectations from a CEPC dataset of $\abih$, whereas the solid blue curves are the fit results. The dashed curves are the signal and background components.}
\end{center}
\end{figure*}

Perhaps the most striking difference between hadron-hadron and $\ee$ collisions is that electrons are fundamental particles whereas hadrons are composite. Consequently the energy of $\ee$ collisions is known. Therefore through conservation laws, the energy and momentum of a Higgs boson can be inferred from other particles in an event without examining the Higgs boson itself. For a Higgsstrahlung event where the $Z$ boson decays to a pair of visible fermions ($ff$),  the mass of the system recoiling against the $Z$ boson, commonly known as the recoil mass, can be calculated assuming the event has a total energy $\sqrt{s}$ and zero total momentum:
\begin{equation} \label{eq:mrecoil}
\mrecoil^2 = (\sqrt{s}-E_{ff})^2 - p^2_{ff} = s-2E_{ff}\sqrt{s}+m_{ff}^2.
\end{equation}
Here, $E_{ff}$, $p_{ff}$ and $m_{ff}$ are, respectively, the total energy, momentum and invariant mass of the fermion pair. The $\mrecoil$ distribution should show a peak at the Higgs boson mass $m_H$ for $\ZH$ and $\ee\to \ee H$ processes, and is expected to be smooth without a resonance structure for other processes in the mass region around 125~GeV.

Two important measurements of the Higgs boson can be performed from the $\mrecoil$ mass spectrum. The Higgs boson mass can be measured from the peak position of the resonance. The width of the resonance is dominated by the beam energy spread (including ISR effects) and energy/momentum resolution of the detector if the  Higgs boson width is only 4.07~MeV as predicted in the SM. The best precision of the mass measurement can be achieved from the leptonic $\Zll\, (\ell=e,\mu)$ decays. The height of the resonance is proportional to the Higgs boson production cross section $\sigma(ZH)$\footnote{For the $Z\to\ee$ decay, there will be a small contribution from the $\ee\to\ee H$ production.}. By fitting the $\mrecoil$ spectrum, the $\ZH$ event yield, and therefore $\sigma(ZH)$, can be extracted, independent of the Higgs boson decays. The Higgs boson branching ratios can then be determined by studying Higgs boson decays in selected $\ZH$ candidates. The recoil mass spectrum has been investigated for both leptonic and hadronic $Z$ boson decays as presented below.
  
\subsection{$\Zll$} 

The leptonic $Z$ boson decay is ideal for studying the recoil mass spectrum of the $\ZX$ events. The decay is easily identifiable and the lepton momenta can be precisely measured.  Figure~\ref{fig:mrecoil_ll} shows the reconstructed recoil mass spectra of $\ZX$ candidates for the $\Zmm$ and $\Zee$ decay modes. The analyses are based on the full detector simulation for the signal events and on the fast detector simulation for background events. They are performed with event selections entirely based on the information of the two leptons, independent of the final states of Higgs boson decays. This approach is essential for the measurement of the inclusive $\ZH$ production cross section and the model-independent determination of the Higgs boson branching ratios. The SM processes with at least 2 leptons in their final states are considered as backgrounds.

The event selection of the $\Zmm$ decay mode starts with the requirement of a pair of identified muons with opposite charges. Events must have the dimuon invariant mass in the range of 80--100~GeV and the recoil mass between 120~GeV and 140~GeV. The muon pair is required to have its transverse momentum larger than 20~GeV, and its opening angle smaller than $175^\circ$. A Boosted Decision Tree~(BDT) technique is employed to enhance the separation between signal and background events. 
The BDT is trained using the invariant mass, transverse momentum, polar angle and acollinearity of the dimuon system. 
Leading background contributions after the selection are from $ZZ$, $WW$ and $Z\gamma$ events. As shown in Fig.~\ref{fig:mrecoil_ll}(a), the analysis has a good signal-to-background ratio. The long high-mass tail is largely due to the initial-state radiation. 

Compared to the analysis of the $\Zmm$ decay, the analysis of the $\Zee$ decay suffers from additional and large background contributions from Bhabha scattering and single boson production. A cut based event selection is performed for the $\Zee$ decay. The electron-positron pair is required to have its invariant mass in the range of 86.2--96.2~GeV and its recoil mass between 120~GeV and 150~GeV. Additional selections based on the kinematic variables of the electron-positron system, the polar angles and the energies of the selected electron and positron, are applied. Events from $e^{+}e^{-}\to \ee(\gamma),\, e^+\nu W^-\,(e^-\bar{\nu}W^+),\, \ee Z$ production are the dominant backgrounds after the selection. 
The recoil mass distribution of the selected events is shown in Fig.~\ref{fig:mrecoil_ll}(b). 

While event selections independent of the Higgs boson decays are essential for the model-independent measurement of $\sigma(ZH)$, additional selection criteria using the Higgs boson decay information can, however, be applied to improve the Higgs boson mass measurement. This will be particularly effective in suppressing the large backgrounds from Bhabha scattering and single $W$ or $Z$ boson production for the analysis of the $\Zee$ decay. These improvements are not implemented in the current study.

\subsection{$\Zqq$}
The recoil mass technique can also be applied to the hadronic $Z$ boson decays ($\Zqq$) of the $\ZX$ candidates. This analysis benefits from a larger $\Zqq$ decay branching ratio, but suffers from the fact that jet energy resolution is worse than the track momentum and electromagnetic energy resolutions. In addition,  ambiguity in selecting jets from the $\Zqq$ decay, particularly in events with hadronic decays of the Higgs boson, can degrade the analysis performance and also introduce model-dependence to the analysis. Therefore, the measurement is highly dependent on the performance of the particle-flow reconstruction and the jet clustering algorithm.

Following the same approach as the ILC study~\cite{Haddad:2014fma}, an analysis based on the fast simulation has been performed. After the event selection,  main backgrounds arise from $Z\gamma$ and $WW$ production. 
Compared with the leptonic decays, the signal-to-background ratio is considerably worse and the recoil mass resolution is significantly poorer. 

\subsection{Measurements of $\sigma(ZH)$ and $m_H$}
\label{sec:ZH}

Both the inclusive $\ZH$ production cross section $\sigma(ZH)$ and the Higgs boson mass $m_H$ can be extracted from fits to the recoil mass distributions of the $e^+e^-\to Z+X \to\ell^+\ell^-/\qq + X$ candidates. For the leptonic $\Zll$ decays, the recoil mass distribution of the signal process $\ee\to ZH$ (and $\ee\to \ee H$ in case of the $Z\to\ee$ decay) is modeled with a Crystal Ball function~\cite{Oreglia:1980cs} whereas the total background is modeled with a polynomial function. As noted above, the recoil mass distribution is insensitive to the intrinsic Higgs boson width should it be as small as predicted by the SM. The Higgs boson mass can be determined with precision of 6.5~MeV and 14~MeV from the $\Zmm$ and $\Zee$ decay modes, respectively. After combining all channels, an uncertainty of 5.9~MeV can be achieved.

The process $e^+e^-\to Z+X\to \qq+X$ contributes little to the precision of the $m_H$ measurement due to the poor $\Zqq$ mass resolution, but dominates the sensitivity of the $\ZH$ cross section measurement because of the large statistics. A simple event counting analysis shows that the expected relative precision on $\sigma(ZH)$ is 0.61\%. In comparison, the corresponding relative precision from the $\Zee$ and $\Zmm$ decays is estimated to be 1.4\% and 0.9\%, respectively. The combined relative precision of the three measurements is 0.5\%. Table~\ref{tab:ZH} summarizes the expected precision on $m_H$ and $\sigma(ZH)$ from a CEPC dataset of 5.6~ab$^{-1}$. 

\begin{center} \footnotesize
\tabcaption{\label{tab:ZH}\small Estimated measurement precision for the Higgs boson mass $m_H$ and the $\ee\to ZH$ production cross section $\sigma(ZH)$ from a CEPC dataset of 5.6~ab$^{-1}$.}
\begin{tabular}{ccc}\hline\hline
\hcp $Z$ decay mode \hcp & $\Delta m_H$ (MeV) & \hcp $\Delta\sigma(ZH)/\sigma(ZH)$ \hcp  \\ \hline
$\ee$           &  14     &   1.4\%               \\
$\mm$           &  6.5    &   0.9\%               \\ 
$\qq$           &  $-$    &   0.6\%      \\ \hline
Combination     &  5.9    &   0.5\%      \\ \hline\hline
\end{tabular}
\end{center}



\section{Analyses of Individual Decay Modes}
\label{sec:decays}

Different decay modes of the Higgs boson can be identified through their unique signatures, leading to the measurements of production rates for these decays. For the $\ZH$ production process in particular, the candidate events can be tagged from the visible decays of the $Z$ bosons, the Higgs boson decays can then be probed by studying the rest of the events. Simulation studies of the CEPC baseline conceptual detector have been performed for the Higgs boson decay modes of $\Hbcg$, $\HWW$, $\HZZ$, $\HZy$, $\Htt$, $\Hmm$ and Higgs boson to invisible particles ($\Hinv$). The large number of the decay modes of the $H$, $W$ and $Z$ boson as well as the $\tau$-lepton leads to a very rich variety of event topologies. This complexity makes it impractical to investigate the full list of final states stemming from the Higgs boson decays. Instead, a limited number of final states of individual Higgs boson decay modes have been considered. For some decay modes, the chosen final states may not be the most sensitive ones, but are nevertheless representatives of the decay mode. In most cases, the dominant backgrounds come from the SM diboson production and the single $Z$ production with the initial and final state radiation.

The studies are optimized for the dominant $ZH$ process, however, the $\vvH$ and $\eeH$ processes are included whenever applicable. The production cross sections of the individual decay modes, $\sigma(ZH)\times\BR$, are extracted. These measurements combined with the inclusive $\sigma(ZH)$ measurement discussed in Section~\ref{sec:massXS}, will allow the determination of the Higgs boson decay branching ratios in a model-independent way. 

In this section, the results of the current CEPC simulation studies of different Higgs boson decay modes are summarized. The studies are based on the CEPC-v1 detector concept and $\ee$ collisions at $\sqrt{s}=250$~GeV.  
The expected relative precision from a CEPC dataset of 5.6~ab$^{-1}$ on the product of the $ZH$ cross section and the Higgs boson decay branching ratio, $\sigma(ZH)\times {\rm BR}$, are presented. Detailed discussions of individual analyses are beyond the scope of this paper and therefore only their main features are presented. 
For the study of a specific Higgs boson decay mode, the other decay modes of the Higgs boson often contribute as well. Those contributions are fixed to their SM expectations and are included as backgrounds unless otherwise noted. However, for the combination of all the decay modes under study, they are allowed to vary within the constraints of the measurements of those decays, see Section~\ref{sec:combinations}.

In addition to the invariant and recoil mass, 
two other mass observables, visible mass and missing mass, are often used in analyses described below. They are defined, respectively, as the invariant mass and recoil mass of all visible experimental objects such as charged leptons, photons and jets, {\it i.e.} practically all particles other than neutrinos.


\vspace*{0.3cm}
\subsection{$\Hbcg$}  
\label{sec:hbcg}

For a SM Higgs boson with a mass of 125 GeV, nearly 70\% of all Higgs bosons decay  into a pair of jets: $b$-quarks (57.7\%), $c$-quarks (2.9\%) and gluons (8.6\%). 
While the $\Hbb$ decay has recently been observed at the LHC~\cite{Aaboud:2018zhk,Sirunyan:2018kst}, the $\Hcc$ and $\Hgg$ decays are difficult, if not impossible, to be identified there due to large backgrounds. In comparison, all these three decays can be isolated and studied at the CEPC. The $\Hcc$ decay is likely the only process for studying Higgs boson coupling to the second-generation quarks at collider experiments. 
The identification of $\Hbcg$ decays poses critical challenges to the CEPC detector performance, particularly the ability to tag $b$- and $c$-quark jets against light-flavored jets ($u,d,s,g$). Thus they are good benchmarks for the design and optimization of the jet flavor tagging performance of the CEPC detector.

Studies are performed in detail for $\ZH$ production with the leptonic decays of the $Z$ bosons. The contribution from the $Z$-fusion process of $\eeH$ is included in the $\ee\to ZH\to \ee H$ study. The analysis is based on full simulation for the Higgs boson signal samples and fast simulation for the $\ell^+\ell^- \qq$ background samples. After selecting two leading leptons with opposite charge, the rest of the reconstructed particles are clustered into two jets to form a hadronically decaying Higgs boson candidate, whose invariant mass
is required to be between  75~GeV and 150~GeV.  
The dilepton invariant mass is required to be within 70--110~GeV for the $\ee$ channel and 81--101~GeV for the $\mm$ channel. Moreover, the dilepton system must have its transverse momentum in the range  10--90~GeV and its recoil mass between 120~GeV and 150~GeV. In addition, a requirement on the polar angle of the Higgs boson candidate, $\vert \cos\theta_{H}\vert < 0.8$, is applied.

\begin{figure*}[t!]
	\begin{center}
		\subfigure[]{\includegraphics[width=0.35\textwidth]{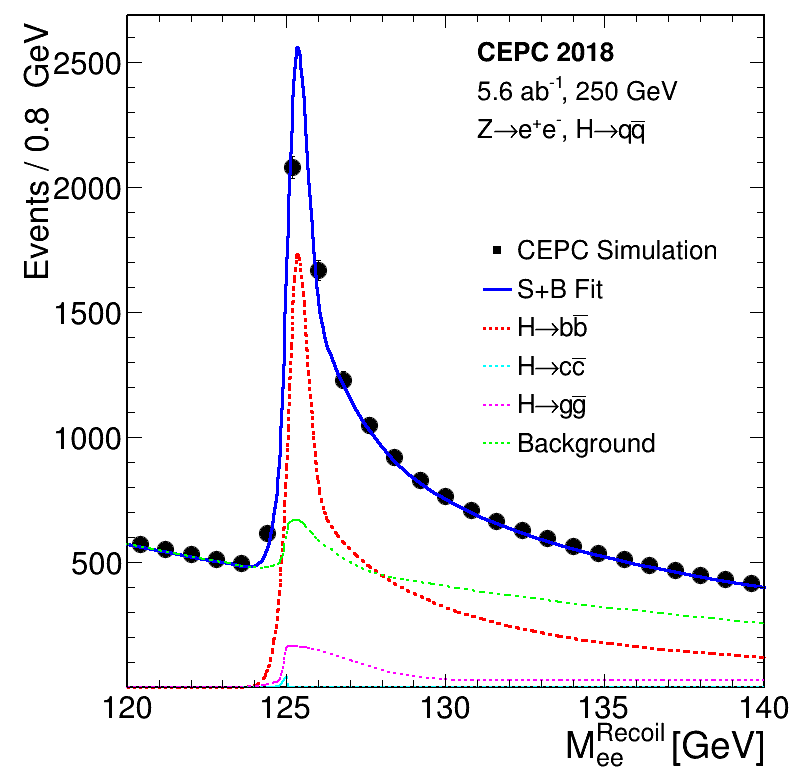}}
		\subfigure[]{\includegraphics[width=0.35\textwidth]{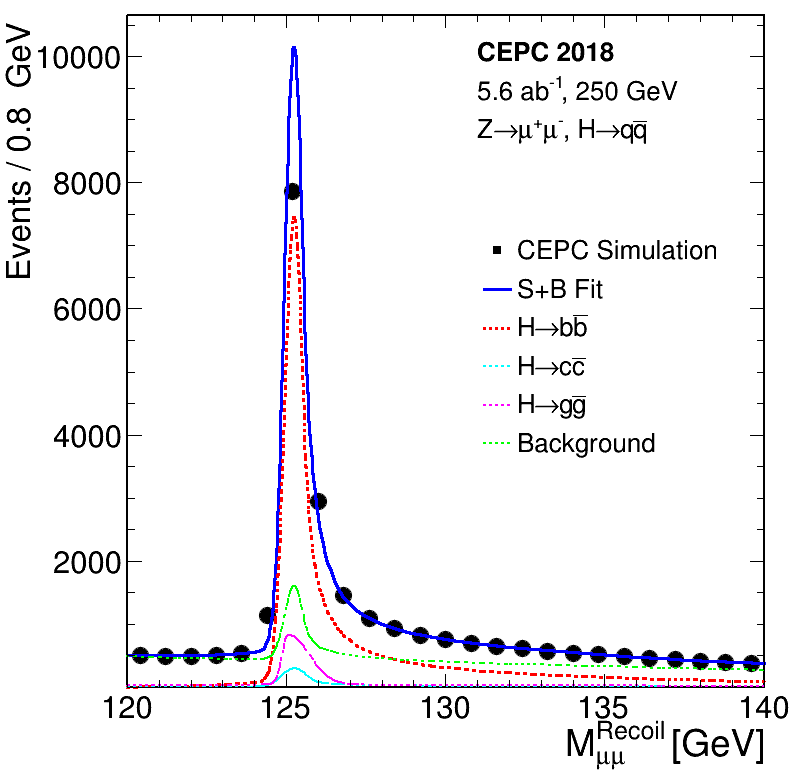}}
		\subfigure[]{\includegraphics[width=0.35\textwidth]{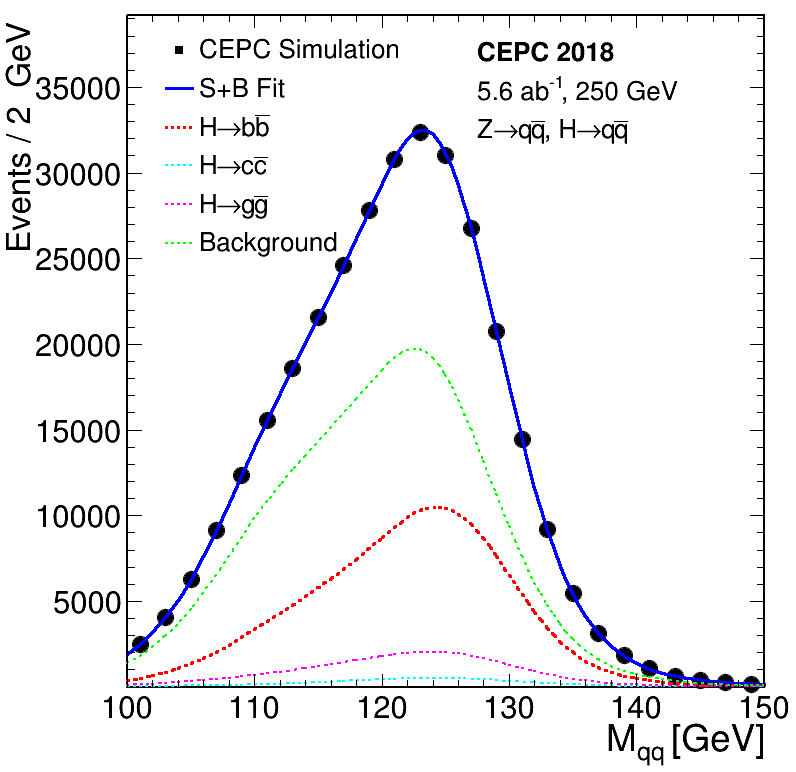}}
		\subfigure[]{\includegraphics[width=0.35\textwidth]{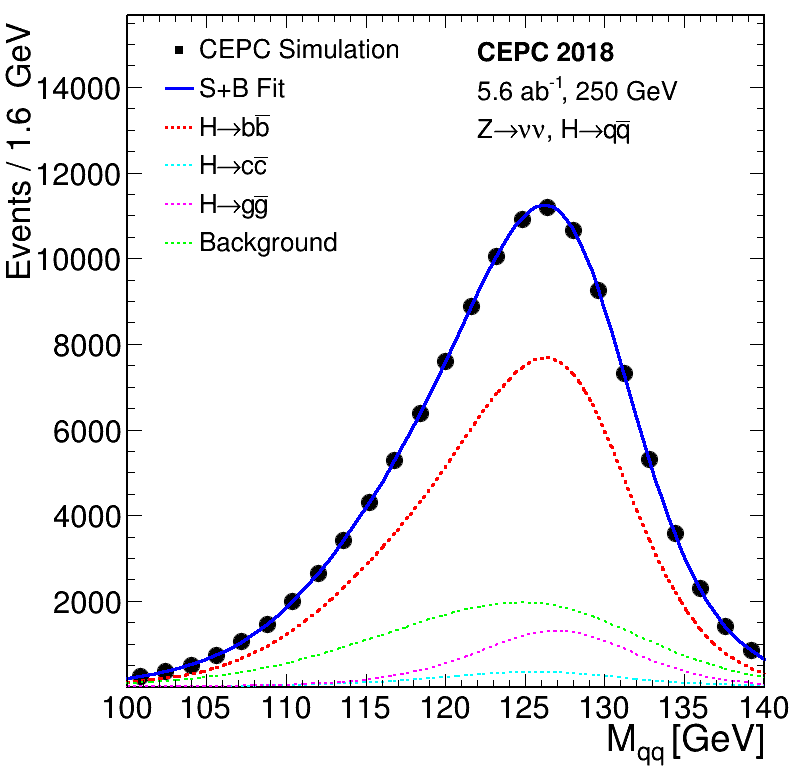}}
		\figcaption{\small $ZH$ production with $\Hbcg$: the recoil mass distributions of (a) $\Zee$ and (b) $\Zmm$; the dijet mass distributions of Higgs boson candidates for (c) $\Zqq$ and (d) $\Zvv$. The markers and their uncertainties represent expectations from a CEPC dataset of $\abih$ whereas the solid blue curves are the fit results. The dashed curves are the signal and background components. Contributions from other decays of the Higgs boson are included in the background.}
		\label{fig:hbb_new1}
	\end{center}
\end{figure*}

In order to identify the flavors of the two jets of the Higgs boson candidate, variables $L_B$ and $L_C$ are constructed using information such as those from LCFIPlus jet flavor tagging algorithm.  
The values of $L_{B}$ ($L_{C}$) are close to one if both jets are originated from $b\,(c)$ quarks and are close to zero if both have light-quark or gluon origins. 
An unbinned maximum likelihood fit  to the $M_{\rm recoil}$, $L_B$ and $L_C$ distributions of candidate events is used to extract the individual signal yields of the $\Hbb$, $\Hcc$ and $\Hgg$ decay modes.  The total probability density function (PDF) is the sum of signal and background
components. For signals, their $M_{\rm recoil}$ PDFs are modeled by Crystal Ball functions~\cite{Oreglia:1980cs} with small exponential tails.  The background PDF is taken as a sum of two components: a background from Higgs boson decays to other final states such as $WW$ and $ZZ$, and a combinatorial background from other sources, dominated by the $\ee\to ZZ\to \ell\ell \qq$ production. The background from other Higgs boson decay channels has the same $M_{\rm recoil}$ PDF
as the signals. The $M_{\rm recoil}$ distribution of the combinatorial background is modeled by a second order polynomial. 
The PDFs of the signal $L_B$ and $L_C$ distributions are described by two dimensional histograms, taken from the MC simulated events. The $L_B$ and $L_C$ distributions of both background components are modeled by 2-dimensional histogram PDFs based on the MC simulation. The dilepton recoil mass distributions of the simulated data and the fit results  
are shown in Fig.~\ref{fig:hbb_new1}(a,b).  
The estimated relative statistical precision of the measurements of $\sigma(ZH)\times \BR(\Hbcg)$ are listed in Table~\ref{tab:chan_2j}. 

		\begin{center}
			\tabcaption{\label{tab:chan_2j}\small Expected relative precision on $\sigma(ZH)\times{\rm BR}$ for the $H\to b\bar{b},\,c\bar{c}$ and $gg$ decays from a CEPC dataset of $\abih$. }
			\begin{tabular}{clccc}  \hline\hline
				$Z$ decay mode & $H \to b\bar{b}$  & $H \to c\bar{c}$ &$H \to gg$ \\
				\hline
				$\Zee$    & 1.3\%  & 12.8\%  & 6.8\%  \\
				$\Zmm$    & 1.0\%  & 9.4\%   & 4.9\%  \\
				$\Zqq$    & 0.5\%  & 10.6\%  & 3.5\%   \\
				$\Zvv$    & 0.4\%  & 3.7\%   & 1.4\%   \\   \hline
				Combination  & 0.3\%  & 3.1\%  & 1.2\%  \\
				\hline\hline
			\end{tabular}
		\end{center}

Table~\ref{tab:chan_2j} also includes the results of the $\Zvv$ and $\Zqq$ decays. For the $\Zqq$ final state, events are clustered into four jets and the mass information of jet pairs are used to select the Higgs and $Z$ boson candidates. 
In addition to $ZZ$, $WW$ is also a major background for this analysis, particularly for the $H\to\cc$ and $H\to gg$ decays.  
As for the $\Zvv$ final state, events are clustered into two jets are to form the Higgs boson candidate, the invisibly decaying $Z$ boson is inferred from the missing mass of the event. Fits similar to the one used in the analysis of the $Z\to \ell^+\ell^-$ channel  is subsequently performed to statistically separate the $H\to\bb,\,\cc$ and $gg$ decay components.  
The simulated data and the fitted dijet mass distributions of the Higgs boson candidates are shown in Fig.~\ref{fig:hbb_new1}(c,d) for $\Zqq$ and $\Zvv$.

Combining all $Z$ boson decay modes studied, a relative statistical precision for $\sigma(ZH)\times\mathrm{BR}$ of 0.3\%, 3.3\% and 1.3\% can be achieved for the $H\to\bb,\,\cc$ and $gg$ decays, respectively.

\vspace*{0.3cm}
\subsection{$H\to WW^*$} 
\label{sec:hww}

For a 125~GeV SM Higgs boson, the $\HWW$ decay has the second largest branching ratio of 21.5\%~\cite{Heinemeyer:2013tqa}.  The sensitivity of the $\sigma(ZH)\times{\rm BR}(\HWW)$ measurement is estimated by combining results from the studies of a few selected final states (Table~\ref{tab:hww}) 
of the $\HWW$ decay of $ZH$ production. SM diboson production is the main background source in all cases. 

\begin{figure*}[t!]
\begin{center}
\subfigure[]{\includegraphics[width=0.40\textwidth]{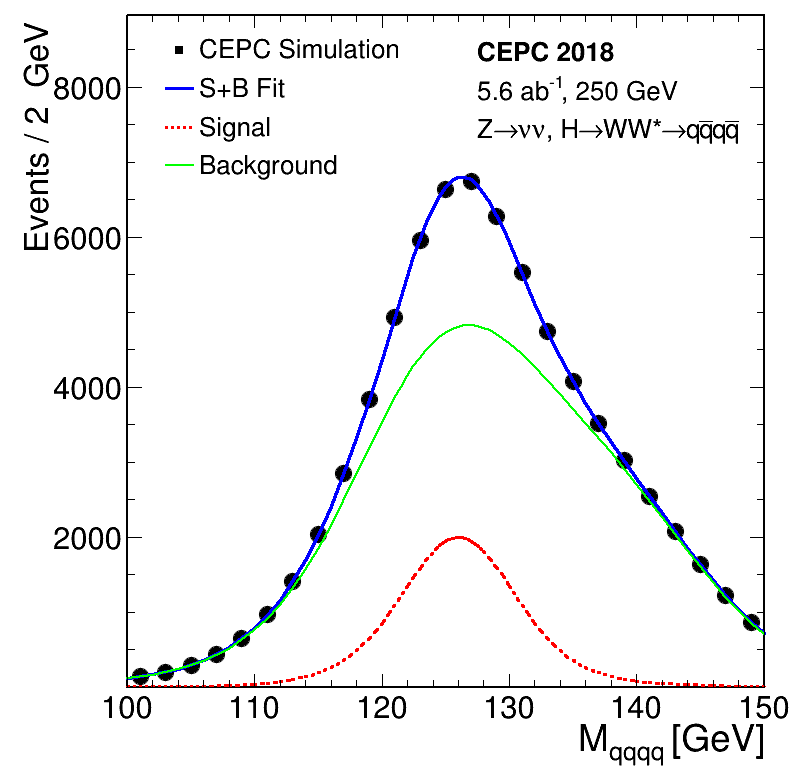}}
\subfigure[]{\includegraphics[width=0.40\textwidth]{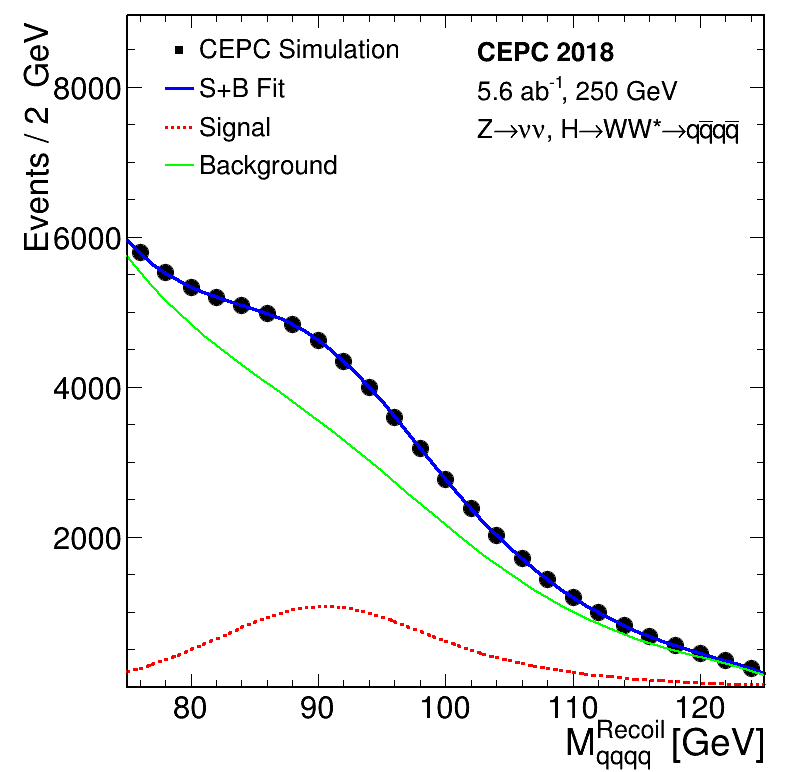}}
\figcaption{\small $ZH$ production with $\Zvv$ and $\HWW\to \qq\qq$: distributions of (a) the visible mass and (b) the missing mass of selected events. The markers and their uncertainties represent the expected number of events in a CEPC dataset of $\abih$, whereas the solid blue curves are the fit results. The dashed curves are the signal and background components. Contributions from other decays of the Higgs boson are included in the background.}
		\label{fig:hww}
	\end{center}
\end{figure*}

For $\Zll$, the $\HWW$ decay final states studied are $\ell\nu\ell'\nu$
and $\ell\nu \qq$. The $ZH$ candidate events are selected by requiring the dilepton 
invariant mass in the range of 80--100~GeV and their recoil mass in 120--150~GeV. 
For $\Zvv$, the $\ell\nu \qq$ and $\qq\qq$ final states are considered for the $\HWW$ decay. The presence of neutrinos in the event results in large missing mass, which is required to be in the range of 75--140 (75--150) GeV for the $\ell\nu \qq$ ($\qq\qq$) final state. The total visible mass of the event must be in the range of 100--150 GeV for both $\ell\nu \qq$ and $\qq\qq$ final states. In addition, the total transverse momentum of the visible particles must be in the range of 20--80 GeV. Additional requirements are applied to improve the signal-background separations. 
For $\Zqq$, the $\HWW\to\qq\qq$ decay is studied. Candidate events are reconstructed into 6 jets. Jets from $\Zqq$, $\Wqq$ and $\HWW\to\qq\qq$ decays are selected by minimizing the $\chi^2$ of their mass differences to the masses of $Z$, $W$ and $H$ boson. Figure~\ref{fig:hww} shows the visible and missing mass distributions after the selection of the $\Zvv$ and $\HWW\to \qq\qq$ final state.

The relative precision on $\sigma(ZH)\times\mathrm{BR}(\HWW)$ from the decay final states studied is summarized in Table~\ref{tab:hww}. 
 
The combination of these decay final states leads to a precision of 0.9\%. This is likely a conservative estimate as many of the final states of the $\HWW$ decay remain to be explored. Including these missing final states will no doubt improve the precision.
\begin{center}
	\begin{minipage}{0.45\textwidth}
		\begin{center}
			\tabcaption{\label{tab:hww}\small Expected relative precision on the $\sigma(ZH)\times{\rm BR}(\HWW)$ measurement from a CEPC dataset of $\abih$.}
			\begin{tabular}{llrr} \hline\hline
				\multicolumn{2}{c}{$ZH$ final state}   & Precision          \\    \hline
				$\Zee$      & $\HWW\to \ell\nu\ell'\nu,\, \ell\nu\qq$   &  2.6\%   \\
				$\Zmm$      & $\HWW\to \ell\nu\ell'\nu,\, \ell\nu\qq$   &  2.4\%   \\
				$\Zvv$      & $\HWW\to \ell\nu \qq, \qq\qq$             &  1.5\%   \\
				$\Zqq$      & $\HWW\to \qq\qq$                          &  1.7\%   \\
				\hline
				\multicolumn{2}{c}{Combination}                         &  0.9\%   \\ 
				\hline\hline
			\end{tabular}
		\end{center}
	\end{minipage}
\end{center}

\vspace*{0.3cm}
\subsection{$\HZZ$}
\label{sec:hzz}

\begin{figure*}[t!]
\begin{center}
\subfigure[]{\includegraphics[width=0.4\textwidth]{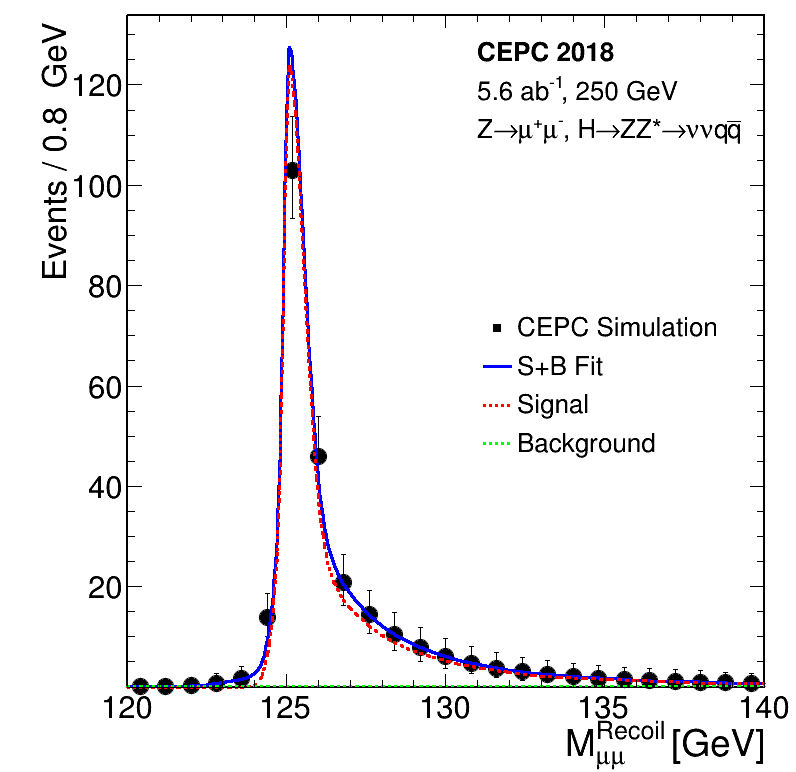}}
\subfigure[]{\includegraphics[width=0.4\textwidth]{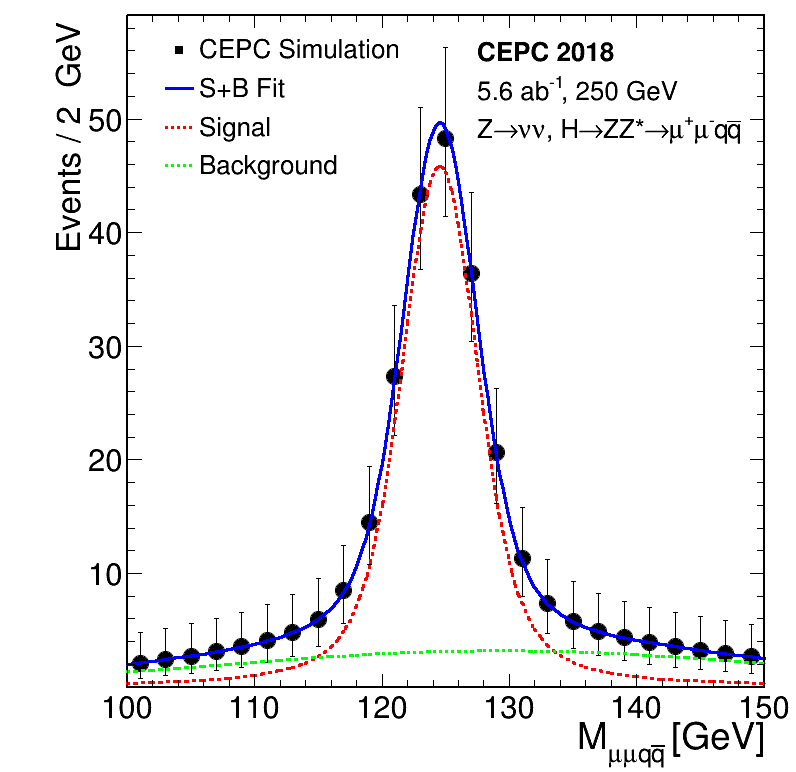}}
\figcaption{\small\label{fig:hzz} $ZH$ production with $\HZZ$: a) the recoil mass distribution of the $\mm$ system for $\Zmm, \HZZ\to\vv \qq$; b) the invariant mass distribution of the $\mm \qq$ system  for $\Zvv,\,\HZZ\to\mm \qq$. The markers and their uncertainties represent expectations from a CEPC dataset of $\abih$, whereas the solid blue curves are the fit results. The dashed curves are the signal and background components. Contributions from other decays of the Higgs boson are included in the background. }
	\end{center} 
\end{figure*}
The $\HZZ$ decay has a branching ratio 2.64\%~\cite{Heinemeyer:2013tqa} for a 125~GeV Higgs boson in the SM.
Events from $\ZH$ production with the $\HZZ$ decay have three $Z$ bosons in their final states with one of them 
being off-shell. $Z$ bosons can decay to all lepton and quark flavors, with the exception of the top quark. Consequently, the $\ee\to ZH\to ZZZ^*$ process has a very rich variety of topologies.

Studies are performed for a few selected $ZH$ final states:  $\Zmm$ and $\HZZ\to \vv \qq$; $\Zvv$ and $\HZZ\to \ell^+\ell^- \qq$. The $W$ and $Z$ boson fusion processes, $\ee\to \ee H$ and $\ee\to \vv H$, are included in the $Z(\ee)H$ and $Z(\vv)H$ studies assuming their SM values for the production rates. For the final states studied, the SM $ZZ$ production is the main background. 
For $\Zmm$ and $\HZZ\to\vv \qq$, the muon pairs must have their invariant masses between 80--100~GeV, recoil masses between 120--160 GeV and transverse momenta larger than 10 GeV. The jet pairs of the $Z^*\to \qq$ decay candidates are required to have their invariant masses in the range of 10--38~GeV. Figure~\ref{fig:hzz}(a) shows the recoil mass distribution of $\Zmm$ after the selection. The background is negligible in this final state.

The candidates of $\Zvv$ and $\HZZ\to\ell^+\ell^- \qq$ are selected by requiring a same-flavor lepton pair and two jets. The total visible energy must be smaller than 180~GeV and the missing mass in the range of 58--138~GeV. Additional requirements are applied on the mass and transverse momenta of the lepton and jet pairs. After the selection, the background is about an order of magnitude smaller than the signal as shown in Fig.~\ref{fig:hzz}(b).

Table~\ref{tab:hzz} summarizes the expected precision on $\sigma(ZH)\times{\rm BR}(\HZZ)$  from the final states considered. The combination of these final states results in a precision of about 4.9\%. The sensitivity can be significantly improved considering that many final states are not included in the current study. In particular, the final state of $\Zqq$ and $\HZZ\to \qq\qq$ which accounts for a third of all $ZH\to ZZZ^*$ decay is not studied. Moreover, there are further potential improvements by using multivariate techniques.
\begin{center}
	\begin{minipage}{0.48\textwidth}
		\begin{center}
			\tabcaption{\label{tab:hzz}\small Expected relative precision for the $\sigma(ZH)\times \mathrm{BR}(\HZZ)$ measurement with an integrated luminosity $\abih$. }
			\begin{tabular}{llrrl}
				\hline\hline
				\multicolumn{2}{c}{$ZH$ final state}   & Precision     \\ \hline
				$\Zmm$     &  $\HZZ\to \vv \qq $             &       7.2\%    \\ 
				$\Zvv$     &  $\HZZ\to \ell^+\ell^- \qq $    &       7.9\%    \\ \hline
				\multicolumn{2}{c}{Combination}                 &     4.9\%       \\
				\hline\hline
			\end{tabular}
		\end{center}
	\end{minipage}
\end{center}

\vspace*{0.3cm}
\subsection{$H\to\gamma\gamma$} 
\label{sec:hgamgam}

The diphoton decay of a 125 GeV Higgs boson has a small branching ratio of 0.23\% in the SM due to its origin involving massive $W$ boson and top quark in loops. However, photons can be identified and measured well, thus the decay can be fully reconstructed with a good precision. The decay also serves as a good benchmark for the performance of the electromagnetic calorimeter. 

\begin{center}
\includegraphics[width=0.42\textwidth]{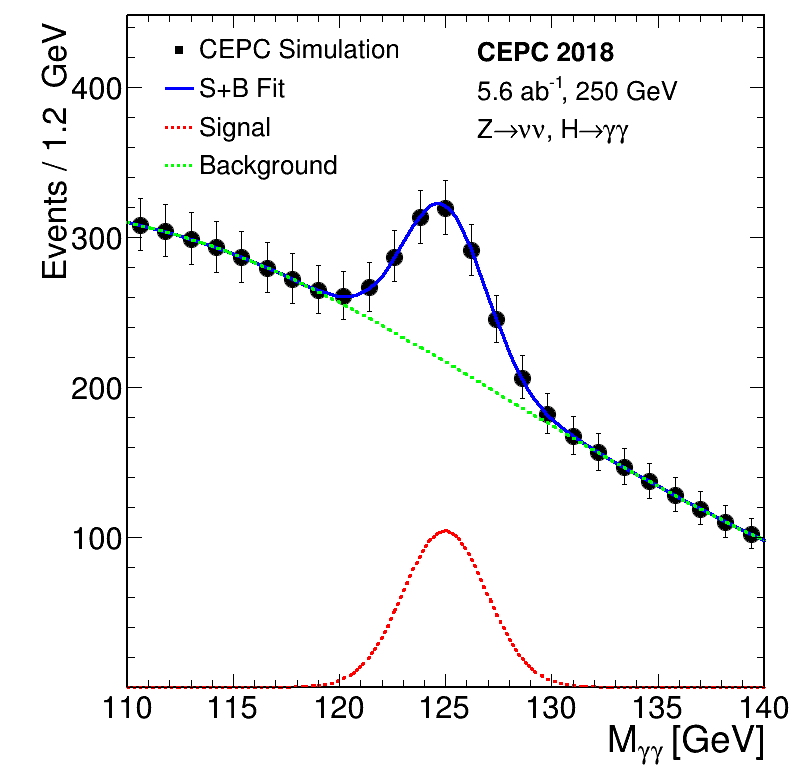}
\figcaption{\small $ZH$ production with $\Hyy$: the diphoton invariant mass distribution for the $\Zvv$ decay. The markers and their uncertainties represent expectations from a CEPC dataset of $\abih$, whereas the solid blue curve is the fit result. The dashed curves are the signal and background components.}
\label{fig:hyy}
\end{center}

Studies are performed for the $ZH$ production with $H\to\gamma\gamma$ and four different $Z$ boson decay modes: $\Zmm,\, \tau^+\tau^-,\, \vv$ and $\qq$. The $\Zee$ decay is not considered because of the expected large background from the Bhabha process.  The studies are based on the full detector simulation for the $\Zqq$ decay channel and the fast simulation for the others. 
Photon candidates are required to have energies greater than 25 GeV and polar angles of $|\cos\theta|<0.9$. The photon pair with the highest invariant mass is retained as the $H\to\gamma\gamma$ candidate and its recoil mass must be consistent with the $Z$ boson mass. For the $\Zmm$ and $\Ztt$ decays,  a minimal angle of 8$^\circ$ between any selected photon and lepton is required to suppress backgrounds from final state radiations. After the selection, the main SM background is the $\ee\to (Z/\gamma^*)\gamma\gamma$ process where the $\gamma$'s arise from the initial and final state radiation.

The diphoton mass is used as the final discriminant for the separation of signal and backgrounds. The distribution for the $\Zvv$ decay mode is shown in Fig.~\ref{fig:hyy}. A relative precision of 6.2\% on $\sigma(ZH)\times{\rm BR}(H\to\gamma\gamma)$ can be achieved.

\vspace*{0.3cm}
\subsection{$\HZy$}
\label{sec:hzg}
 
Similar to the $\Hyy$ decay, the $\HZy$ decay in the SM is mediated by $W$-boson and top-quark loops and has a branching ratio of 0.154\%. The $\HZy$ analysis targets the signal process of $ZH\to ZZ\gamma\to\vv\qq\gamma$, in which one of the $Z$ bosons decays into a pair of quarks and the other decays into a pair of neutrinos.


The candidate events are selected by requiring exactly one photon with transverse energy between 20--50~GeV and at least two jets, each with transverse energy greater than 10~GeV. The dijet invariant mass and the event missing mass must be within windows of $\pm 12$~GeV and $\pm 15$~GeV of the $Z$ boson mass, respectively. Additional requirements are applied on the numbers of tracks and  calorimeter clusters as well as on the transverse and longitudinal momenta of the $Z$ boson candidates. The backgrounds are dominated by the processes of single boson, diboson, $\qq$, and Bhabha production.

After the event selection, the photon is paired with each of the two $Z$ boson candidates to form Higgs boson candidates and the mass differences, $\Delta M=M_{\qq\gamma}-M_{\qq}$ and $\Delta M=M_{\vv\gamma}-M_{\vv}$, are calculated. Here the energy and momentum of the $\vv$ system are taken to be the missing energy and momentum of the event. For signal events, one of the mass differences is expected to populate around $M_H-M_Z\sim 35$~GeV whereas the other should be part of the continuum background.  Figure~\ref{fig:hzg} shows the $\Delta M$ distribution expected from an integrated luminosity of $\abih$. Modeling the signal distribution of the correct pairing with a Gaussian and the background (including wrong-pairing contribution of signal events) with a polynomial, a likelihood fit results in a relative precision of 13\% on $\sigma(ZH)\times {\rm BR}(\HZy)$.  

This analysis can be improved with further optimizations and the use of multivariate techniques. Other decay modes such as $ZH\to ZZ\gamma\to \qq\,\qq\gamma$ should further improve the precision on the $\sigma(ZH)\times{\rm BR}(\HZy)$ measurement.

\begin{center}
	\begin{minipage}{0.48\textwidth}
		\begin{center}
			\includegraphics[width=\textwidth]{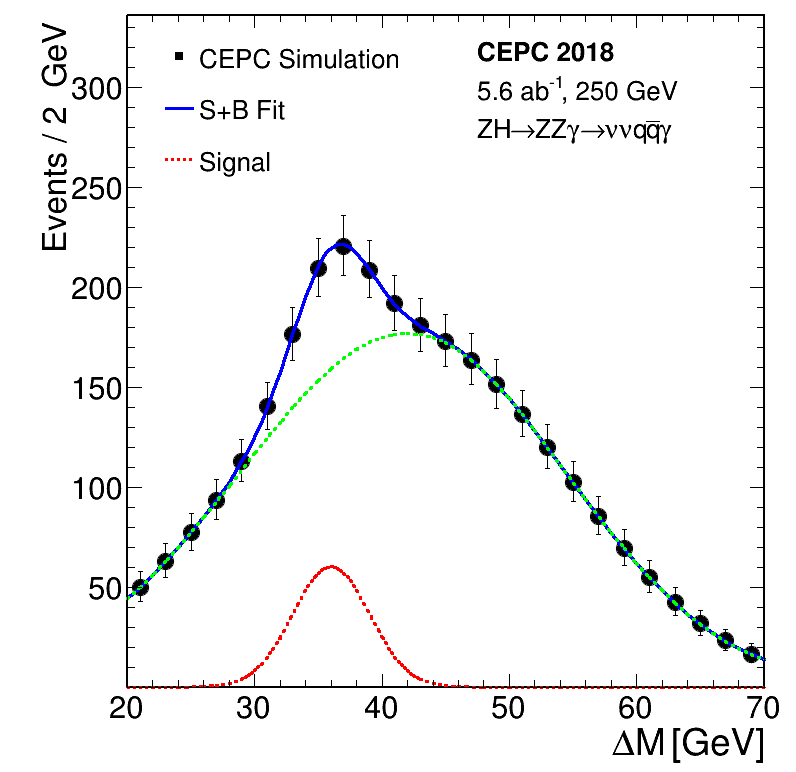}
			\figcaption{\small The distribution of the mass difference $\Delta M$  ($M_{\qq\gamma}-M_{\qq}$ and $M_{\vv\gamma}-M_{\vv}$) of the selected $\ee\to ZH\to ZZ\gamma\to\vv\qq\gamma$ candidates expected in a dataset with an integrated luminosity of $\abih$. The signal distribution shown is for the correct pairings of the Higgs boson decays. }
			\label{fig:hzg}
		\end{center}
	\end{minipage}
\end{center}

\vspace*{0.3cm}
\subsection{$\Htt$}
\label{sec:htautau}

The $\Htt$ decay has a branching ratio of 6.32\%~\cite{Heinemeyer:2013tqa} at $m_H=125$~GeV in the SM. The $\tau$-lepton is short-lived and decays to one or three charged pions along with a number of neutral pions. The charged and neutral pions, as well as the two photons from the decay of the latter,  can be well resolved and measured by the CEPC detector.

Simulation studies are performed for $\ee\to ZH$ production with $\Htt$ and $Z\to \mm,\vv$ and $\qq$ decays. For $\Zmm$, candidates are first required to have a pair of oppositely charged muons with their invariant mass between 40--180 GeV and their recoil mass between 110--180 GeV. For $\Zvv$, candidates are preselected by requiring a missing mass in the range of 65--225 GeV, a visible mass greater than 50~GeV and an event visible transverse momentum between 10--100 GeV.  For both decays,  a BDT selection is applied after the preselection to identify ditau candidates. The BDT utilizes information such as numbers of tracks and photons and the angles between them. After these selections, the $ZH$ production with the non-tau decays of the Higgs boson is the dominant ($>95\%$) background for $\Zmm$ and contributes to approximately 40\% of the total background for $\Zvv$. The rest of the background in the $\Zvv$ channel comes from  diboson production. For $\Zqq$, candidates are required to have a pair of tau candidates with their invariant mass between 20--120 GeV, a pair of jets with their mass between 70--110 GeV and their recoil mass between 100--170 GeV. The main background is again from  $ZH$ production originating from the decay modes other than the intended $ZH\to \qq\tau^+\tau^-$ decay. The rest of the background is primarily from $ZZ$ production. 

\begin{figure*}[ht!]
\begin{center}
\subfigure[]{\includegraphics[width=0.4\textwidth]{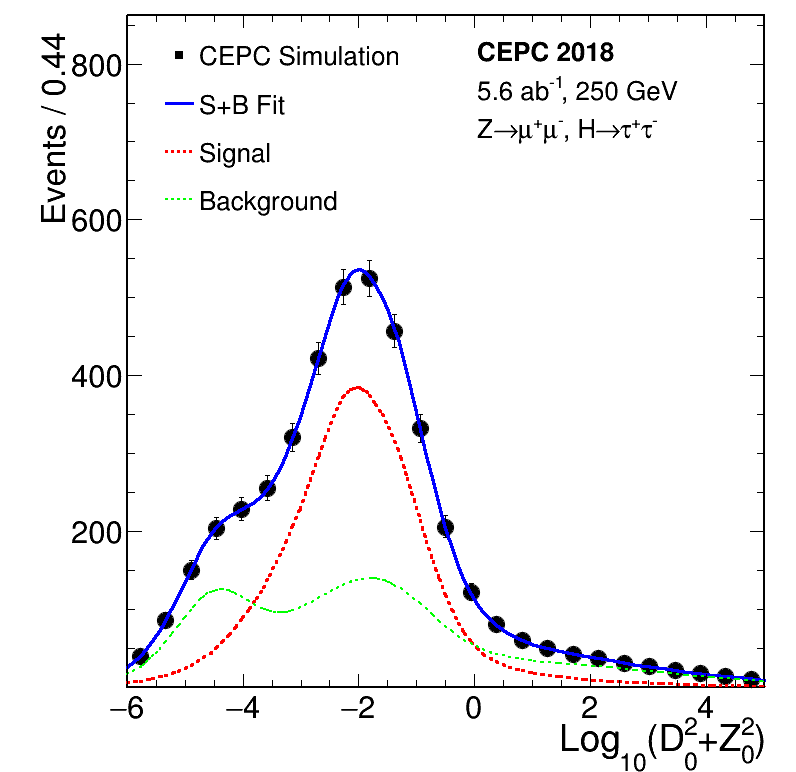}}
\subfigure[]{\includegraphics[width=0.4\textwidth]{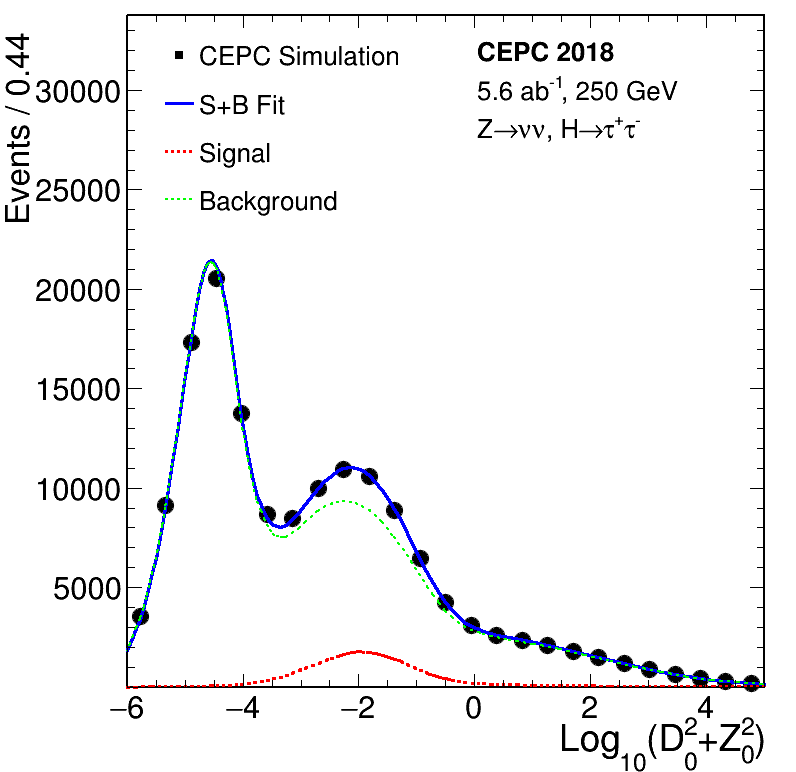}}
\figcaption{\label{fig:htautau}\small Distributions of the impact parameter variable of the leading tracks from the two tau candidates in the $Z$ decay mode:  (a) $\Zmm$ and  (b) $\Zvv$. Here $D_0$ and $Z_0$ are the transverse and longitudinal impact parameters, respectively. The markers and their uncertainties represent expectations from a CEPC dataset of $\abih$, whereas the solid blue curves are the fit results. The dashed curves are the signal and background components. Contributions from other decays of the Higgs boson are included in the background.}
	\end{center}
\end{figure*}

The final signal yields are extracted from fits to the distributions of variables based on the impact parameters of the leading tracks of the two tau candidates as shown in Fig. \ref{fig:htautau}. Table~\ref{tab:htt} summarizes the estimated precision on $\sigma(ZH)\times {\rm BR}(\Htt)$ expected from a CEPC dataset of $\abih$ for the three $Z$ boson decay modes studied. The precision from the $\Zee$ decay mode extrapolated from the $\Zmm$ study is also included. The $\ee\to\ee H$ contribution from the $Z$ fusion process is fixed to its SM value in the extrapolation.  In combination, the relative precision of 0.8\% is expected for $\sigma(ZH)\times{\rm BR}(\Htt)$.

\begin{center}
\begin{minipage}{0.5\textwidth}
\begin{center}
\tabcaption{\label{tab:htt}\small Expected relative precision for the $\sigma(ZH)\times \mathrm{BR}(\Htt)$ measurement from a CEPC dataset of $\abih$.}
\begin{tabular}{llrr}
\hline\hline
\multicolumn{2}{c}{$ZH$ final state}   & Precision  &   \\ \hline
$\Zmm$     &  $\Htt$    &        2.6\%   \\ 
$\Zee$     &  $\Htt$    &        2.7\%   \\
$\Zvv$     &  $\Htt$    &        2.5\%   \\
$\Zqq$     &  $\Htt$    &        0.9\%   \\ \hline
\multicolumn{2}{c}{Combination}     &     0.8\%             \\
\hline\hline
\end{tabular}
\end{center}
\end{minipage}
\end{center}

\vspace*{0.3cm}
\subsection{$H\to\mm$} 
\label{sec:hmm}

The dimuon decay of the Higgs boson, $\Hmm$, is sensitive to
the Higgs boson coupling to the second-generation fermions with a clean final-state signature.
In the SM, the branching ratio of the decay is $2.18\times 10^{-4}$~\cite{Heinemeyer:2013tqa} for $m_{H}=125$ GeV.

To estimate CEPC's sensitivity for the $\Hmm$ decay, studies are performed for the $ZH$ production with the $Z$ decay modes: $\Zll$, $\Zvv$, and $\Zqq$. In all cases, the SM production of $ZZ$ is the dominant background source. Candidate events are selected by requiring a pair of muons with its mass between 120--130~GeV and their recoiling mass consistent with the $Z$ boson mass (in the approximate range of 90--93~GeV, depending on the decay mode). Additional requirements are applied to identify specific $Z$ boson decay modes. For $\Zll$, candidate events must have another lepton pair with its mass consistent with $m_Z$. In the case of $\Zmm$, the muon pairs of the $\Zmm$ and $\Hmm$ decays are selected by minimizing a $\chi^2$ based on their mass differences with $m_Z$ and $m_H$.  For the $\Zvv$ decay, a requirement on the missing energy is applied. For the $\Zqq$ decay, candidate events must have two jets with their mass consistent with $m_Z$. 
To further reduce the $ZZ$ background, differences between the signal and background in kinematic variables, such as the polar angle, transverse momentum and energy of the candidate $\Hmm$ muon pair, are exploited. Simple criteria on these variables are applied for the $\Zll$ and $\Zvv$ decay mode whereas a BDT is used for the $\Zqq$ decay. 

\begin{center}
	\begin{minipage}{0.48\textwidth}
		\begin{center}
			\includegraphics[width=\textwidth]{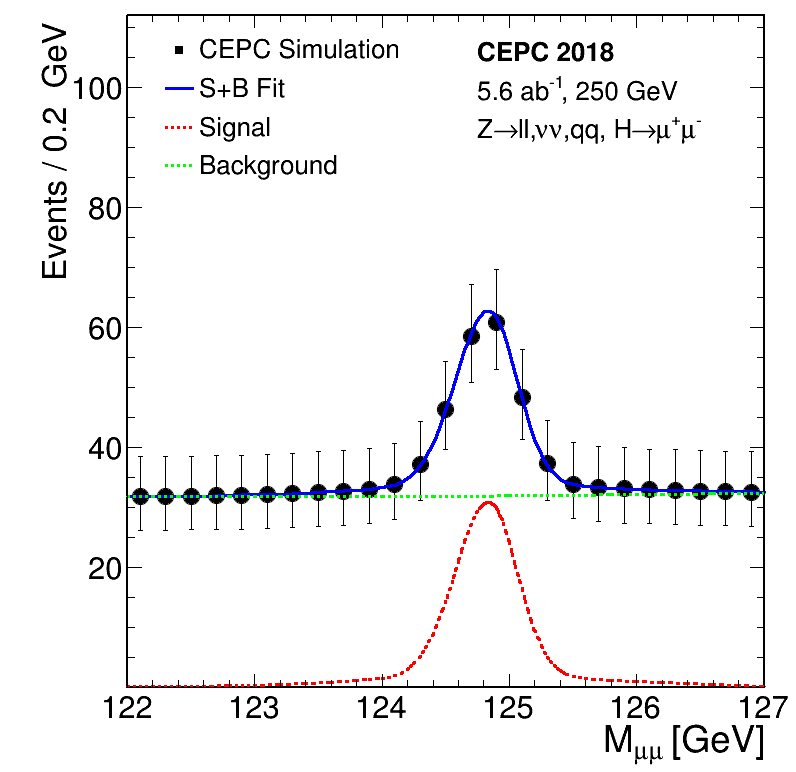}
			\figcaption{\label{fig:hmm}\small $ZH$ production with the $\Hmm$ decay: dimuon invariant mass distribution of the selected $\Hmm$ candidates expected from  an integrated luminosity of $\abih$ at the CEPC. The distribution combines contributions from $\Zll$, $\Zvv$, and $\Zqq$ decays. The markers and their uncertainties represent expectations whereas the solid curve is the fit result. The dashed curves are the signal and background components.}
		\end{center}
	\end{minipage}
\end{center}

In all analyses, the signal is extracted through unbinned likelihood fits to the $M_{\mm}$ distributions in the range of 120--130~GeV with a signal-plus-background model. Analytical functions are used model both the signal and background distributions.  The signal model is a Crystal Ball function while the background model is described by a second-order Chebyshev polynomial. The dimuon mass distribution combining all $Z$ boson decay modes studied is shown in Fig.~\ref{fig:hmm} with the result of the signal-plus-background fit overlaid.  The combined relative precision on the $\sigma(ZH)\times {\rm BR}(\Hmm)$ measurement is estimated to be about 16\% for data corresponding to an integrated luminosity of $\abih$.

\vspace*{0.3cm}
\subsection{The invisible decay of the Higgs boson: $\Hinv$}

In the SM, the Higgs boson can decay invisibly via $\HZZ\to \vv\vv$. For a Higgs boson mass of 125~GeV, this decay has a branching ratio of $1.06\times 10^{-3}$. In many extensions to the SM, the Higgs boson can decay directly to invisible particles~\cite{Shrock:1982kd,Griest:1987qv,Englert:2011yb,Bonilla:2015uwa}. In this case, the branching ratio can be significantly enhanced.

The sensitivity of the ${\rm BR}(\Hinv)$ measurement is studied for the $\Zll$ and $\Zqq$ decay modes. The $\HZZ\to\vv\vv$ decay is used to model the $\Hinv$ decay both  in the context of the SM and its extensions. This is made possible by the fact that the Higgs boson is narrow scalar so that its production and the decay can be treated separately. The main background is  SM $ZZ$ production with one of the $Z$ bosons decay invisibly and the other decays visibly.
Candidate events in the $\Zll$ decay mode are selected by requiring a pair of lepton with its mass between 70--100 GeV and event visible energy in the range 90--120 GeV. Similarly, candidate events in $\Zqq$ are selected by requiring two jets with its mass between 80--105 GeV and event visible energy in the range 90--130 GeV. Additional selections, including using a BDT to exploit the kinematic differences between signal and background events, are also implemented.

\begin{center}
\begin{minipage}{0.48\textwidth}
\begin{center}\footnotesize
\tabcaption{\small Expected relative precision on $\sigma(ZH)\times {\rm BR}(\Hinv)$ and 95\% CL upper limit on ${\rm BR}(\Hinv)$ from a CEPC dataset of $\abih$.}
\begin{tabular}{llcclcl} \hline\hline
\multicolumn{2}{c}{$ZH$ final}     & Relative precision    & Upper limit on \\ 
\multicolumn{2}{c}{state studied}  &  on $\sigma\times\BR$ & $\BR(\Hinv)$    \\ \hline
$\Zee$   & $\Hinv$  & $339\%$   & 0.82\%   \\
$\Zmm$   & $\Hinv$  & $232\%$   & 0.60\%   \\
$\Zqq$   & $\Hinv$  & $217\%$   & 0.57\%   \\ \hline
\multicolumn{2}{c}{Combination}     & $143\%$ & 0.41\%  \\
\hline\hline
\end{tabular}
\label{tab:hinv}
\end{center}
\end{minipage}
\end{center}

Table~\ref{tab:hinv} summarizes the expected precision on the measurement of $\sigma(ZH)\times{\rm BR}(\Hinv)$ and the 95\% confidence-level~(CL) upper limit on ${\rm BR}(\Hinv)$ from a CEPC dataset of $\abih$. Subtracting the SM $\HZZ\to\vv\vv$ contribution, a 95\% CL upper limit of 0.30\% on $\BRinvBSM$, the BSM contribution to the $\Hinv$ decay can be obtained.

\vspace*{0.3cm}
\subsection{Measurement of $\sigma(\vvH)\times{\rm BR}(\Hbb)$ }
\label{sec:vvh}

\begin{figure*}[t!]
	\begin{center}
		\subfigure[]{\includegraphics[width=0.4\textwidth]{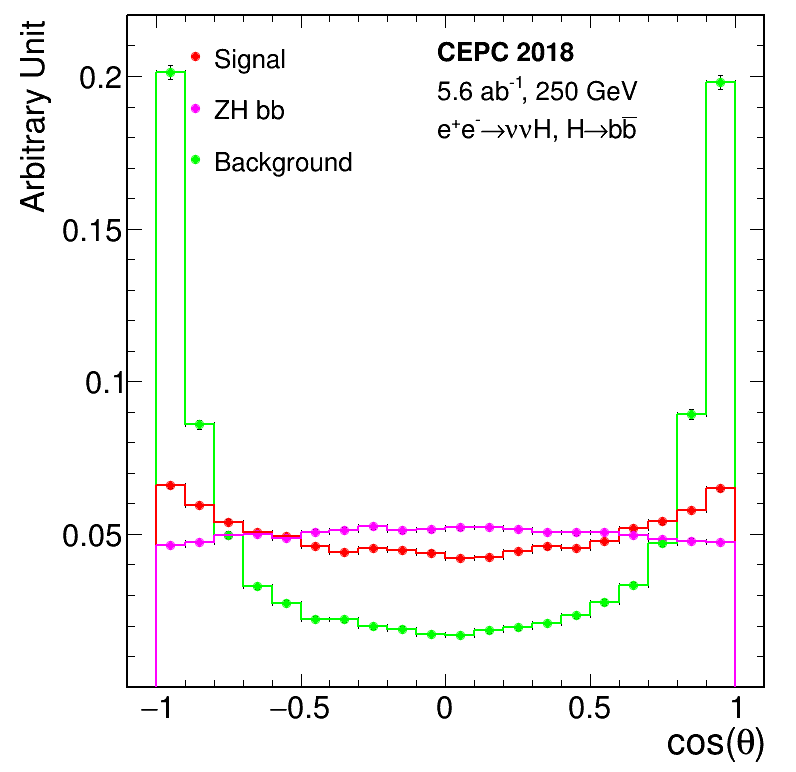}}
		\subfigure[]{\includegraphics[width=0.4\textwidth]{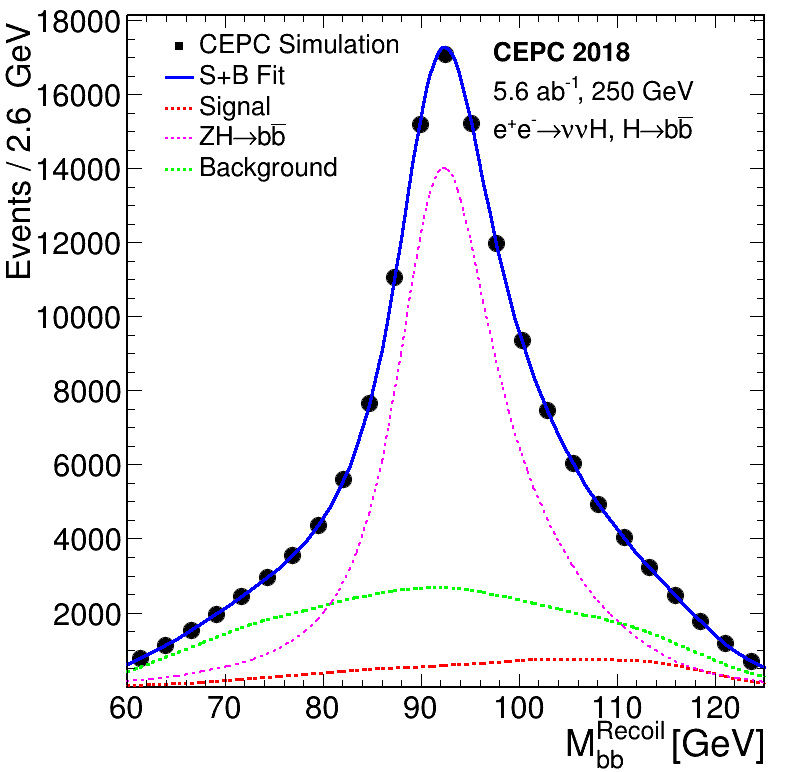}}
		\figcaption{\small Distributions of the $\bb$ system of the $\ee\to\vv\bb$ events:  (a) cosine of the polar angle $\theta$ before the event selection and (b) the recoil mass after the event selection. Contributions from $\vvH$, $ZH$ and other SM processes are shown. The $\cos\theta$ distributions are normalized to unity and therefore only shapes are compared.}
		\label{fig:vvh}
	\end{center}
\end{figure*}

The $W$-fusion $\vvH$ process has a cross section of 3.3\% of that of the $ZH$ process at $\sqrt{s}=250$~GeV.  The product of its cross section and ${\rm BR}(\Hbb)$, $\sigma(\vv H)\times\mathrm{BR}(\Hbb)$, is a key input quantity to one of the two model-independent methods for determining the Higgs boson width at the CEPC, see Section~\ref{sec:combinations}. The $\ee\to \vv H\to \vv\bb$ process has the same final state as the $\ee\to ZH\to\vv\bb$ process, but has a rate that is approximately one sixth of $\ee\to ZH\to\vv\bb$ at $\sqrt{s}=250$~GeV. The main non-Higgs boson background is the SM $ZZ$ production.

The $Z(\vv)H$ background is irreducible and can also interfere with $\vv H$ in the case of $Z\to\nu_e\bar{\nu}_e$. However, the interference effect is not considered in the current study.  The $\vv H$ and $Z(\vv)H$ contributions can be separated through the exploration of their kinematic differences. While the invariant mass distributions of the two $b$-quark jets are expected to be indistinguishable, the recoil mass distribution should exhibit a resonance structure at the $Z$ boson mass for $Z(\vv)H$ and show a continuum spectrum for $\vv H$. Furthermore, Higgs bosons are produced with different polar angular distributions, see Fig.~\ref{fig:vvh}(a).

Candidate events are selected by requiring their visible energies between  105~GeV and 155~GeV, visible masses within 100--135~GeV, and missing masses in the range of 65--135~GeV. The two $b$-quark jets are identified using the variable $L_B$ described in Section~\ref{sec:hbcg}.
To separate $\vv H$ and $Z(\vv)H$ contributions, a 2-dimensional simultaneous fit in the plane of the recoil mass and polar angle of the $\bb$ system is performed. The recoil mass resolution is improved through a kinematic fit by constraining the invariant mass of the two $b$-jets within its resolution to that of the Higgs boson mass. Figure~\ref{fig:vvh}(b) shows the recoil mass distribution of the $\bb$ system after the kinematic fit. A fit to the $M_{\bb}-\cos\theta$ distribution with both rates of $\vv H$ and $Z(\vv) H$ processes as free parameters leads to relative precision of 2.9\% for $\sigma(\vv H)\times{\rm BR}(\Hbb)$ and 0.30\% for $\sigma(ZH)\times {\rm BR}(\Hbb)$. The latter is consistent with the study of the $\Hbcg$ decay described in Section~\ref{sec:hbcg}. Fixing the $Z(\vv)H(\bb)$ contribution to its SM expectation yields a relative precision of 2.6\% on $\sigma(\vvH)\times {\rm BR}(\Hbb)$.

\section{Combinations of Individual Measurements}
\label{sec:combinations}

\subsection{Combined measurements of $\sigmaBR$ and ${\rm BR}$}

With the measurements of the inclusive cross section $\sigma(ZH)$ and the cross section times the branching ratio  $\sigma(ZH)\times\mathrm{BR}$ for the individual Higgs boson decay modes, the branching ratio $\mathrm{BR}$ can be extracted. Most of the systematic uncertainties associated with the measurement of $\sigma(ZH)$ cancel in this procedure. A maximum likelihood fit is used to estimate the precision on the $\mathrm{BR}$s. For a given Higgs boson decay mode, the likelihood has the form:
 
\begin{equation}
L(\mathrm{BR},\theta) = \mathrm{Poisson}\left.\left[N^{\mathrm{obs}}\right|N^{\mathrm{exp}}(\mathrm{BR},\theta)\right]\cdot G(\theta),
\label{eq:likelihood}
\end{equation}
where $\mathrm{BR}$ is the parameter of interest and $\theta$ represents nuisance parameters associated with systematic uncertainties.  The number of observed events is denoted by $N^{\rm obs}$, $N^{\rm exp}(\mathrm{BR},\theta)$ is the expected number of events, and $G(\theta)$ is a set of constraints on the nuisance parameters due to the systematic uncertainties.  The number of expected events is the sum of signal and background events. The number of signal events is 
calculated from the integrated luminosity, the $\ZH$ cross section $\sigma(ZH)$ measured from the recoil method, Higgs boson branching ratio $\mathrm{BR}$, the event selection efficiency $\epsilon$.  The number of the expected background events, $N^b$, is estimated using Monte Carlo samples. Thus:

\begin{equation}
N^{\rm exp}(\mathrm{BR},\theta) =\mathrm{Lumi}(\theta^{\rm lumi})\times \sigma_{ZH} (\theta^{\sigma})\times \mathrm{BR} \times \epsilon(\theta^\epsilon) + N^b(\theta^b)
\label{eq:nexp},
\end{equation}
where $\theta^X\ (X={\rm lumi}, \sigma$, $\epsilon$ and $b$) are the nuisance parameters of their corresponding parameters or measurements. Even with $10^6$ Higgs boson events, statistical uncertainties are expected to be dominant and thus systematic uncertainties are not taken into account for the current studies. The nuisance parameters are fixed to their nominal values.

For the individual analyses discussed in Section~\ref{sec:decays}, contamination from Higgs boson production or decays other than the one under study are fixed to their SM values for simplicity. In the combination, however, these constraints are removed and the contamination are constrained only by the analyses targeted for their measurements. For example, the $\Hbcg$ analysis suffers from contamination from the $H\to WW^*,ZZ^*\to\qq\qq$ decays. For the analysis discussed in Section~\ref{sec:hbcg}, these contaminations are estimated from SM. In the combination fit, they are constrained by the $H\to WW^*$ and $H\to ZZ^*$ analyses described in Sections~\ref{sec:hww} and~\ref{sec:hzz}, respectively. Taking into account these across-channel contaminations properly generally leads to small improvements in precision. For example, the precision on $\sigma(ZH)\times {\rm BR}(H\to ZZ^*)$ is improved from 5.3\% of the standalone analysis to 4.9\% after the combination.  

Table~\ref{tab:HiggsSigmaBR} summarizes the estimated precision of Higgs boson property measurements, combining all studies described in this paper. 
For the leading Higgs boson decay modes, namely $\bb$, $\cc$, $gg$, $WW^*$, $ZZ^*$ and $\tau^+\tau^-$, percent level precision is expected. 
The best achievable statistical uncertainties for a dataset of $\abih$ are 0.26\% for $\sigma(e^+e^-\to ZH)\times \mathrm{BR}(\Hbb)$ and 0.5\% for $\sigma(e^+e^-\to ZH)$.
Even for these measurements, statistics is likely to be the dominant uncertainty source. Systematic uncertainties due to the acceptance of the detector, the efficiency of the object reconstruction/identification, the luminosity and the beam energy determination are expected to be small. The integrated luminosity can be measured with a 0.1\% precision, a benchmark already achieved at the LEP~\cite{Schael:2013ita}, and can be potentially improved in  the future. The center-of-mass energy will be known better than 1~MeV, resulting negligible uncertainties on the theoretical cross section predictions and experimental recoil mass measurements.

The estimated precision is expected to improve as more final states are explored and analyses are improved. This is particularly true for $ZH\to ZWW^*$ and $ZH\to ZZZ^*$ with complex final states. Therefore, Table~\ref{tab:HiggsSigmaBR} represents conservative estimates for many Higgs boson observables.

\begin{table*} [t!]
	\begin{center}
		\caption{\label{tab:HiggsSigmaBR}\small Estimated precision of Higgs boson property measurements for the CEPC-v1 detector concept operating at $\sqrt{s}=250$~GeV. All precision are relative except for $m_H$ and $\BRinvBSM$ for which $\Delta m_H$ and 95\% CL upper limit are quoted respectively. The extrapolated precision for the CEPC-v4 concept operating at $\sqrt{s}=240$~GeV are included for comparisons, see Section~\ref{sec:v1tov4}.} 
		\begin{tabular}{clcccccl} \hline\hline
			&   	         &  \multicolumn{5}{c}{Estimated Precision} & \hspace*{3mm}  \\ \cline{3-7}
			& Property       &  \multicolumn{2}{c}{CEPC-v1}   &\hspace*{1.5cm}&  \multicolumn{2}{c}{CEPC-v4}  \\ \hline
			& $m_H$          &  \multicolumn{2}{c}{5.9 MeV}   &&  \multicolumn{2}{c}{5.9 MeV}  \\ 
			& $\Gamma_H$     &  \multicolumn{2}{c}{2.7\%}     &&  \multicolumn{2}{c}{2.8\%}  \\
			& $\sigma(ZH)$   &  \multicolumn{2}{c}{0.5\%}    &&  \multicolumn{2}{c}{0.5\%}  \\
			& $\sigma(\vv H) $  & \multicolumn{2}{c}{3.0\%}  &&  \multicolumn{2}{c}{3.2\%}  \\ \hline
			\\ 
			& Decay mode      &  $\sigmaBR$ &  $\BR$   &&  $\sigmaBR$  & $\BR$        \\ \hline    
			& $\Hbb$          &   0.26\%   &   0.56\% &&   0.27\%  &  0.56\%  \\
			& $\Hcc$          &   3.1\%    &   3.1\%  &&   3.3\%   &  3.3\%  \\
			& $\Hgg$          &   1.2\%    &   1.3\%  &&   1.3\%   &  1.4\%  \\
			& $\HWW$          &   0.9\%    &   1.1\%  &&   1.0\%   &  1.1\%  \\
			& $\HZZ$          &   4.9\%    &   5.0\%  &&   5.1\%   &  5.1\%  \\
			& $\Hyy$          &   6.2\%    &   6.2\%  &&   6.8\%   &  6.9\%  \\
			& $\HZy$          &   13\%     &   13\%   &&   16\%    &  16\%   \\
			& $\Htt$          &   0.8\%    &   0.9\%  &&   0.8\%   &  1.0\%   \\
			& $\Hmm$          &   16\%     &   16\%   &&   17\%    & 17\%     \\
			& $\BRinvBSM$     & $-$    &   $<0.28\%$   && $-$ &  $<0.30\%$  	\\ \hline\hline
		\end{tabular}
	\end{center}
\end{table*}

\subsection{Extrapolation to CEPC-v4}
\label{sec:v1tov4}

As discussed in Section~\ref{sec:detop}, the CEPC conceptual detector design has evolved from CEPC-v1 to CEPC-v4 with the main change being the reduction of the solenoidal field from 3.5~Tesla to 3.0~Tesla. In the meantime, the nominal CEPC center-of-mass energy for the Higgs boson factory has been changed from 250~GeV to 240~GeV. The results presented above are based on CEPC-v1 operating at $\sqrt{s}=250$~GeV. 
However, given the relative small differences in the performance of the two detector concepts and in $\sqrt{s}$, the results for CEPC-v4 operating at $\sqrt{s}=240$~GeV can be estimated through extrapolation taking into account changes in signal and background cross sections as well as track momentum resolution. From 250~GeV to 240~GeV, the $\ZH$ and $\vvH$ cross sections are reduced, respectively, by approximate 5\% and 10\% while cross sections for background processes are increased by up to 10\%. The change in magnetic field affects the $\Hmm$ analysis the most whereas its effect on other analyses are negligible. 
The extrapolated results for CEPC-v4 at 240~GeV are included in Table~\ref{tab:HiggsSigmaBR}. In most cases, small relative degradations of a few percent are expected. For the following analyses, the extrapolated results for CEPC-v4 at $\sqrt{s}=240$~GeV are used.

\subsection{Measurement of Higgs boson width}
\label{sec:HiggsWidth}

The Higgs boson width ($\Gamma_H$) is of special interest as it is sensitive to BSM physics in Higgs boson decays that are not directly detectable or searched for. However, the 4.07~MeV width predicted by the SM is too small to be measured with a reasonable precision from the distributions of either the invariant mass of the Higgs boson decay products or the recoil mass of the system produced in association with the Higgs boson. In a procedure that is unique to lepton colliders, the width can be determined from the measurements of Higgs boson production cross sections and its decay branching ratios. This is because the inclusive $\ZH$ cross section $\sigma(ZH)$ can be measured from the recoil mass distribution, independent of the Higgs boson decays. 

Measurements of $\sigma(ZH)$ and $\mathrm{BR}$'s have been discussed in Sections~\ref{sec:massXS} and~\ref{sec:decays}. By combining these measurements, the Higgs boson width can be calculated in a model-independent way:
\begin{equation}
\Gamma_H = \frac{\Gamma(\HZZ)}{{\rm BR}(\HZZ)} \propto \frac{\sigma(ZH)}{{\rm BR}(\HZZ)} 
\label{eq:width1}
\end{equation}
where $\Gamma(\HZZ)$ is the partial width of the $\HZZ$ decay. Because of the small expected ${\rm BR}(\HZZ)$ value for a 125~GeV Higgs boson (2.64\% in the SM), the precision of $\Gamma_H$ is limited by the $H\to ZZ^*$ analysis statistics. 
It can be improved including the decay final states with larger branching ratios, e.g. the $\Hbb$ decay:
\begin{equation}
\Gamma_H = \frac{\Gamma(\Hbb)}{{\rm BR}(\Hbb)}
\label{eq:hbb}
\end{equation}
where the partial width $\Gamma(\Hbb)$ can be independently extracted from the cross section of the $W$ fusion process $\ee\to \vv H\to \vv\,\bb$:
\begin{eqnarray}
\sigma(\vv H\to \vv\,\bb) \propto \Gamma(\HWW)\cdot{\rm BR}(\Hbb) \\
= \Gamma(\Hbb)\cdot {\rm BR}(\HWW).
\label{eq:width3}
\end{eqnarray}
Thus, the Higgs boson total width is:
\begin{equation}
\Gamma_H = \frac{\Gamma(\Hbb)}{{\rm BR}(\Hbb)} \propto \frac{\sigma(\vvH)}{{\rm BR}(\HWW)}
\label{eq:width2}
\end{equation}
where ${\rm BR}(\Hbb)$ and ${\rm BR}(\HWW)$ are measured from the $\ZH$ process. The limitation of this method is the precision of the $\sigma(e^+e^-\to\vv H \to \vv\, \bb)$ measurement. 

The expected precision on $\Gamma_H$ is {\color{black} 5.1\%} from the measurements of $\sigma(ZH)$ and ${\rm BR}(\HZZ)$ and is {\color{black} 3.5\%} from the measurements of $\sigma(\vv H\to \vv\bb)$, ${\rm BR}(\Hbb)$ and ${\rm BR}(\HWW)$. The quoted precision is dominated by the  ${\rm BR}(\HZZ)$ measurement for the former case and the $\sigma(\vv H\to \vv\bb)$ measurement for the latter case. The combined $\Gamma_H$ precision of the two measurements is {\color{black} 2.8\%}, taking into account the correlations between the two measurements.

\section{Higgs Boson Coupling Measurements }\label{sec:kappaEFT}

To understand the implications of the estimated CEPC precision shown in Table~\ref{tab:HiggsSigmaBR} on possible new physics models, the results need to be interpreted in terms of constraints on the parameters in the Lagrangian. This is often referred to as the ``Higgs boson coupling measurements'', even though the term can be misleading as discussed below. 

There is no unique way to present the achievable precision on the couplings. Before going into the discussion of the CEPC results, we briefly comment on the choices made here.  
The goal of the theory interpretation here is to obtain a broad idea of the CEPC sensitivity to the Higgs couplings. The interpretation should be simple with intuitive connections between the models and the experimental observables. Ideally,  it should have as little  model assumptions as possible. Furthermore, it would be convenient if the results can be interfaced directly with the higher order theoretical calculations, renormalization group equation evolutions, etc. Unfortunately, it is impossible to achieve all of these goals simultaneously. 

Two popular frameworks are, instead, chosen for the interpretation of the CEPC results: the so-called $\kappa$-framework 
\cite{Dawson:2013bba,deFlorian:2016spz,Han:2013kya,Peskin:2013xra,Giardino:2013bma,Belanger:2013xza,Bechtle:2014ewa,Cheung:2014noa, Fichet:2015xla, Lafaye:2017kgf} and the effective field theory (EFT) frameworks~\cite{Barger:2003rs, Corbett:2012ja, Elias-Miro:2013mua, Pomarol:2013zra, Amar:2014fpa, Ellis:2014jta, Falkowski:2015fla, Butter:2016cvz, Craig:2014una, Beneke:2014sba, Craig:2015wwr, Ellis:2015sca, Ge:2016zro, Ellis:2017kfi, Durieux:2017rsg, Barklow:2017suo, Barklow:2017awn, DiVita:2017vrr, Chiu:2017yrx, Ellis:2018gqa, Durieux:2018ggn}. As discussed in more detail later, none of these is perfect. But neither of these is wrong as long as one is careful not to over interpret the results. Another important aspect of making projections on the physics potential of a future experiment is that they need to be compared with other experiments. The choices made here follows the most commonly used approaches to facilitate such comparisons. In the later part of this section, Higgs physics potential beyond coupling determination is also discussed.

\subsection{Coupling Fits in the $\kappa$-framework}

The Standard Model makes specific predictions for the Higgs boson couplings to the SM fermions, $g_{\rm SM}(Hff)$, and to the SM gauge bosons, $g_{\rm SM}(HVV)$. In the $\kappa$-framework,  the potential deviations from the SM are parametrized using the $\kappa$ parameters defined as:
\begin{equation}
 \kappa_{f} = \frac{g(Hff)}{g_{\rm SM}(Hff)},\ \  \kappa_{V} = \frac{g(HVV)}{g_{\rm SM}(HVV)},
\end{equation}
with $\kappa_i=1$ being the SM prediction. The rates of the Higgs boson production and decays are modified accordingly. For example, 
\begin{equation}
\begin{alignedat}{3}
  \sigma(ZH)&=&  \kappa_Z^2\,\cdot\, \sigma_{\rm SM}(ZH)& \\  
  \sigma(ZH)\times\BR(H\to ff)  &=& 
  \frac{\kappa_Z^2\kappa_f^2}{\kappa_\Gamma^2}\,\cdot\,\sigma_{\rm SM}(ZH)&\times\BR_{\rm SM}(H\to ff) 
\end{alignedat}
\end{equation}
Here $\kappa_\Gamma^2(\equiv \Gamma_H/\Gamma_H^{\rm SM})$ parametrizes the change in the Higgs boson width due to both the coupling modifications and the presence of BSM decays.

Apart from the tree-level couplings,  there are also loop-level couplings of $Hgg$, $H\gamma\gamma$ and $HZ\gamma$ in the SM. In the absence of new physics, these couplings, often referred to as the effective couplings, can be expressed using the $\kappa$ parameters, described previously. 
However, new physics states in the loops can  alter these couplings. For this reason, three additional $\kappa$ parameters:  $\kappa_g$, $\kappa_\gamma$ and $\kappa_{Z\gamma}$ are introduced to parametrize the potential deviations from the SM for the three effective $Hgg$, $H\gamma\gamma$ and $HZ\gamma$ couplings, respectively.

It is possible that the Higgs boson can decay directly into new particles or have BSM decays to SM particles. In this case,  two types of new decay channels should be distinguished:
\begin{enumerate}
 \item   Invisible decay. This is a specific channel in which Higgs boson decay into new physics particles that are ``invisible'' in the detector. Such decays can be specifically searched for.  If detected,  its rate can be measured. The CEPC sensitivity to this decay channel is quantified by the upper limit on $\BRinvBSM$.
    
 \item   Exotic decays. These include all the other new physics channels. Whether they can be observed, and, if so, to what precision, depends sensitively on the  final states. In one extreme, the final states can be very distinct, and the rate can be well measured. In the another extreme, they can be completely swamped by the background. Without the knowledge of the final states and the corresponding expected CEPC sensitivity, the exotic decays are accounted for by treating the Higgs boson width $\Gamma_H$ as an independent free parameter in the interpretation. 
\end{enumerate}

In general, possible deviations of all SM Higgs boson couplings should be considered.  
However, in the absence of obvious light new physics states with large couplings to the Higgs boson or to other SM particles, a very large deviation ($> \mathcal{O}(1)$) is unlikely.
For smaller deviations, the Higgs  phenomenology is not sensitive to
the deviations of $\kappa_e$, $\kappa_u$, $\kappa_d$ and $\kappa_s$ as the Higgs boson couplings to these particles are negligible compared with the couplings to other particles~\cite{Gao:2016jcm}.
Therefore, these $\kappa$ parameters are set to unities.

The CEPC will not be able to directly measure the Higgs boson coupling to top quarks. A deviation
of this coupling from its SM value does enter the $Hgg$, $H\gamma\gamma$ and $HZ\gamma$ amplitudes. However, this effect is parametrized by $\kappa_g$, $\kappa_\gamma$ and $\kappa_{Z\gamma}$ already. Therefore, $\kappa_t$ is not considered as an independent parameter. For simplicity, previous studies often do not include $\kappa_{Z\gamma}$ in the fit\footnote{Adding $\kappa_{Z\gamma}$ back in the decay process would only lead to completely negligible changes in the projection for other parameter and the precision on $\kappa_{Z\gamma}$ itself would be 8\%.}. We will follow this approach here. This leaves the following set of 10 
independent parameters: 
\begin{equation}
  \kappa_b,\ \kappa_c,\ \kappa_g,\ \kappa_W,\ \kappa_\tau,\ \kappa_Z,\ \kappa_\gamma,\ \kappa_\mu,\ \BRinvBSM, \ \Gamma_H.
\label{eq:10parameters}
\end{equation}
Additional assumptions can be made to reduce the number of parameters~\cite{LHCHiggsCrossSectionWorkingGroup:2012nn,Heinemeyer:2013tqa}. For example, it can be reduced to a 7-parameter set, by assuming lepton universality, and the absence of exotic and invisible decays (excluding $\HZZ\to\vv\vv$) \cite{Dawson:2013bba,LHCHiggsCrossSectionWorkingGroup:2012nn}:
\begin{equation}
  \kappa_b,\ \kappa_c,\ \kappa_g,\ \kappa_W,\ \kappa_Z,\ \kappa_\gamma,\ \kappa_\tau =\kappa_\mu.
\label{eq:yparameters}
\end{equation}
This is useful for studies at hadron colliders as the Higgs boson total width cannot be measured with good precision. The interpretation of the CEPC results is also performed using this reduced set to allow for direct comparisons with the expected HL-LHC sensitivity.

The $\kappa_i$ parameters give a simple and intuitive parametrization of the potential deviations. It has a direct connection with the observables shown in Table~\ref{tab:HiggsSigmaBR} and does cover many possible modifications of the couplings. However, the $\kappa$-framework has its limitations as well. Strictly speaking, it should not be understood as the modification of the SM renormalizable couplings by a multiplicative factor. 
For instance, some of such $\kappa$ modifications violate gauge invariance. Higher order corrections in the $\kappa$-framework cannot be easily defined.
Moreover, the $\kappa_i$ parameters do not include all possible effects of new physics either. For example, apart from the overall size, potential new physics can also introduce form factors which can change the kinematics of particles that couple to a particular vertex. Manifestations of this effect can be seen in the EFT analysis. It is useful to compare with the EFT analysis discussed in the next subsection. The EFT relates $\kappa_Z$ and $\kappa_W$, and further expands them into three different Lorentz structures. Moreover, some of these higher dimensional $HVV$ couplings are also connected with $\kappa_\gamma$ and anomalous trilinear gauge couplings. 
The current EFT analysis does not include any new light degrees of freedom, in contrast to the $\kappa$-framework with independent parameters $\BRinvBSM$ and $\Gamma_H$. 
Overall, $\kappa$-framework does capture the big picture of the CEPC capability in precision Higgs boson measurements. It is useful as long as its limitations are understood. 

The LHC and especially the HL-LHC will provide valuable and complementary information about the Higgs boson properties. For example, the LHC is capable of directly measuring the $t\bar{t} H$ process~\cite{Sirunyan:2018hoz,Aaboud:2018urx}. It can also use differential cross sections to differentiate contributions between the top-quark and other heavy particle states in the loop of the $Hgg$ vertex~\cite{Banfi:2013yoa,Azatov:2013xha,Grojean:2013nya,Buschmann:2014twa}. Moreover, it can separate contributions from different operators in the couplings between the Higgs and vector bosons~\cite{Ellis:2014dva}. For the purpose of the coupling fit in the $\kappa$-framework, the LHC, with its large statistics, improves the precision of rare processes such as $H\to \gamma\gamma$. Note that a large portion of the systematic uncertainties intrinsic to a hadron collider can be canceled by taking ratios of measured cross sections. For example, combining the ratio of the rates of  $pp \to \Hyy$ and $pp \to \HZZ$ at the LHC and the measurement of the $HZZ$ coupling at the CEPC can significantly improve the $\kappa_\gamma$ precision.  These are the most useful inputs from the LHC to combine with the CEPC. Similar studies of combination with the LHC for the ILC can be found in Refs.~\cite{Han:2013kya,Klute:2013cx,Peskin:2013xra,Fujii:2017vwa,Barklow:2017suo}.

The results of the 10-parameter and the 7-parameter fits for the CEPC with an integrated luminosity of $\abih$ are shown in Table~\ref{tab:kappa-fit}.\footnote{Theoretical uncertainties associated with the cross section and Higgs boson property calculations are ignored in these fits as both will be improved and are expected to be smaller than the statistical uncertainties ~\cite{Sun:2016bel,Gong:2016jys,Lepage:2014fla} by the time of the CEPC experiment.}
The combined precision with the HL-LHC estimates (using fit result number 15 of Ref.~\cite{ATL-PHYS-PUB-2014-016}) are also shown. The HL-LHC estimates used assume no theoretical uncertainties and thus represent the aggressive HL-LHC projection.\footnote{Note that the LHC and the CEPC have different sources of theoretical uncertainties, for detailed discussion, see Refs.~\cite{Dawson:2013bba,Heinemeyer:2013tqa,Denner:2011mq,Almeida:2013jfa,Lepage:2014fla}.} It is assumed that the HL-LHC will operate at $\sqrt{s}=14$~TeV and accumulate an  integrated luminosity of 3000~$\fbi$. For the 7-parameter fit, the Higgs boson width is a derived quantity, not an independent parameter. Its precision, derived from the precision of the fitted parameters, is 2.4\% for the CEPC alone and 1.8\% when combined with the HL-LHC projection.

\begin{figure*}
\centering
\includegraphics[width=0.84\textwidth]{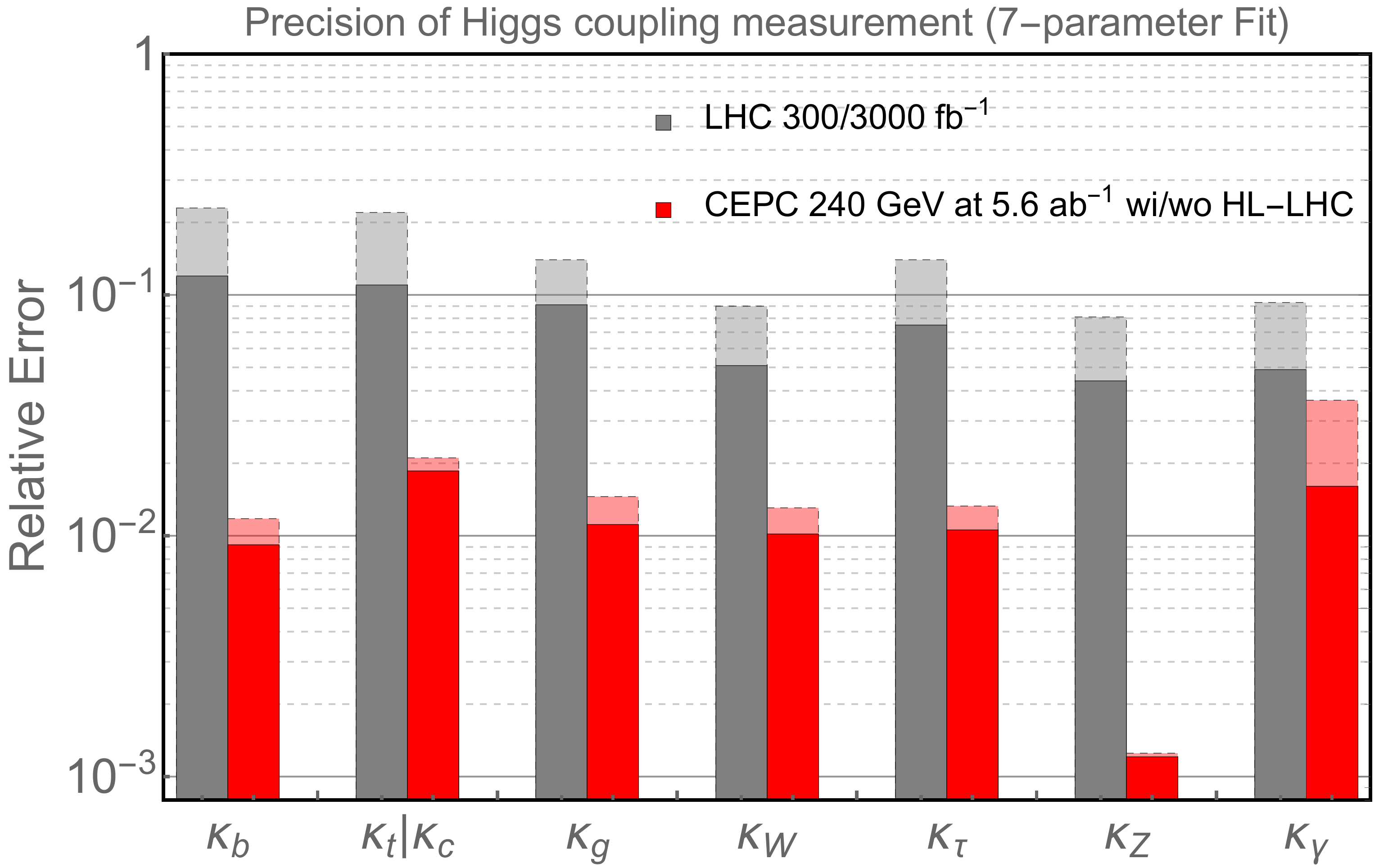} 
\caption{The results of the 7-parameter fit and comparison with the HL-LHC~\cite{ATL-PHYS-PUB-2014-016}. The projections for the CEPC at 240 GeV with an integrated luminosity of $\abih$ are shown. The CEPC results without combination with the HL-LHC input are  shown as light red bars. The LHC projections for an integrated luminosity of 300~$\fbi$ are shown in light gray bars.
}
\label{fig:7para}
\includegraphics[width=0.84\textwidth]{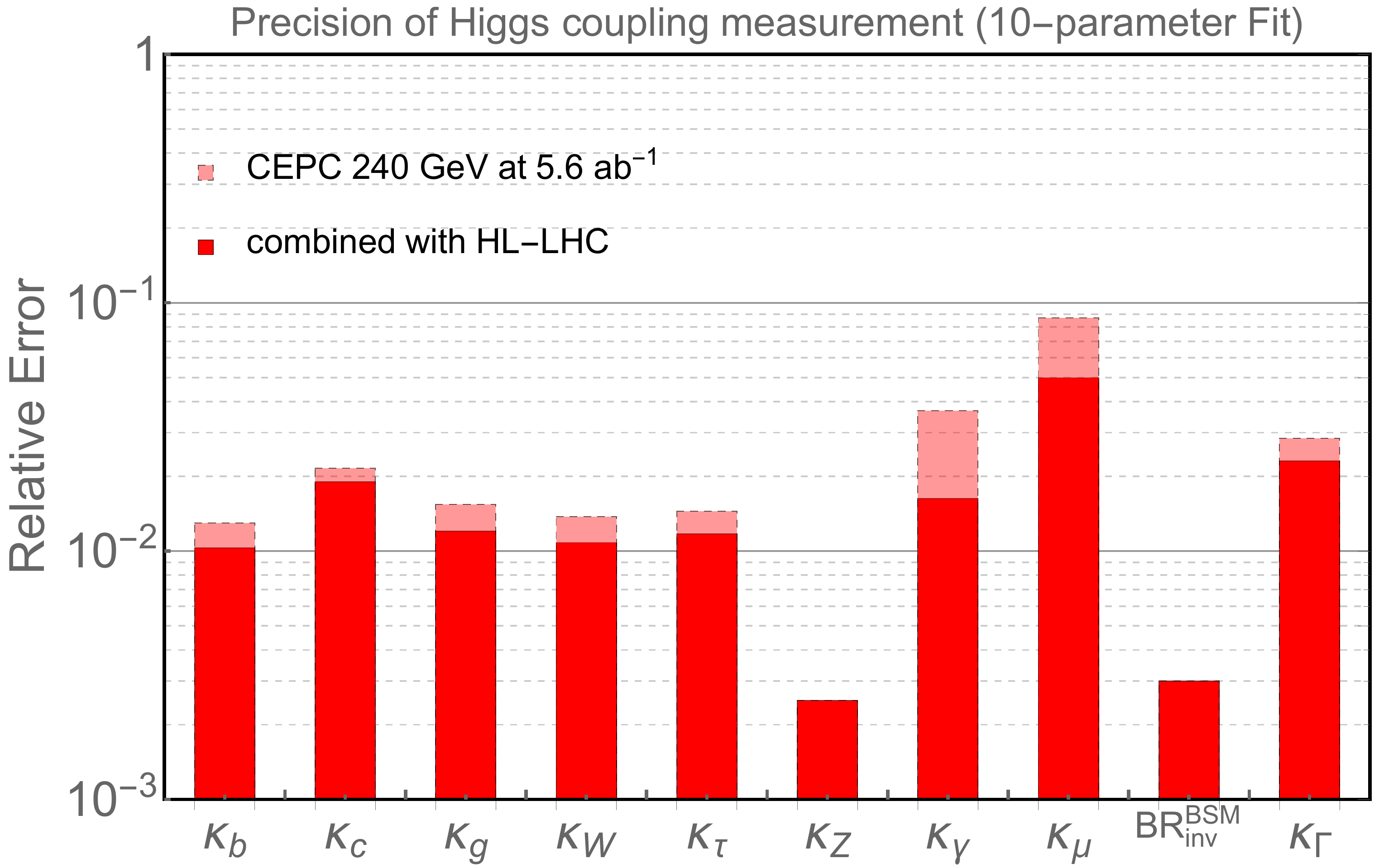}
\caption{
The 10 parameter fit results for the CEPC at 240 GeV with an integrated luminosity of $\abih$ (light red bars) and for the combination with the HL-LHC inputs (dark red bars). All the numbers are relative precision except for $\BRinvBSM$ for which the 95\% CL upper limit are quoted.
}
\label{fig:10para}
\end{figure*}

The CEPC Higgs boson property measurements mark a giant step beyond the HL-LHC. First of all, in contrast to the LHC, a lepton collider Higgs factory is capable of measuring the Higgs boson width and the absolute coupling strengths to other particles.  A comparison with the HL-LHC is only possible with model dependent assumptions. One of such comparisons is within the framework of the 7-parameter fit, shown in Fig.~\ref{fig:7para}.  Even with this set of restrictive assumptions, the advantage of the CEPC is still significant. The measurement of $\kappa_Z$ is more than a factor of 10 better. The CEPC can also improve significantly the precision on a set of $\kappa$ parameters that are affected by large backgrounds at the LHC, such as $\kappa_b$, $\kappa_c$, and $\kappa_g$. Note that this is in comparison with the HL-LHC projection with large systematic uncertainties. Such uncertainties are typically under much better control at lepton colliders. Within this 7-parameter set, the only coupling that the HL-LHC can give a competitive measurement is $\kappa_\gamma$, for which the CEPC sensitivity is statistically limited. This is also the most valuable input that the HL-LHC can give to the Higgs boson coupling measurements at the CEPC, which underlines the importance of combining the results from these two facilities. 

The direct search for the Higgs boson decay to invisible particles from BSM physics is well motivated and closely connected to the dark sectors. The CEPC with an integrated luminosity of  $\abih$ has a sensitivity of 0.30\% expressed in terms of the 95\% CL upper limit on the decay branching ratio, as shown in Table~\ref{tab:kappa-fit}. The HL-LHC, on the other hand, has a much lower sensitivity of 6--17\%~\cite{Dawson:2013bba} while optimistically may reach 2--3.5\%~\cite{Bernaciak:2014pna}. 

\begin{table*}
\caption{\label{tab:kappa-fit}Coupling measurement precision from the 10-parameter fit and 7-parameter fit described in the text for the CEPC, and corresponding results after combination with the HL-LHC.  All the numbers refer to are relative precision except for $\BRinvBSM$ for which the 95\% CL upper limit are quoted respectively. Some entries are left vacant for the 7-parameter fit as they are not dependent parameters under the fitting assumptions.
}
\begin{center}
\begin{tabular}{c|cc|cc}\hline\hline
\multicolumn{5}{c}{Relative coupling measurement precision and the 95\% CL upper limit on $\BRinvBSM$} \\ \hline
&  \multicolumn{2}{c}{10-parameter fit}  &  \multicolumn{2}{|c}{7-parameter fit} \\ \cline{2-5}
Quantity& \hcp CEPC \hcp   & CEPC+HL-LHC  & \hcp CEPC \hcp   & CEPC+HL-LHC  \\ \hline
$\kappa_b$ 	&    1.3\%  &    1.0\%  &  1.2\%  &   0.9\% \\
$\kappa_c$ 	&    2.2\%  &    1.9\%  &  2.1\%  &   1.9\% \\
$\kappa_g$ 	&    1.5\%  &    1.2\%  &  1.5\%  &   1.1\% \\
$\kappa_W$ 	&    1.4\%  &    1.1\%  &  1.3\%  &   1.0\% \\
$\kappa_\tau$ 	&    1.5\%  &    1.2\%  &  1.3\%  &   1.1\% \\
$\kappa_Z$ 	&   0.25\%  &   0.25\%  & 0.13\%  &  0.12\% \\
$\kappa_\gamma$ &    3.7\%  &    1.6\%  &  3.7\%  &   1.6\% \\
$\kappa_\mu$ 	&    8.7\%  &    5.0\%  & --  & --  \\
$\BRinvBSM$     & $<0.30\%$ & $<0.30\%$ & --  & -- \\
$\Gamma_H$ 	&    2.8\%  &    2.3\%  & --  & --  \\
\hline\hline
\end{tabular}
\end{center}
\end{table*}

As discussed above, one of the greatest advantages of a lepton collider Higgs factory is its capability to measure the Higgs boson width and couplings in a {\it model-independent} way. The projection of such a determination at the CEPC is shown in Fig.~\ref{fig:10para}. 
For most of the measurements, an order of magnitude improvements over the HL-LHC are expected.  The CEPC has a clear advantage in the measurement of $\kappa_Z$. It can also set a much stronger constraint on  $\BRinvBSM$.

\subsection{Effective-Field-Theory Analysis}
\label{subsec:eft}

With the assumption that the scale of new physics is higher than the relevant energy directly accessible at the Higgs factory,  the effect of new physics can be characterized within the EFT framework. In this framework, operators with dimension greater than four supplement the SM Lagrangian.   Imposing baryon and lepton numbers conservation, all higher dimensional operators are of even dimension:
\begin{equation}
\mathcal{L}_{\rm EFT} =
    \mathcal{L}_{\rm SM} + 
    \sum_i \frac{c^{(6)}_i}{\Lambda^2} \mathcal{O}^{(6)}_i + 
    \sum_j \frac{c^{(8)}_j}{\Lambda^4} \mathcal{O}^{(8)}_j +         \cdots   \label{eq:smeft}
\end{equation}
%
%
where $\Lambda$ is the new physics scale. The leading new physics effects at the electroweak scale would be from the dimension-six operators.  To obtain robust constraints on the Wilson coefficients, $c_i$, a global analysis is required, which includes contributions from all possible dimension-six operators.  While a large number of dimension-six operators can be written down, only a subset of them contribute to the Higgs boson processes at the leading order.  Among these operators, some are much better constrained by other measurements.  It is thus reasonable to focus on the operator that primarily contribute to the Higgs boson processes and therefore reduce the parameter space by making appropriate assumptions, as done in the recent studies for future lepton colliders~\cite{Ellis:2015sca, Ellis:2017kfi, Durieux:2017rsg, Barklow:2017suo, Barklow:2017awn, DiVita:2017vrr, Chiu:2017yrx}.  Following these studies, the $\CP$-violating operators as well as the ones that induce fermion dipole interactions are discarded.  At the leading order, $\CP$-violating operators do not have linear contributions to the rates of the Higgs boson processes.  While they do contribute to  angular observables at the leading order~\cite{Beneke:2014sba, Craig:2015wwr}, these operators are usually much better constrained by the Electric Dipole Moment~(EDM) experiments~\cite{Barr:1990vd, Fan:2013qn, Baron:2013eja}, though some parameter space is still available for the $\CP$-violating couplings of the Higgs boson to heavy flavor quarks and leptons~\cite{Chien:2015xha, Harnik:2013aja}.  The interference between the fermion dipole interactions with SM terms are suppressed by the fermion masses.  
The corresponding operators also generate dipole moments, which are stringently constrained, especially for light fermions.  For the operators that modify the Yukawa coupling matrices, only the five diagonal ones that correspond to the top, charm, bottom, tau, and muon Yukawa couplings are considered, which are relevant for the Higgs boson measurements at the CEPC.  

Before presenting the projections, some brief comments on the EFT framework are in order. In comparison with the $\kappa$-framework, a significant advantage of the EFT is that it gives physical parametrization of potential new physics effects. EFT operators can be used directly in computations. The EFT framework also allow for a natural inclusion of new observables, with possible correlations automatically taken into account. At the same time, the connections with experimental observables are less direct and intuitive. Sometimes, the EFT approach is referred to as model-independent. This is only accurate to a certain extent. It assumes that there are no new light degrees of freedom. In practice, assumptions are often made to simplify the set of EFT operators, as also done here. 

The electroweak precision observables are already tightly constrained by the LEP $Z$-pole and $W$ mass measurements.  The CEPC $Z$-pole run can further improve the constraints set by the LEP, thanks to the enormous amount ($\sim 10^{11}$--$10^{12}$) of $Z$ bosons. The $W$ mass can also be measured with a precision of a few MeVs at the CEPC even without a dedicated $WW$ threshold run.  Given that the expected precision of the $Z$-pole observables and the $W$ mass are much higher than the ones of Higgs boson observables, it is assumed that the former ones are perfectly constrained, which significantly simplifies the analysis.  In particular, in a convenient basis all the contact interaction terms of the form $HVf\bar{f}$ can be discarded since they also modify the fermion gauge couplings.  Realistic $Z$-pole constraints have also been considered in recent studies~\cite{Barklow:2017suo, Barklow:2017awn, Chiu:2017yrx}, but certain assumptions (such as flavor-universality) and simplifications are made.  Future studies with more general frameworks are desired to fully determine the impact of the $Z$-pole measurements on the Higgs boson analysis.

The measurements of the triple gauge couplings (TGCs) from the diboson process $e^+e^- \to WW$ play an important role in the Higgs boson coupling analysis under the EFT framework.  Focusing on $\CP$-even dimension-six operators, the modifications to the triple gauge vertices from new physics can be parametrized by three anomalous TGC parameters (aTGCs),  conventionally denoted as $\delta g_{1, Z}$, $\delta \kappa_\gamma$ and $\lambda_Z$~\cite{Hagiwara:1993ck, Gounaris:1996rz}.  Among them, $\delta g_{1,Z}$ and $\delta \kappa_\gamma$ are generated by operators that also contribute to the Higgs boson processes.  At 240\,GeV, the $e^+e^- \to WW$ process cross section is almost two orders of magnitude larger than that of the $\ZH$ process.  The measurements of the diboson process thus provide strong constraints on the operators that generate the aTGCs.  
A dedicated study on the TGC measurements at the CEPC is not currently available.  A simplified analysis is thus performed to estimate the aTGC sensitivity.  The results are shown in Table~\ref{tab:tgccepc}.  The analysis roughly follows the methods in Refs.~\cite{Durieux:2017rsg, Bian:2015zha}.  Only the $WW$ events in the semi-leptonic (electron or muon) channel are used, which are easier to reconstruct and have a sizable branching ratio ($\approx29\%$).  In particular, the production polar angle, as well as the two decay angles of the leptonically decaying $W$ boson, can be fully reconstructed, which contain important information on the aTGCs.  The two decay angles of the hadronically decaying  $W$ boson can only be reconstructed with a two-fold ambiguity.  
A $\chi^2$ fit of the three aTGC parameters to the binned distribution of all five angles is performed, from which the one-sigma interval for each of the three aTGCs as well as the correlations among them are extracted.  
A signal selection efficiency of 80\% is assumed.  The effects of systematic uncertainties and backgrounds are not considered, assuming they are under control after the selection cuts.

%
\begin{center}
\tabcaption{\label{tab:tgccepc}The estimated constraints on aTGCs from the measurements of the diboson process ($e^+e^- \to WW$) in the semi-leptonic channel at the CEPC 240\,GeV with $\abih$ data and unpolarized beams.  All angular distributions are used in the fit.  Only the statistical uncertainties of the signal events are considered, assuming a selection efficiency of 80\%.
}
\begin{tabular}{|c||c|ccc|} \hline
\multicolumn{5}{|c|}{CEPC 240\,GeV ($\abih$)}  \\  \hline
& uncertainty &   \multicolumn{3}{c|}{correlation matrix}    \\ \cline{2-5}
&  &    $\delta g_{1,Z}$  &  $\delta \kappa_\gamma$  & $\lambda_Z$      \\ \hline
$\delta g_{1,Z}$          &    $1.2\times10^{-3}$    &    1 & 0.08 & -0.90    \\
$\delta \kappa_\gamma$    &    $0.9\times10^{-3}$    &      &  1   &  -0.42    \\
$\lambda_Z$               &     $1.3\times10^{-3}$   &      &      & 1          \\  \hline
\end{tabular}
\end{center}
%

%

Under the assumptions specified above, the dimension-six operator contribution to the Higgs boson and diboson processes consists of a total of twelve degrees of freedom.  While all non-redundant bases are equivalent, it is particularly convenient to choose a basis in which the twelve degrees of freedom can be mapped to exactly twelve operators, whereas the rest are removed by the assumptions.  Two such bases are considered in this analysis. The first is defined by the set of dimension-six operators in Table~\ref{tab:op1}.  Among them, $\mathcal{O}_{3W}$ corresponds to the aTGC parameter $\lambda_Z$, $\mathcal{O}_{HW}$ and $\mathcal{O}_{HB}$ generate the aTGC parameters $\delta g_{1,Z}$ and $\delta \kappa_\gamma$ as well as Higgs boson anomalous couplings, while the rest operators can only be probed by the Higgs boson measurements at the leading order.
The second basis is the so-called ``Higgs basis,'' proposed in Ref.~\cite{Falkowski:2001958}.  In the Higgs basis, the parameters are defined 
in terms of the mass eigenstates after the electroweak symmetry breaking, and can be directly interpreted as the size of the Higgs boson couplings.   Different from the original Higgs basis, this analysis follows Ref.~\cite{Durieux:2017rsg}, with the parameters associated with the $Hgg$, $H\gamma\gamma$ and $HZ\gamma$ vertices normalized to the SM one-loop contributions, and denoted as $\bar{c}_{gg}$, $\bar{c}_{\gamma\gamma}$ and $\bar{c}_{Z\gamma}$ (as opposed to $c_{gg}$, $c_{\gamma\gamma}$ and $c_{Z\gamma}$ in Ref.~\cite{Falkowski:2001958}).  The parameter $\bar{c}^{\rm \,eff}_{gg}$ is further defined to absorb all contributions to the $Hgg$ vertex.  
With these redefinitions, the set of twelve parameters is given by
\begin{equation}\footnotesize
    \delta c_Z        \,,\,
    c_{ZZ}            \,,\,
    c_{Z\square}        \,,\,
    \bar{c}_{\gamma\gamma}    \,,\,
    \bar{c}_{Z\gamma}        \,,\,
    \bar{c}^{\rm \,eff}_{gg}       \,,\,
    \delta y_t        \,,\,
    \delta y_c        \,,\,
    \delta y_b        \,,\,
    \delta y_\tau        \,,\,
    \delta y_\mu        \,,\,
    \lambda_Z        \,.
\label{eq:para12}
\end{equation}

These parameters can be conveniently interpreted as the precision of the Higgs boson couplings analogous to those in the $\kappa$-framework.  In particular, $\delta c_Z$, $\bar{c}_{\gamma\gamma}$, $\bar{c}_{Z\gamma}$, $\bar{c}^{\rm \,eff}_{gg}$ and $\delta y_{t,\,c,\,b,\,\tau,\,\mu}$ modifies the sizes of the SM Higgs boson couplings to $ZZ$, $\gamma\gamma$, $Z\gamma$, $gg$ and fermions, respectively.  $c_{ZZ}$ and $c_{Z\square}$ parametrize the anomalous $HZZ$ couplings: 
\begin{equation}
\mathcal{L} =  \frac{H}{v} \bigg[ c_{ZZ} \, \frac{g^2+g'^2}{4} Z_{\mu\nu} Z^{\mu\nu} + c_{Z\square} \, g^2 Z_\mu \partial_\nu Z^{\mu\nu}  \bigg] + ... \,,
\end{equation}
which are not present in the SM at the leading order.  The $HWW$ couplings are written in terms of the parameters shown in Eq.~\ref{eq:para12} via gauge invariance and are not shown explicitly.  For the three aTGC parameters, $\lambda_Z$ is kept in Eq.~\ref{eq:para12}, while $\delta g_{1,Z}$ and $\delta \kappa_\gamma$ are written in terms of the linear combinations of $ c_{ZZ}$, $c_{Z\square}$, $\bar{c}_{\gamma\gamma}$ and $\bar{c}_{Z\gamma}$.  The exact definitions of the Higgs basis and the translation to the basis in Table~\ref{tab:op1} can be found in Ref.~\cite{Durieux:2017rsg}.

\begin{table*}
\begin{center}
\caption{\label{tab:op1} A complete set of $\CP$-even dimension-six operators that contribute to the Higgs boson and TGC measurements, assuming there is no correction to the $Z$-pole observables and the $W$~mass, and also no fermion dipole interaction. $G^A_{\mu\nu}$, $W^a_{\mu\nu}$ and $B_{\mu\nu}$ are the field strength tensors for the SM $SU(3)_c$, $SU(2)_L$ and $U(1)_Y$ gauge fields, respectively.  For $\mathcal{O}_{y_u}$, $\mathcal{O}_{y_d}$ and $\mathcal{O}_{y_e}$, only the contributions to the diagonal elements of the Yukawa matrices that corresponds to the top, charm, bottom, tau, and muon couplings are considered.}
\begin{tabular}{l|l} \hline\hline
$\mathcal{O}_H = \frac{1}{2} (\partial_\mu |H^2| )^2$ &  $\mathcal{O}_{GG} =  g_s^2 |H|^2 G^A_{\mu\nu} G^{A,\mu\nu}$  \\ 
$\mathcal{O}_{WW} =  g^2 |H|^2 W^a_{\mu\nu} W^{a,\mu\nu}$  & $\mathcal{O}_{y_u} = y_u |H|^2 \bar{Q}_L \tilde{H} u_R \,+\, {\rm h.c.}$ \hspace{0.25cm} {\scriptsize $(u \to t, c)$}  \\
$\mathcal{O}_{BB} =  g'^2 |H|^2 B_{\mu\nu} B^{\mu\nu}$ &  $\mathcal{O}_{y_d} = y_d |H|^2 \bar{Q}_L H d_R \,+\, {\rm h.c.}$ \hspace{0.3cm} {\scriptsize $(d \to b)$}  \\
$\mathcal{O}_{HW} =  ig(D^\mu H)^\dagger \sigma^a (D^\nu H) W^a_{\mu\nu}$  &  $\mathcal{O}_{y_e} = y_e |H|^2 \bar{L}_L H e_R \,+\, {\rm h.c.}$ \hspace{0.36cm} {\scriptsize $(e \to \tau, \mu)$}  \\
$\mathcal{O}_{HB} =  ig'(D^\mu H)^\dagger  (D^\nu H) B_{\mu\nu}$ &      $\mathcal{O}_{3W} = \frac{1}{3!} g \epsilon_{abc} W^{a\,\nu}_\mu W^b_{\nu \rho} W^{c\,\rho\mu} $  \\   \hline\hline
\end{tabular}
\end{center}
\end{table*}

\begin{figure*}
\centering
\includegraphics[width=.95\textwidth]{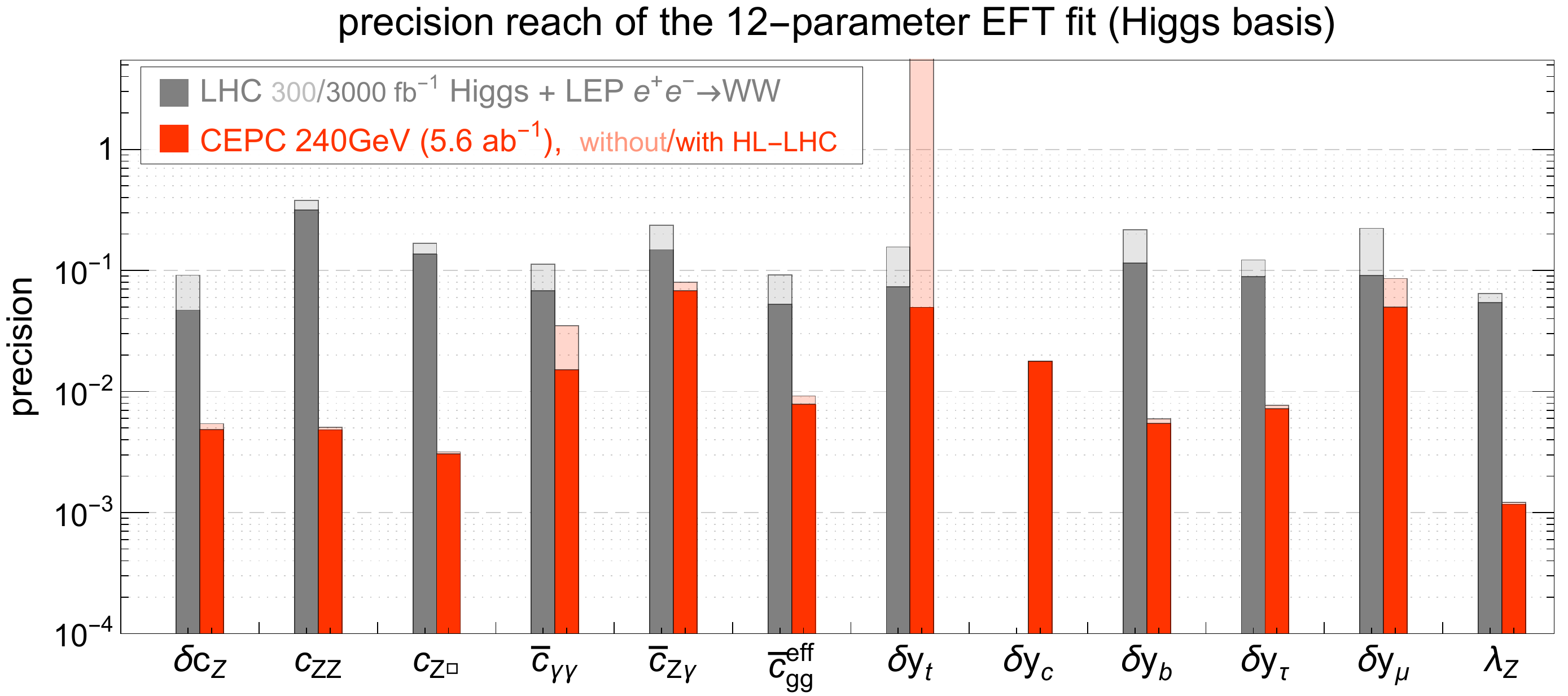}
\caption{One-sigma precision of the twelve parameters in the Higgs basis.  The first column shows the results from the LHC Higgs boson measurements with $300\,{\rm fb}^{-1}$ (light gray bar) and $3000\,{\rm fb}^{-1}$ (dark gray bar) combined with the LEP diboson ($e^+e^- \to WW$) measurement.  The second column shows the results from the CEPC with $\abih$ data collected at 240\,GeV with unpolarized beam.  The results from the CEPC alone are shown in light red bars, and the ones from a combination of the CEPC and the HL-LHC are shown in dark red bars. For the LHC fits, $\delta y_c$ is fixed to zero.
}
\label{fig:fit1h}

\includegraphics[width=.95\textwidth]{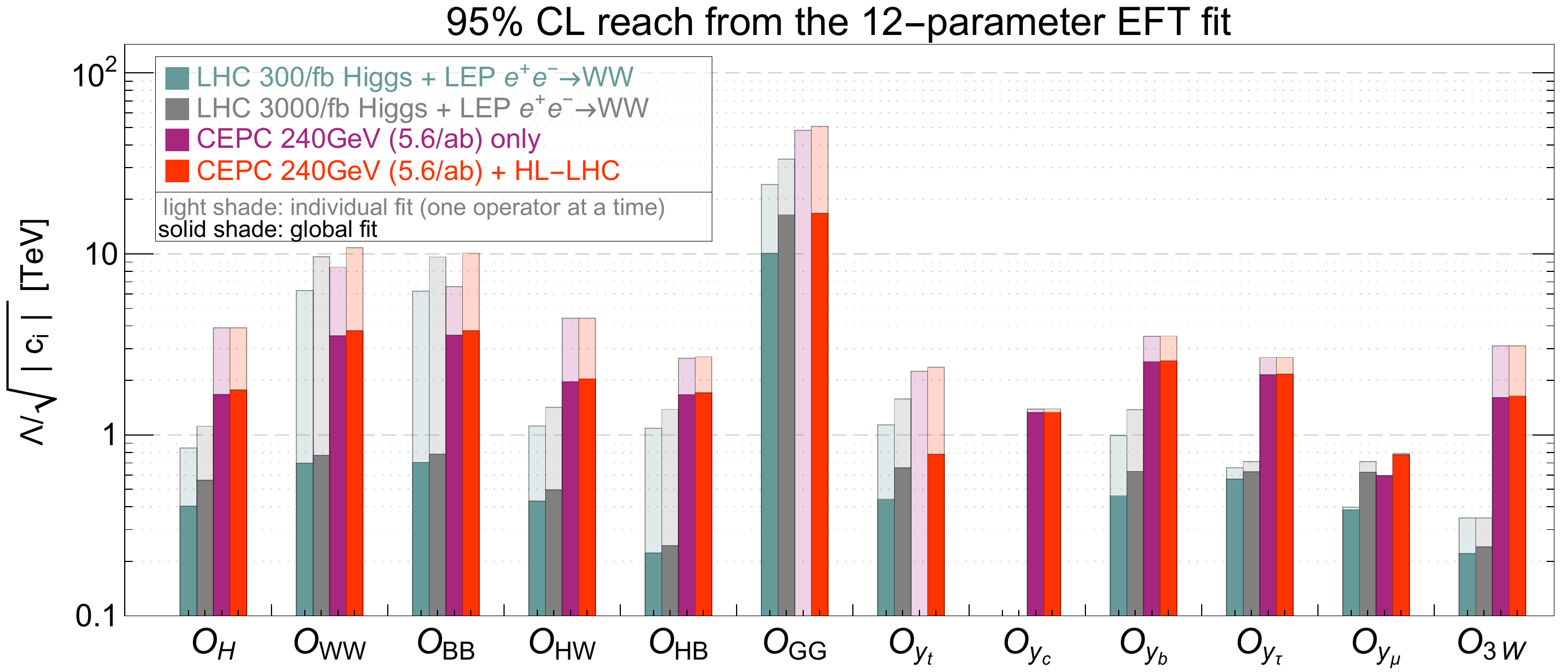}%
\caption{The 95\%\,CL sensitivity to $\Lambda/\sqrt{|c_i|}$ for the operators in the basis defined in Table~\ref{tab:op1}. The first two columns show the results from the LHC Higgs boson measurements with $300\,{\rm fb}^{-1}$ and $3000\,{\rm fb}^{-1}$ combined with the LEP diboson ($e^+e^- \to WW$) measurement.  The last two columns  show the results from the CEPC alone and the combination of the CEPC and the HL-LHC ($3000\,{\rm fb}^{-1}$).  The results of the global fits are shown with dark colored bars.  The results from individual fits (by switching on one operator at a time) are shown with light colored bars. 
For the LHC fits, $\delta y_c$ is fixed to zero.
}
\label{fig:fit1s}
\end{figure*}

The estimated precision of all the Higgs boson rate measurements in Section~\ref{sec:combinations} (Table~\ref{tab:HiggsSigmaBR}), along with their correlations, are included as inputs to the EFT global analysis.  In addition, the angular observables of the $\ZH, \, Z\to \ell^+\ell^-, \, \Hbb$ channel are included, following the studies in Refs.~\cite{Beneke:2014sba, Craig:2015wwr}.  This channel is almost background-free after the selection, with a signal selection efficiency of about 40\%. For the TGC measurements, the results in Table~\ref{tab:tgccepc} are used as inputs. 
The global $\chi^2$ is obtained by summing over the $\chi^2$ of all the measurements.  Due to the high precision of the measurements, it is shown that for all observables, keeping only the linear terms of all EFT parameters gives a very good approximation~\cite{Durieux:2017rsg}.  This greatly simplifies the fitting procedure, as the total $\chi^2$ can be written as
\begin{equation}
    \chi^2 = \sum_{ij} (c-c_0)_i  \, \sigma^{-2}_{ij} \, (c-c_0)_j
    \,, \mbox{where}\ \vspace{1cm}
    \sigma^{-2}_{ij} \equiv \left( \delta c_i \: \rho_{ij} \: \delta c_j  \right)^{-1}
    \,,  \label{eq:chipara}
\end{equation}
where $c_i$'s are the EFT parameters, $c_0$'s are the corresponding central values which are zero by construction, as the measurements are assumed to be SM-like.  The one-sigma uncertainties $\delta c_i$ and the correlation matrix $\rho$ can be obtained from $\sigma^{-2}_{ij} = {\partial^2\,\chi^2} \big/ {\partial c_i \partial c_j}$.

For comparison, the sensitivities of the LHC 14\,TeV with total luminosities of $300\,{\rm fb}^{-1}$ and $3000\,{\rm fb}^{-1}$ are also considered. These are combined with the diboson ($e^+e^- \to WW$) measurements at the LEP as well as the LHC 8\,TeV Higgs boson measurements.  For the LHC 14\,TeV Higgs boson measurements, the projections by the ATLAS collaboration~\cite{ATL-PHYS-PUB-2014-016} are used, while the composition of each channel is obtained from Refs.~\cite{ATL-PHYS-PUB-2013-014, ATL-PHYS-PUB-2014-012, ATL-PHYS-PUB-2014-006, ATL-PHYS-PUB-2014-011, ATL-PHYS-PUB-2014-018}.  The constraints from the LHC 8\,TeV Higgs boson measurements and the diboson measurements at the LEP are obtained directly from Ref.~\cite{Falkowski:2015jaa}.  While the LHC diboson measurements can potentially improve the constraints on aTGCs set by the LEP \cite{Butter:2016cvz}, they are not included in this analysis due to the potential issues related to the validity of the EFT \cite{Contino:2016jqw, Falkowski:2016cxu} and the assumption that the TGCs dominated by the non-anomalous terms~\cite{Zhang:2016zsp}.

The results of the 12-parameter fit at the CEPC are shown in Fig.~\ref{fig:fit1h} for the Higgs basis and Fig.~\ref{fig:fit1s} for the basis in Table~\ref{tab:op1}.  The results from the LHC Higgs boson measurements (both $300\,{\rm fb}^{-1}$ and $3000\,{\rm fb}^{-1}$) combined with the LEP diboson measurements are shown in comparison.  The results of the combination of the CEPC with the HL-LHC~($3000\,{\rm fb}^{-1}$) are also shown in addition to the ones from the CEPC alone.  
In Fig.~\ref{fig:fit1h}, the results are shown in terms of the one-sigma precision of each parameter.  The LHC results are shown with gray columns with $300\,{\rm fb}^{-1}$ ($3000\,{\rm fb}^{-1}$) in light (dark) bars, while the CEPC ones are shown with the red columns, with the CEPC-alone (combination with the HL-LHC) results shown in light (dark) bars.

In Fig.~\ref{fig:fit1s}, the results are presented in terms of the sensitivity to $\Lambda/\sqrt{|c_i|}$ at 95\% CL for each operator as defined in Eq.~\ref{eq:smeft}, where $\Lambda$ is the scale of new physics and $c_i$ is the corresponding Wilson coefficient.  Four columns are shown separately for the LHC $300\,{\rm fb}^{-1}$, the HL-LHC $3000\,{\rm fb}^{-1}$, the CEPC alone and the CEPC combined with the HL-LHC.  The results of the global fits, i.e. simultaneous fits to the 12 parameters, are shown with dark colored bars.  The results from individual fits are shown with light colored bars, which are obtained by switching on one operator at a time with the rest fixed to zero. 

%

It is transparent from Fig.~\ref{fig:fit1h} that the CEPC can measure the Higgs boson couplings with  precision that is one order of magnitude better than the LHC~\cite{ATL-PHYS-PUB-2014-016,CMS:2013xfa}.  For the parameters $\bar{c}_{\gamma\gamma}$, $\bar{c}_{Z\gamma}$ and $\delta y_\mu$, the clean signal and small branching ratios of the corresponding channels ($H\to \gamma\gamma/Z\gamma/\mu\mu$) makes the HL-LHC precision comparable to the CEPC.   The combination with the LHC measurements thus provides non-negligible improvements, especially for those parameters. 
It should be noted that, while $\delta y_t$ modifies the $Hgg$ vertex via the top-quark loop contribution, the CEPC alone cannot discriminate it from the $Hgg$ contact interaction 
obtained from integrating out a heavy new particle in the loop.  The parameter $\bar{c}^{\rm \,eff}_{gg}$ absorbs both contributions and reflects the overall precision of the $Hgg$ coupling.  The combination with the LHC $t\bar{t}H$ measurements can resolve this flat direction.  The CEPC measurements, in turn, can improve the constraint on $\delta y_t$ set by the LHC by providing much better constraints on the other parameters that contribute to the $t\bar{t}H$ process.
%
It should also be noted that the measurement of the charm Yukawa coupling is not reported in Ref.~\cite{ATL-PHYS-PUB-2014-016}, while the projection of its constraint has a large variation among different studies and can be much larger than one~\cite{ATL-PHYS-PUB-2015-043, Bodwin:2013gca, Perez:2015aoa, Brivio:2015fxa, Bishara:2016jga, Carpenter:2016mwd}.  Therefore, $\delta y_c$ is fixed to be zero for the LHC-only fits, as treating $\delta y_c$ as an unconstrained free parameter generates a flat direction in the fit which makes the overall sensitivity much worse.  The CEPC, on the other hand, provides excellent measurements of the charm Yukawa coupling and can constrain $\delta y_c$ to about $\sim 2\%$.

Regarding the sensitivity to $\Lambda/\sqrt{|c_i|}$ in Fig.~\ref{fig:fit1s}, it is also clear that the CEPC has a significantly better performance than the LHC.  If the couplings are na\"ively assumed to be of order one ($c_i \sim 1$), the Higgs boson measurements at the CEPC would be sensitive to new physics scales at several TeV.  While the individual sensitivity to some of the operators at the LHC can be comparable to the CEPC ({\it e.g.}, $O_{WW}$ and $O_{BB}$ from the measurement of $H\to \gamma\gamma$), the CEPC sensitivity is much more robust under a global framework. This is due to its comprehensive measurements of both the inclusive $ZH$ cross section and the exclusive rates of many Higgs boson decay channels.
Operators $O_{GG}$ and $O_{y_t}$ both contribute to the $Hgg$ vertex.  While the CEPC can provide strong constraints on either of them if the other is set to zero, they can only be constrained in a global fit if the $t\bar{t}H$ measurements at the LHC are also included.  It is also important to note that the validity of EFT can be a potential issue for the LHC measurements~\cite{Contino:2016jqw}.  Depending on the size of the couplings, the inferred bounds on the new physics scale $\Lambda$ can be comparable with or even smaller than the energy scale probed by the LHC.  The CEPC has a smaller center of mass energy and much better precision, which ensures the validity of EFT for most new physics scenarios.

%
In Table~\ref{tab:deltacepc}, the numerical results of the global fit are presented for the CEPC in terms of the one-sigma band of the 12 parameters and the correlations among them.  The results assume an integrated luminosity of $5.6\,{\rm ab}^{-1}$ at 240\,GeV with unpolarized beams, both without and with the combination with the HL-LHC ($3000\,{\rm fb}^{-1}$) Higgs boson measurements.  With both the one-sigma bounds and the correlation matrix, the corresponding $\chi^2$ can be reconstructed, which can be used to derive the constraints in any other EFT basis or any particular model that can be matched to the EFT.  This offers a convenient way to study the sensitivity to new physics models, as detailed knowledge of the experimental measurements are not required.

%

In the EFT framework, it is explicitly assumed that the Higgs boson width is the sum of all partial widths of its SM decay channels.  
This is because the EFT expansion in Eq.~\ref{eq:smeft} relies on the assumption that the new physics scale is sufficiently high, while any potential Higgs boson exotic decay necessarily introduces light BSM particles, thus in direct conflict with this assumption.  One can nevertheless treat the Higgs boson total width as a free parameter in the EFT global fit and obtain an indirect constraint of it, as done in Ref.~\cite{Barklow:2017suo}.  With this treatment, the CEPC can constrain the Higgs boson width to a precision of $1.7\%$ ($1.6\%$ if combined with the HL-LHC).  This result is significantly better than the one from the 10-parameter coupling fit in Table~\ref{tab:kappa-fit} ($3.4\%/2.6\%$).  The improvement is mainly because the $HWW$ and $HZZ$ couplings are treated as being independent in the 10-parameter coupling fit, while in the EFT framework they are related to each other under gauge invariance and custodial symmetry.
It should also be noted that the Higgs boson width determined using Eqs.~\ref{eq:width1} and~\ref{eq:width2} explicitly assumes that the $HWW$ and $HZZ$ couplings are independent of the energy scale.  Such an assumption is not valid in the EFT framework with the inclusion of the anomalous couplings.

\begin{table*}
\begin{center}\scriptsize
\caption{\label{tab:deltacepc}The one-sigma uncertainties for the 12 parameters from the CEPC (240\,GeV, $\abih$)  in the Higgs basis and the basis of dimension-six operators.  For both cases, the upper (lower) row correspond to results without (with) the combination of the HL-LHC Higgs boson measurements..  Note that, without the $t\bar{t}H$ measurements, $\delta y_t$ can not be constrained in a global fit, thus $c_{GG}$ and $c_{y_t}$ can not be resolved.}
\begin{tabular}{|c|c|c|c|c|c|c|c|c|c|c|c|} \hline
\multicolumn{12}{|c|}{Higgs basis} \\  \hline
$\delta c_Z$ & $c_{ZZ}$ & $c_{Z\square}$ & $\bar{c}_{\gamma\gamma}$ & $\bar{c}_{Z\gamma}$ & $\bar{c}^{\rm \,eff}_{gg}$ & $\delta y_t$ & $\delta y_c$ & $\delta y_b$ & $\delta y_\tau$ & $\delta y_\mu$  & $\lambda_Z$ \\ \hline
	0.0054 & 0.0051 & 0.0032 & 0.035 & 0.080 & 0.0092 & -- & 0.018 & 0.0060 & 0.0077 & 0.086 & 0.0012 \\
	0.0048 & 0.0048 & 0.0030 & 0.015 & 0.068 & 0.0079 & 0.050 & 0.018 & 0.0055 & 0.0072 & 0.050 & 0.0012 \\ \hline\hline
\multicolumn{12}{|c|}{$c_i/\Lambda^2 \,[{\rm TeV}^{-2}]$  of dimension-six operators} \\  \hline
$c_H$ & $c_{WW}$ & $c_{BB}$ & $c_{HW}$ & $c_{HB}$ & $c_{GG}$ & $c_{y_t}$ & $c_{y_c}$ & $c_{y_b}$ & $c_{y_\tau}$ & $c_{y_\mu}$ & $c_{3W}$  \\ \hline
	0.18 & 0.040 & 0.040 & 0.13 & 0.18 & -- & -- & 0.28 & 0.077 & 0.11 & 1.4 & 0.19 \\
	0.16 & 0.035 & 0.035 & 0.12 & 0.17 & 0.0018 & 0.82 & 0.28 & 0.076 & 0.11 & 0.83 & 0.19 \\  \hline 
\end{tabular}
\end{center}
\end{table*}

\subsection{The Higgs boson self-coupling}
\label{subsec:hhh}

The Higgs boson self-coupling is a critical parameter governing the dynamics of the electroweak symmetry breaking.  In the SM, the Higgs boson trilinear and quadrilinear couplings are fixed once the values of the electroweak vacuum expectation value and the Higgs boson mass are known.  Any deviation from the SM prediction is thus clear evidence of new physics beyond the SM.  The Higgs trilinear coupling is probed at the LHC by the measurement of the di-Higgs production.  Current bounds on the Higgs trilinear coupling is at the $\mathcal{O}(10)$ level, while the HL-LHC is expected to improve the precision to the level of $\mathcal{O}(1)$ \cite{ATL-PHYS-PUB-2017-001}.
The prospects for extracting the Higgs boson quadrilinear coupling are much less promising, even for a 100\,TeV hadron collider~\cite{Contino:2016spe}.


\begin{table*}[ht!]
\begin{center}
\caption{\label{tab:cepchhh1}The $\Delta\chi^2=1$ (one-sigma) and $\Delta\chi^2=4$ (two-sigma) bounds of $\delta\kappa_\lambda$ for various scenarios, obtained in a global fit by profiling over all other EFT parameters.  } 
\begin{tabular}{c|cc} \hline
Bounds on $\delta\kappa_\lambda$ &  $\Delta\chi^2=1$ & $\Delta\chi^2=4$  \\ \hline
CEPC 240\,GeV ($\abih$)   & $[-3.0, \, +3.1]$  &  $[-5.9, \, +6.2]$ \\  
HL-LHC   & $[-0.9, \, +1.3]$  &  $[-1.7, \, +6.1]$  \\  
HL-LHC+CEPC 240\,GeV & $[-0.8, \, +1.0]$ & $[-1.5, \, +2.7]$  \\  \hline
\end{tabular}
\end{center}
\end{table*}


To measure the di-Higgs production at a lepton collider, a sufficiently large center of mass energy ($\gtrsim 400\,$GeV) is required, which is likely to be achieved only at a linear collider.  The CEPC, instead, can probe the Higgs boson trilinear coupling via its loop contributions to the single Higgs boson processes.  This indirect approach, nevertheless, provides competitive sensitivity, since the loop suppression is compensated by the high precision of the Higgs boson measurements at the CEPC~\cite{McCullough:2013rea}.  With a precision of $0.5\%$ on the inclusive $ZH$ cross section at 240\,GeV,  the Higgs boson trilinear coupling can be constrained to a precision of $35\%$, assuming all other Higgs boson couplings that contribute to $\ZH$ are SM-like.~\footnote{ A better precision can be obtained by using, in addition,  exclusive channels, such as $\sigma(ZH)\times {\rm BR}(\Hbb)$. However, this will require an even stronger assumption, i.e. that all Higgs boson couplings contributing to the branching ratios are also SM-like except for the Higgs boson trilinear coupling. }  
While this indirect bound is comparable to the direct ones at linear colliders, it relies on strong assumptions which are only applicable to some specific models.

A more robust approach is to include all possible deviations on the Higgs boson couplings simultaneously and constrain the Higgs boson trilinear coupling in a global fit.  The EFT framework presented in Section~\ref{subsec:eft} is ideal for such an analysis.  Under this framework, the one-loop contributions of the trilinear Higgs boson coupling to all the relevant Higgs boson production and decay processes are included, following Ref.~\cite{DiVita:2017vrr}.  The new physics effect is parametrized by the quantity $\delta \kappa_\lambda \equiv \kappa_\lambda -1$, where $\kappa_\lambda$ is the ratio of the Higgs boson trilinear coupling to its SM value,
\begin{equation}
\kappa_\lambda \equiv \frac{\lambda_3}{\lambda_3^{\rm sm}}\,, \qquad \lambda_3^{\rm sm} = \frac{m_H^2}{2 v^2}\,.
\end{equation}



The global fit is performed simultaneously with $\delta \kappa_\lambda$ and all the 12 EFT parameters defined in Section~\ref{subsec:eft}.
The results are presented in Table~\ref{tab:cepchhh1}.  The results for the HL-LHC are also shown, which were obtained in Ref.~\cite{DiVita:2017eyz} under the same global framework.  For the CEPC 240\,GeV, the one-sigma bound on $\delta \kappa_\lambda$ is around $\pm 3$, significantly worse than the $35\%$ in the $\delta \kappa_\lambda$-only fit.  This is a clear indication that it is difficult to resolve the effects of $\delta \kappa_\lambda$ from other Higgs boson couplings.  For the HL-LHC, the reach on $\delta \kappa_\lambda$ is still dominated by di-Higgs production.  However, as a result of the destructive interferences among diagrams, di-Higgs production at the LHC cannot constrain $\delta \kappa_\lambda$ very well on its positive side, even with the use of differential observables~\cite{Azatov:2015oxa}.  The combination of the HL-LHC and the CEPC 240\,GeV thus provides a non-trivial improvement to the HL-LHC result alone, in particular for the two-sigma bound on the positive side, which is improved from $+6.1$ to $+2.7$.  This is illustrated in Fig.~\ref{fig:cepchhh1}, which plots the profiled $\chi^2$ as a function of $\delta \kappa_\lambda$ for the two colliders.
\begin{center}
	\includegraphics[width=0.35\textwidth]{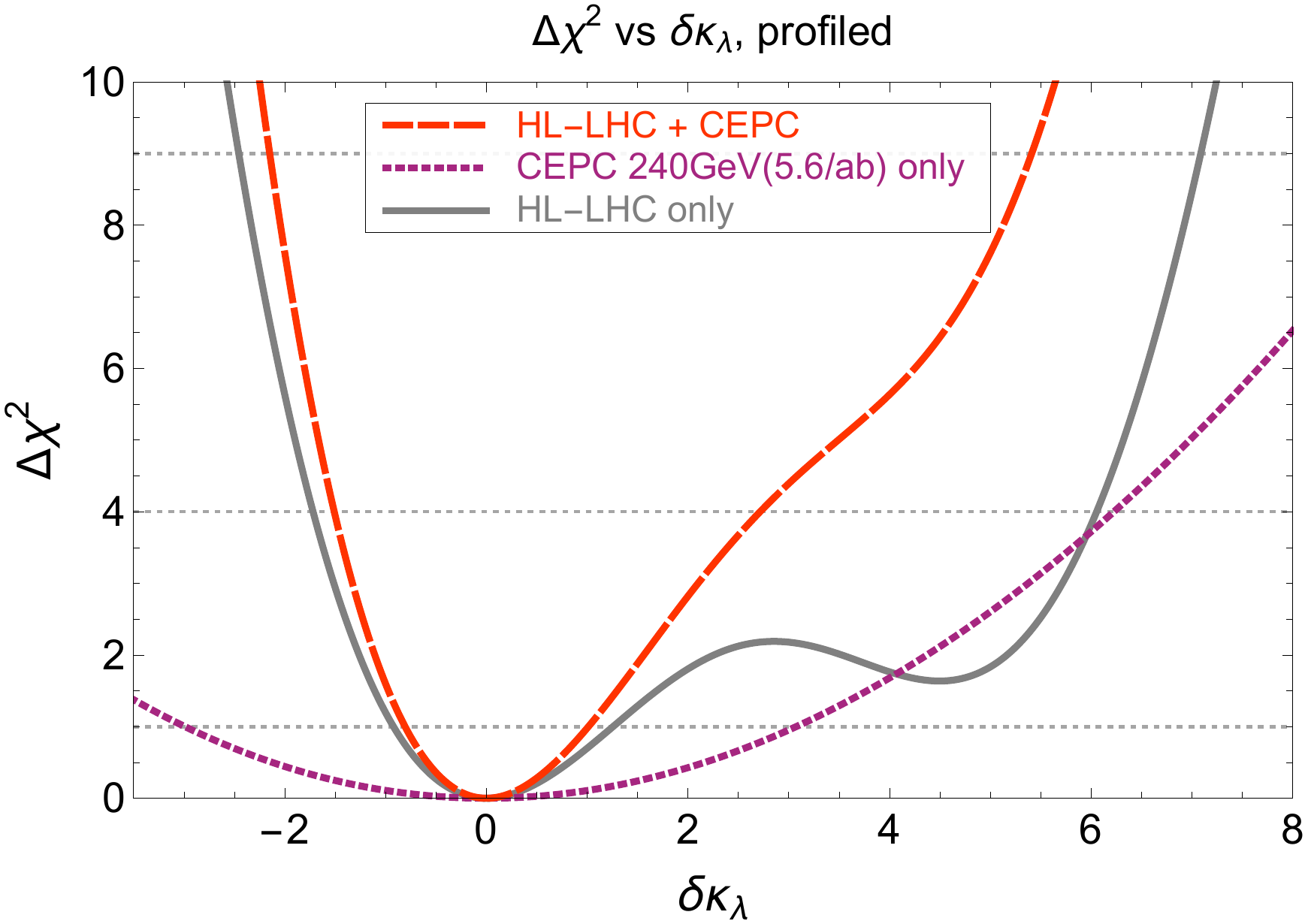}
	\figcaption{\label{fig:cepchhh1} Chi-square as a function of $\delta\kappa_\lambda$ after profiling over all other EFT parameters for the HL-LHC, the CEPC and their combination.  The results for the HL-LHC are obtained from Ref.~\cite{DiVita:2017eyz}.}
\end{center}

\subsection{Higgs boson and top-quark couplings}

Interactions of the Higgs boson with the top quark are widely viewed as a window to new physics beyond the SM. The CEPC potential on the interactions between the Higgs boson and the top quark can be evaluated~\cite{HiggsTop,Li:2015kxc,Vryonidou:2018eyv,Durieux:2018tev, Durieux:2018ggn, Boselli:2018zxr} by parametrizing these interactions in terms of dimension-six gauge-invariant operators~\cite{AguilarSaavedra:2008zc,AguilarSaavedra:2009mx}. This EFT basis enlarges the Higgs basis EFT considered above. Moreover, the $\CP$ violation effects in the third generation Yukawa couplings are reflected in the imaginary parts of the Wilson coefficients of operators $\mathcal{O}_{y_t}$ and $\mathcal{O}_{y_b}$,
\begin{eqnarray}
\Delta y_t &=& y_t^{\rm SM}\left(\Re[C_{y_t}] \frac {v^3} {2 m_t \Lambda^2} + i  \Im[C_{y_t}] \frac {v^3} {2m_t\Lambda^2} \right)\\
\Delta y_b &=& y_t^{\rm SM}\left(\Re[C_{y_b}] \frac {v^3} {2 m_b \Lambda^2} + i  \Im[C_{y_b}] \frac {v^3} {2m_b\Lambda^2} \right).
\end{eqnarray}

In this section, the effect of introducing $\CP$ phases in the Yukawa operators in Higgs boson physics is discussed. For more detailed discussion on a complete set of Higgs boson and Top quark operators, see Ref.~\cite{HiggsTop}. 
The dominant sources of constraints are from $H\to \gamma\gamma$ and $H\to gg$ for $\mathcal{O}_{y_t}$, and $H\to gg$ and $H\to b\bar b$ for  $\mathcal{O}_{y_b}$. Given that $H\to gg$ measurements are sensitive to both operators, a joint analysis of $\mathcal{O}_{y_t}$ and $\mathcal{O}_{y_b}$ will yield a significantly different result comparing to individual operator analysis. A joint analysis for these two operators in terms of Yukawa coupling strengths and the associated $\CP$ phases is performed at the CEPC. The important physics cases for such considerations are highlighted.

\begin{figure*}
\begin{center}
\subfigure[]{\includegraphics[width=0.45\textwidth]{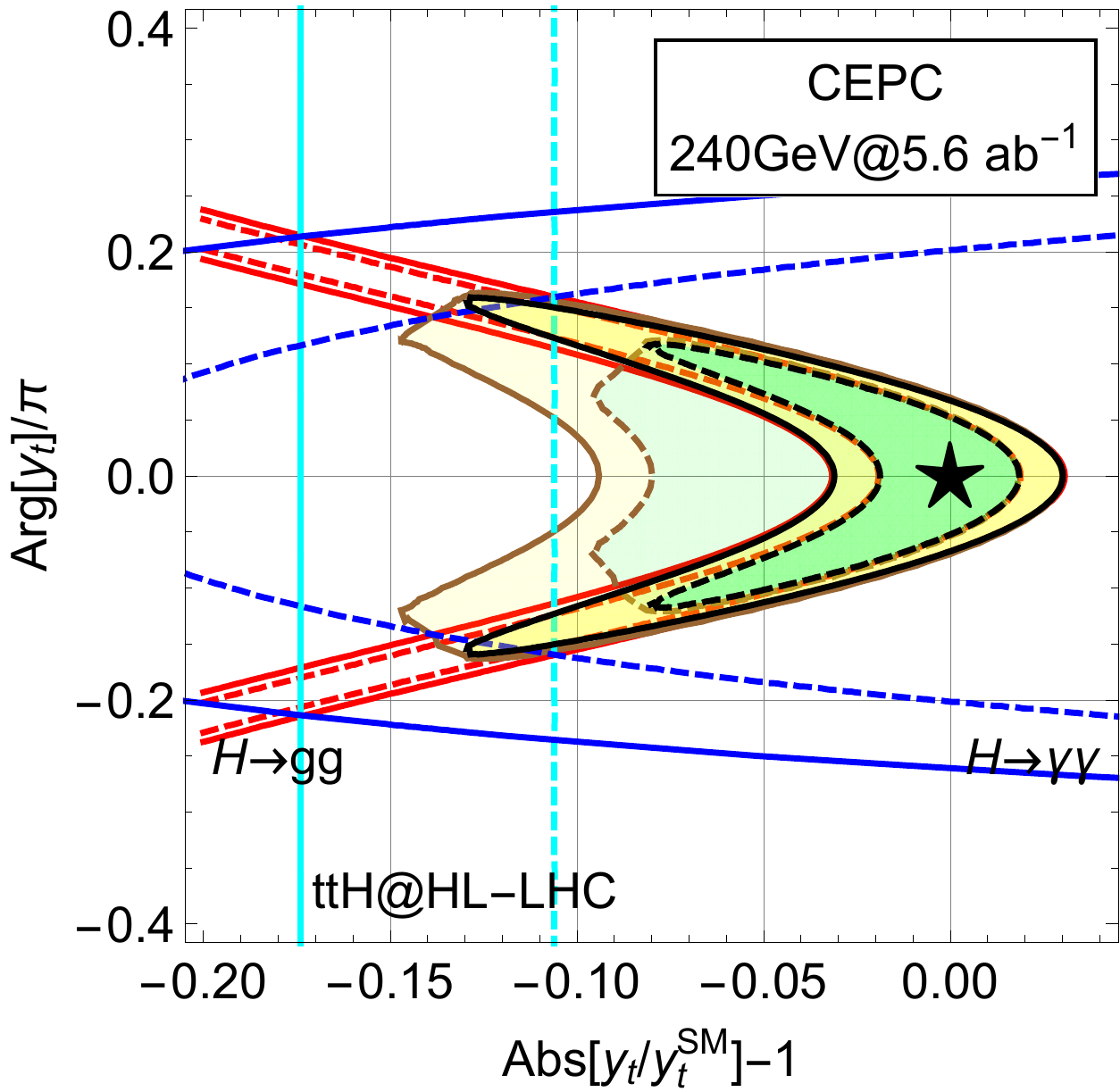}} 
\subfigure[]{\includegraphics[width=0.45\textwidth]{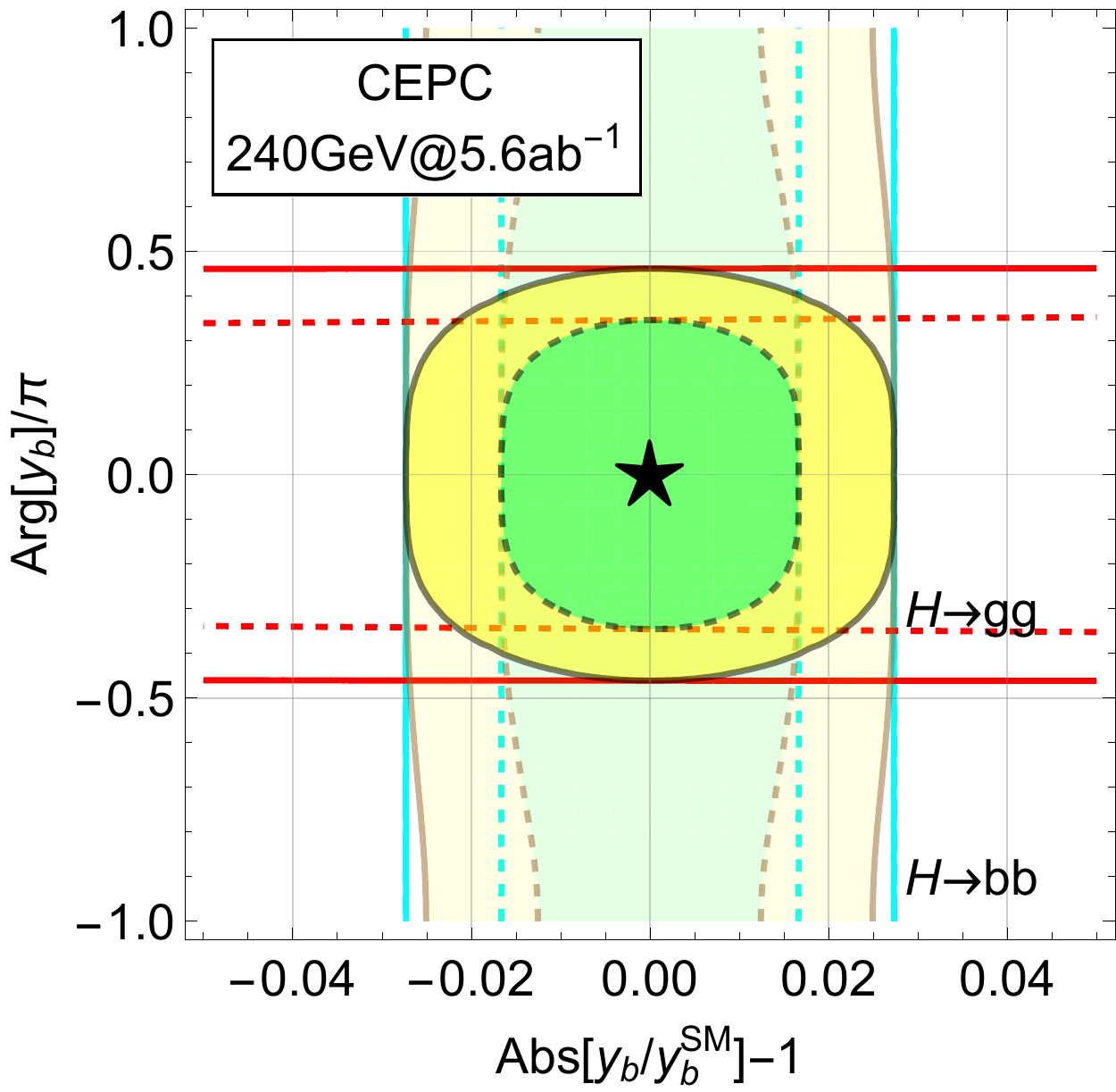}}
\end{center}
\caption{Results for analysis on $C_{y_t}$ and $C_{y_b}$ in the projected allowed regions for modification to the top-quark and bottom-quark Yukawa coupling magnitude and $\CP$ phase at 68\% and 95\% CL. 
The combined results for the CEPC are shown in black curves. 
The source of individual constraints for the single operator analysis are labeled correspondingly. For a joint analysis of simultaneous appearance of both $\mathcal{O}_{y_t}$ and $\mathcal{O}_{y_b}$ operators, the results for the CEPC are shown in the enlarged yellow (95\% CL) and green regions (68\% CL) with thick brown boundary lines.
}
\label{fig:op1op2}
\end{figure*}

Constraints on the top-quark and bottom-quark Yukawa couplings, including their $\CP$ phases,  are presented, respectively, in the left and right panels of Fig.~\ref{fig:op1op2}, respectively. The 68\% and 95\% CL exclusion bands are shown in dashed and solid lines. 
The limits for the CEPC are shown in {\em bright} black and magenta lines for individual operator analysis and the {\em bright} green and yellow shaded regions representing the allowed parameter space at 68\% and 95\% CL, respectively. The {\em dimmed} thick black curves represent the results after turning on both operators ${\cal O}_{tH}$ and ${\cal O}_{bH}$ at the same time, using a profile-likelihood method profiling over other parameters.  
Furthermore, in the left panel the cyan band represents constraints from the HL-LHC $t\bar tH$ measurements, red bands are constraints from the CEPC $H\to gg$ measurements and blue bands are constraints from the CEPC $H\to \gamma\gamma$ measurements. Similarly, in the right panel, the cyan bands are constraints from $H\to b\bar b$ and the red bands are constraints from $H\to gg$ at the CEPC.

The left panel of Fig.~\ref{fig:op1op2} shows that the expected sensitivity on the  modification in the magnitude of top-quark Yukawa coupling is around $\pm3\%$ for the single operator analysis. This is relaxed to $[-9.5\%,+3\%]$ assuming zero $\CP$ phase for the top-quark Yukawa coupling and allowing the bottom-quark Yukawa coupling and its phase to vary freely. The phase of the top-quark Yukawa coupling can be constrained to $\pm 0.16\pi$. This constraint is driven by the $H\to \gamma\gamma$ measurement, where a sizable phase shift will enlarge the $H\to\gamma\gamma$ decay rate via reducing the interference with the SM $W$ boson loop. The constraint on the magnitude of the top-quark Yukawa coupling is driven by the $H\to gg$ measurement which is dominated by the top-quark loop contribution. Note that constraints from the $H\to gg$ measurement are not constant with respect to the Yukawa coupling magnitude. This is due to the different sizes of the top-quark loop contribution to $Hgg$ through scalar and pseudoscalar couplings.
Similarly, as shown in the right panel of Fig.~\ref{fig:op1op2} for the bottom-quark Yukawa coupling, the constraint for the magnitude is $\pm 2.5\%$. For the $\CP$ phase, the constraint changes from $\pm 0.47\pi$ to zero when the top-quark Yukawa coupling is left free.

\section{Higgs boson $\CP$ test and exotic decays}
\label{sec:CPandExotic}

In addition to the studies based on the simulation of the CEPC baseline conceptual detector, the sensitivity of tests  on Higgs boson spin/$C\!P$ properties and in constraining branching ratios of Higgs boson exotic decays are also estimated. These estimates are based on previously published phenomenological studies and are summarized in this section.

\newcommand{\comment}[1]{{\bf [#1]}}
\newcommand{\be}{\begin{equation}}
\newcommand{\ba}{\begin{eqnarray}}
\newcommand{\ea}{\end{eqnarray}}
\newcommand{\bi}[1]{\bibitem{#1}}
\newcommand{\fr}[2]{\frac{#1}{#2}} 
\newcommand{\non}{\nonumber}
\newcommand{\ar}{\mbox{$\rightarrow$}}
\def\ra{\rightarrow}
\newcommand{\re}{{\Re}e}
\newcommand{\im}{{\Im}m}
\newcommand{\Et}{\ensuremath{E_\mathrm{T}}}                                     
\newcommand{\ifb}{\ensuremath{\mathrm{fb^{-1}}}}                                

\def\sss{\scriptscriptstyle}


\subsection{Tests of Higgs boson spin/$C\!P$ property}
The $C\!P$ properties  of the  Higgs boson and, more generally,
its anomalous couplings to gauge bosons in the presence of BSM physics,
can be measured at the CEPC using the $e^+e^-(\to Z^*)\to ZH\to \mu^+\mu^- b\bar{b}$ process.
It is convenient to express the effects of the anomalous couplings 
in terms of the fractions of events from the anomalous contribution relative to the 
SM predictions. 
These fractions are invariant under the independent rescalings of all couplings, see Refs.~\cite{Gao:2010qx,Bolognesi:2012mm,Anderson:2013afp}.

Two of the anomalous $HZZ$ coupling measurements are of particular interest at the CEPC:  
the fraction of the high-order $C\!P$-even contribution due to
either SM contribution or new physics, $f_{a2}$, and
the fraction of a $C\!P$-odd contribution due to new physics, $f_{a3}$.
The following two types of observables can be used to measure these anomalous couplings of the Higgs bosons.
\begin{enumerate} 
\item 
The dependence of the $e^+e^- \to Z^*\to ZH $ cross section on $\sqrt{s}$  is different for different $C\!P$ property of the Higgs boson~\cite{Anderson:2013afp}.
Therefore, measurements of the cross section at several different energies will yield useful information about anomalous $HZZ$ couplings. 
However this has non-trivial implications to the accelerator design and
is not included in this study as a single value of $\sqrt{s}$ is assumed for the CEPC operating
as a Higgs boson factory.

\begin{center}
\includegraphics[width=0.4\textwidth]{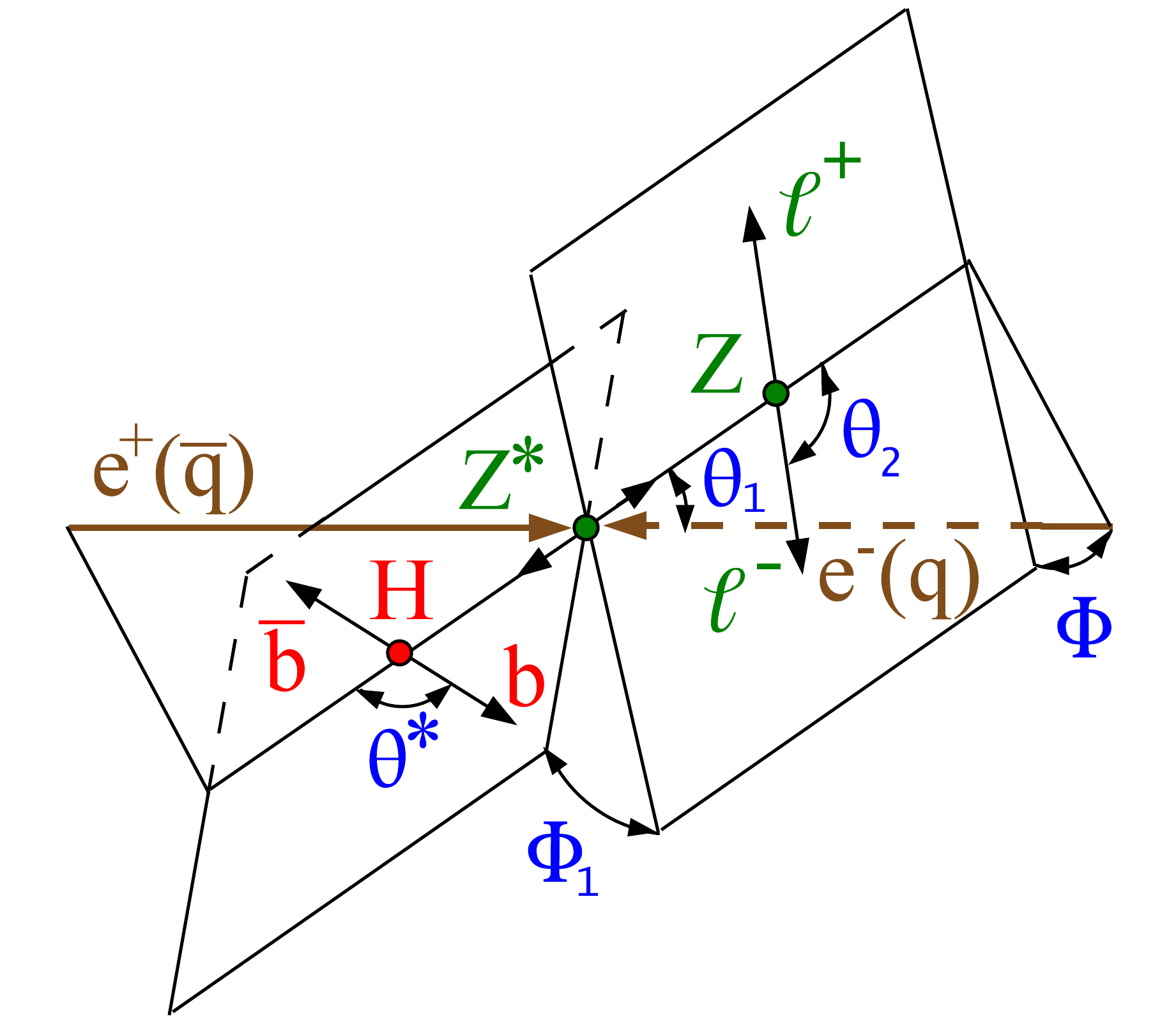}
\figcaption{\small\label{fig:HiggsAngles} The Higgs boson production and decay angles for the $\ee \!\to\! Z^*\!\to\! ZH \!\to\! \mm\bb$ process~\cite{Anderson:2013afp}.}
\end{center}

\item Angular distributions, $\cos\theta_1$ or $\cos\theta_2$ and
$\Phi$ as defined in Fig.~\ref{fig:HiggsAngles}.
These angles are also sensitive to interference between $C\!P$-even and $C\!P$-odd couplings.
In particular forward-backward asymmetry with respect to $\cos\theta_1$ or $\cos\theta_2$ 
and non-trivial phase in the $\Phi$ distributions can lead to an
unambiguous interpretation of $C\!P$ violation. 
\end{enumerate} 
\begin{figure*}
\begin{center}
\includegraphics[width=0.3\textwidth]{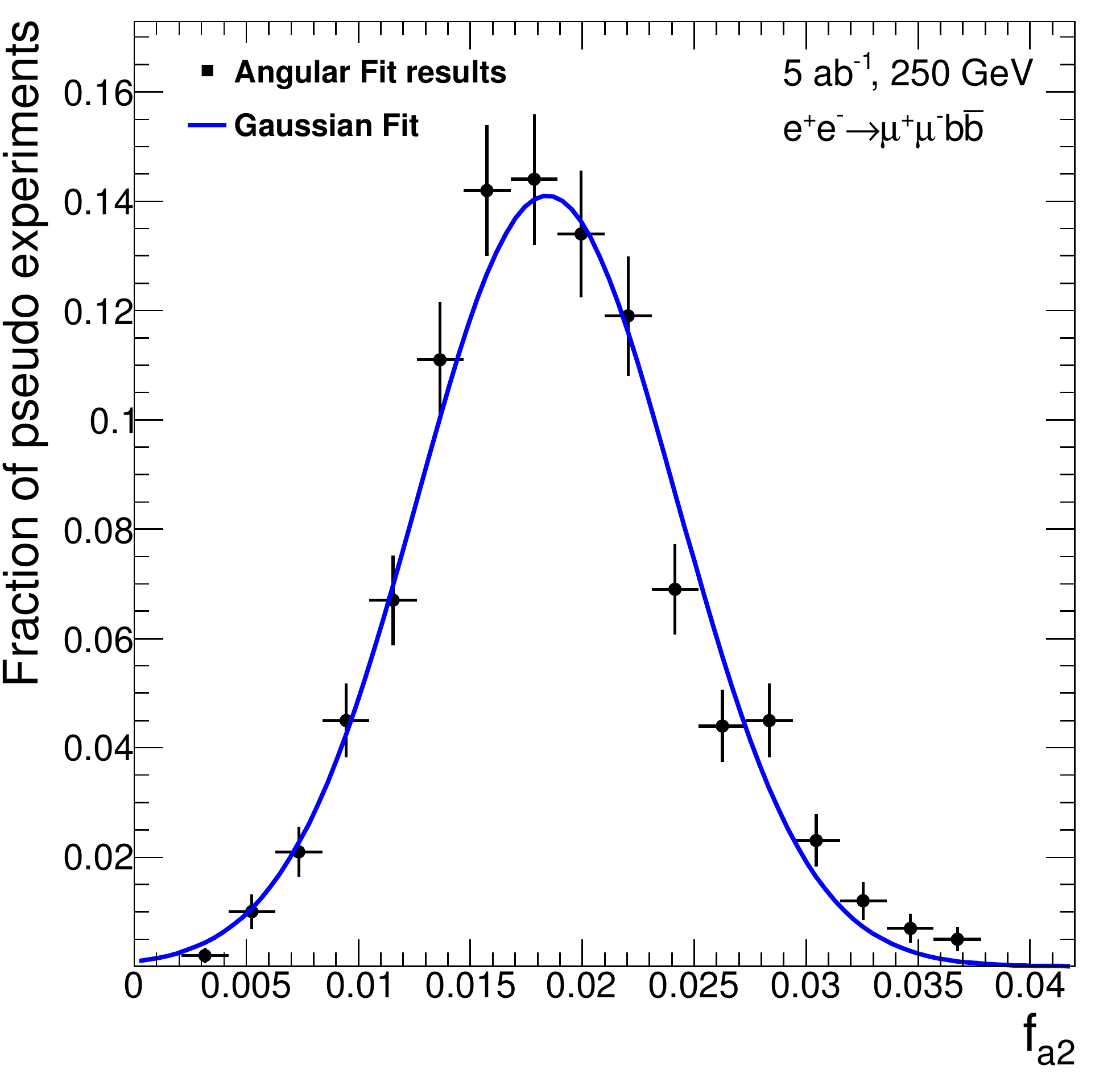}
\includegraphics[width=0.3\textwidth]{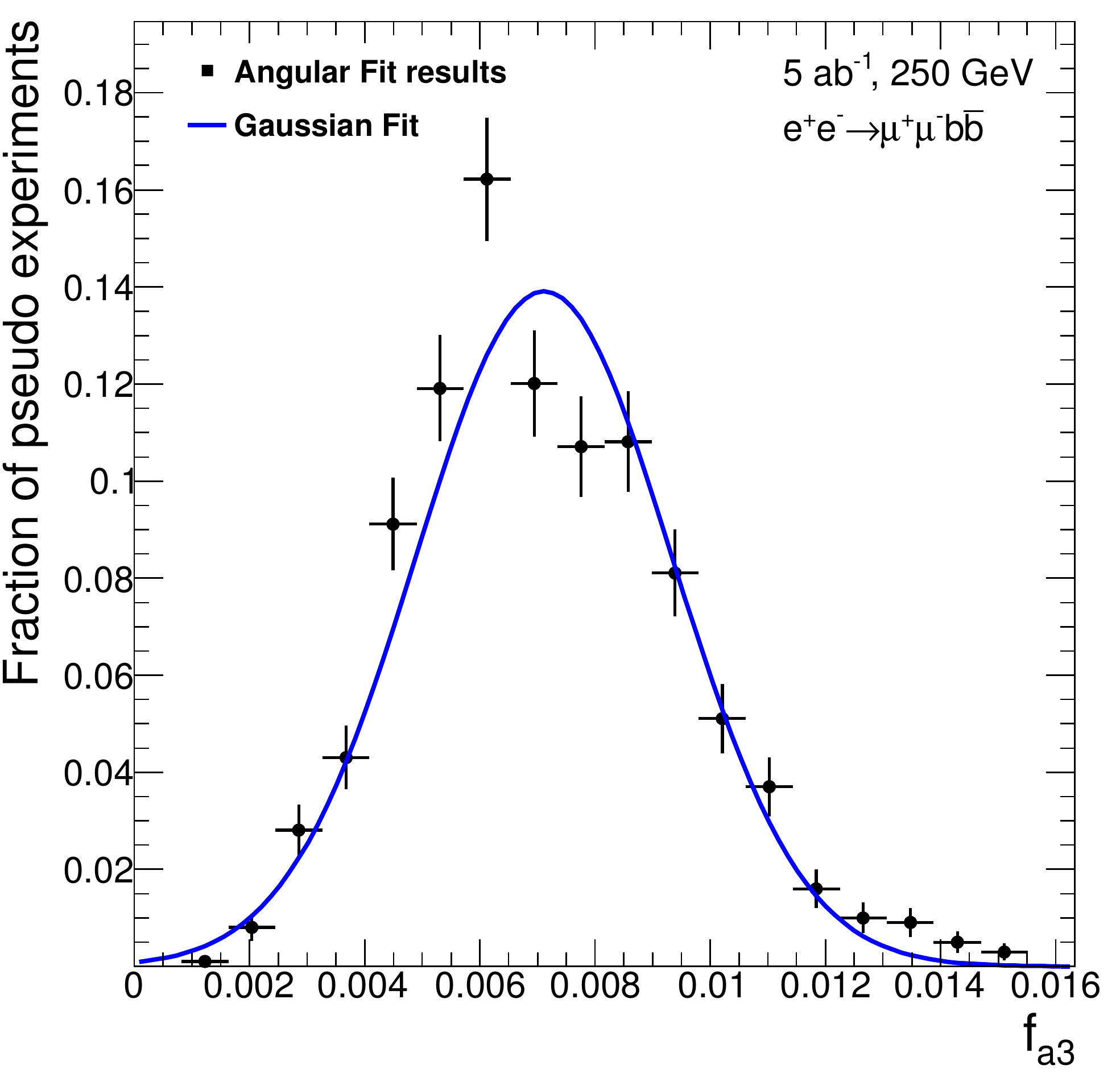}
\includegraphics[width=0.3\textwidth]{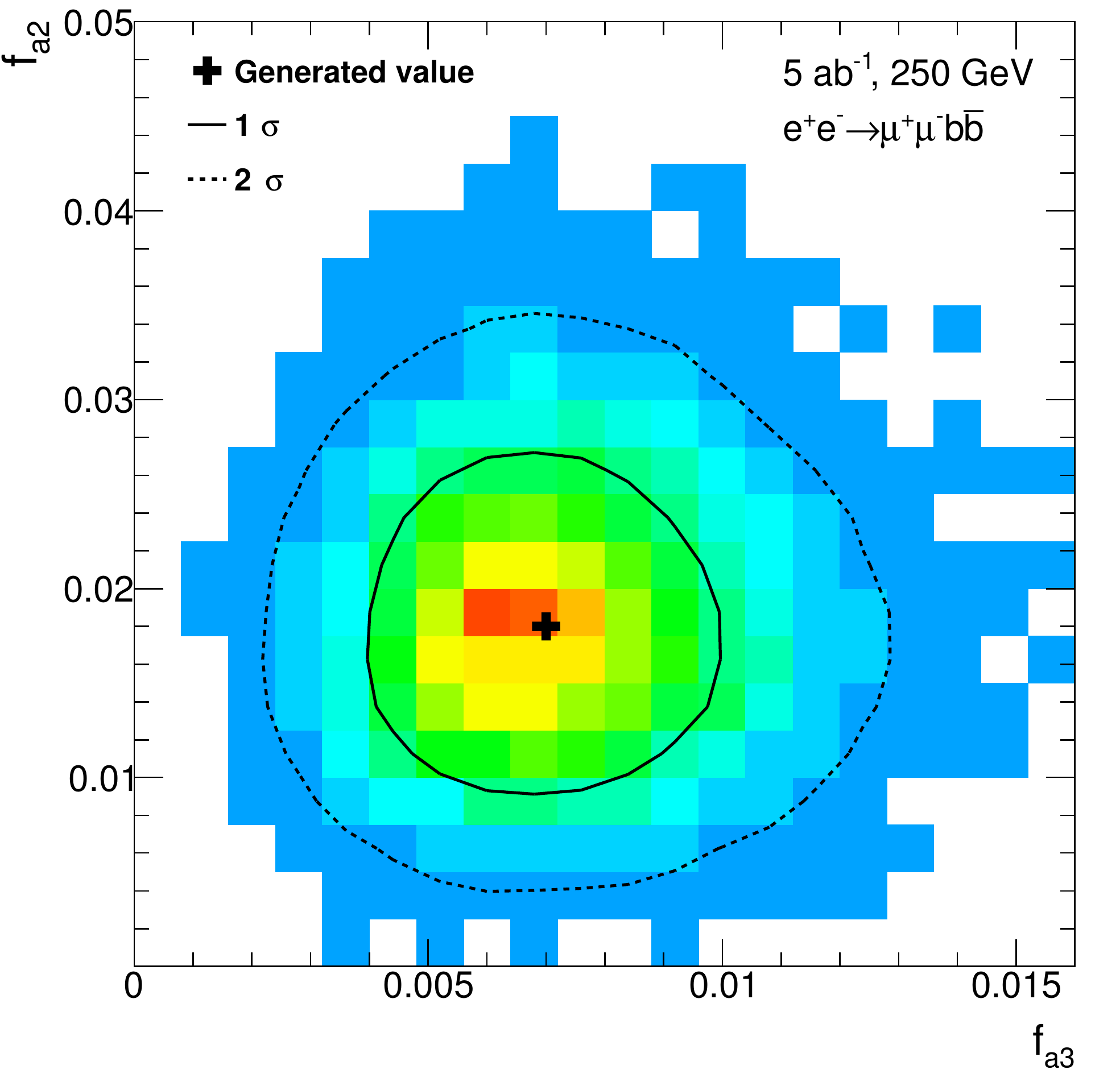}
\figcaption{\small Distribution of fitted values of $f_{a2}$ and $f_{a3}$ in a large number of pseudo-experiments. In the left and middle plots, only the parameter shown is floated. Other parameters are fixed to their SM values. Right plot: simultaneous fit of non-zero $f_{a2}$ and $f_{a3}$, with 68\% and 95\% confidence level contours shown.\label{fig:ilc_fitresults}}
\end{center}
\end{figure*}

To estimate the sensitivity on the anomalous couplings,
a maximum likelihood fit~\cite{Anderson:2013afp} is performed
to quantify the compatibility of the observed angular distributions to the theory predictions, including both signal and background processes. 
In this likelihood fit, the signal probability density functions are taken 
from analytical predictions that are validated using
a dedicated MC program, the JHU generator~\cite{Gao:2010qx,Bolognesi:2012mm},
which incorporates all the anomalous couplings,
spin correlations, the interference of all contributing amplitudes. 
The background probability density function is modeled using simulation
based on $e^+e^-\to ZZ\to \ell^+\ell^- b\bar{b}$ process
in MadGraph~\cite{Alwall:2007st}.

Several thousand statistically independent pseudo-experiments are generated and fitted
to estimate the sensitivity to $f_{a2}$ and $f_{a3}$, defined as the 
smallest values that can be measured with $3\sigma$ away from 0.
All other parameters in the fit, including the number of expected signal and background events, are fixed. The expected sensitivity of $3\sigma$ discovery is estimated to be $0.018$ for $f_{a2}$ and $0.007$ for $f_{a3}$. Figure~\ref{fig:ilc_fitresults}(a,b) show the distributions of the fitted values of $f_{a2}$ and $f_{a3}$ from the pseudo-experiments expected for $f_{a2}=0.018$ and $f_{a3}=0.008$, respectively. A simultaneous fit of $f_{a2}$ and $f_{a3}$ is also performed with
the 68\% and 95\% CL contours shown in Fig.~\ref{fig:ilc_fitresults}(c).

The sensitivities for $f_{a2}$ and $f_{a3}$ are then converted to
the corresponding parameters defined for the on-shell $H\to ZZ^*$ decays, $f_{a2}^{\rm dec}$ and $f_{a3}^{\rm dec}$, in order to compare with the sensitivities from the LHC experiments as described in Ref.~\cite{Anderson:2013afp}. The corresponding sensitivities of $f_{a2}^{\rm dec}$ and $f_{a3}^{\rm dec}$
are $2\times 10^{-4}$ and $1.3\times 10^{-4}$, respectively.
The much smaller values in the $f_{a2,a3}^{\rm dec}$ are
due to the much larger $m_{Z^*}^2$ in the $\ee\to Z^*\to ZH$ process compared to the value from the Higgs boson decays.

Compared to the ultimate sensitivity of HL-LHC as shown in Ref.~\cite{Anderson:2013afp}, 
the sensitivities in the $f_{a2}$ and $f_{a3}$ at the CEPC are better by a factor of
300 and 3.
Further improvements can be achieved by exploring kinematics in the $H\to b\bar{b}$ decays,
including other $Z$ decay final states, and combining with the overall cross-section
dependence of the signal as obtained by a threshold scan in $\sqrt{s}$.



\subsection{Higgs boson exotic decays}
\label{subsec:exotic}

The Higgs boson can be an important portal to new BSM physics. Such new physics could manifest itself through the exotic decays of the Higgs boson if some of the degrees of freedom are light.
The Higgs boson BSM decays have a rich variety of possibilities. The two-body decays of the Higgs boson into BSM particles, $H\to X_1 X_2$, where the BSM particles $X_i$ are allowed to subsequently decay, are considered here. These decay modes are classified into four cases, schematically shown in Fig. \ref{fig:topo}. These processes are well-motivated by BSM models such as singlet extensions of the SM, two-Higgs-doublet-models, SUSY models, Higgs portals, gauge extensions of the SM, and so on~\cite{Curtin:2013fra,deFlorian:2016spz,Liu:2016zki}. In this study, only prompt decays of the BSM particles are considered. For the Higgs boson decaying into long-lived particles, novel search strategies have to be developed in the future, using also the latest advances in the detector development~\cite{Liu:2018wte}.

\begin{center}
  \includegraphics[width=0.48\textwidth]{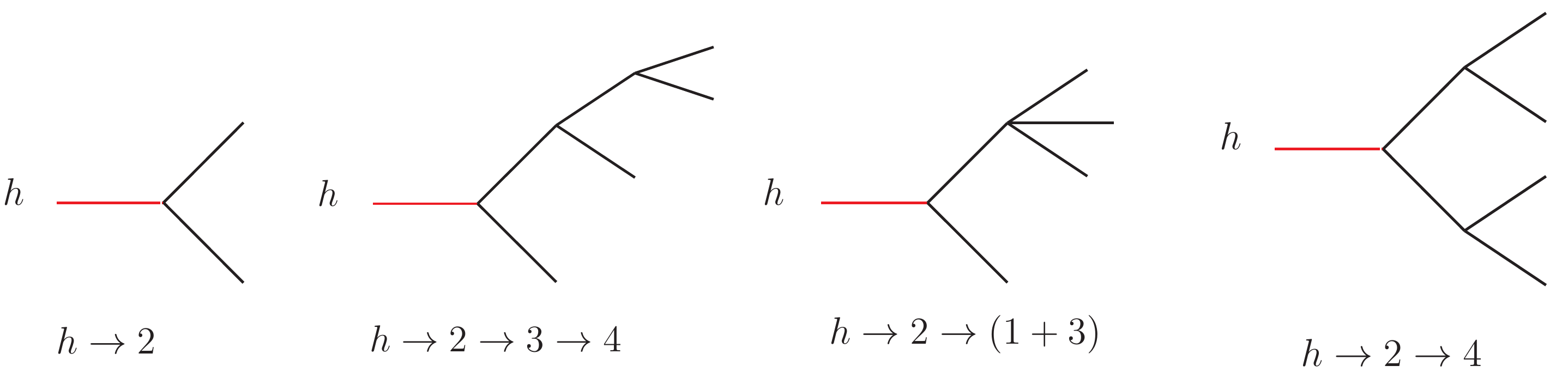}
  \figcaption{\small\label{fig:topo}The topologies of exotic decays of the Higgs boson. }
\end{center}

For the CEPC running at 240\,GeV, the most important Higgs boson production mechanism is the $\ZH$ production. The $Z$ boson with visible decays enables the  Higgs boson tagging using the ``recoil mass'' technique as described in Section~\ref{sec:massXS}. A cut around the peak of the recoil mass spectrum would remove the majority of the SM background. Further selection and tagging on the Higgs boson decay products can ensure that the major background would be from the SM decays of the Higgs bosons. The details of these analysis can be found in Ref.~\cite{Liu:2016zki}.

The set of Higgs boson exotic decays with their projected LHC constraints and limits from the CEPC with an integrated luminosity of $~\abih$ are summarized in Table~\ref{tab:summary}. For the LHC constraints, both the current limits and projected limits on these exotic decay channels from various references are tabulated. The comparison are performed for particular benchmark points to demonstrate the qualitative difference between the (HL-)LHC and CEPC.

A selection of results for channels, which are hard to be constrained at the LHC, is shown in Table~\ref{tab:summary} and Fig.~\ref{fig:summary}.
 The red bars in the figure correspond to the results using the leptonic $Z$ boson decays that are produced in association with the Higgs bosons. The hadronic decaying $Z$-boson provides around ten times more statistics and hence further inclusion will improve the results significantly. Based upon the study of the Higgs bosons decaying into $WW^*$, $ZZ^*$ and invisible particles, hadronic decaying $Z$ bosons are conservatively assumed to provide the same upper limits as the leptonic $Z$ boson decays and, hence, improve the limits by around 40\% when combined. This extrapolated results are shown in yellow bars. 

In comparison with the HL-LHC, the improvement on the Higgs boson exotic decay branching ratios is significant, varying from one to four orders of magnitude for the channels considered. 
For the Higgs boson exotic decays into hadronic final states plus missing energy, $b\bar b+\met$, $jj+\met$ and $\tau^+\tau^-+\met$, the CEPC improves the HL-LHC sensitivity by three to four orders of magnitude. These significant improvements benefit from the low QCD background and the Higgs boson tagging from the recoil mass reconstruction at the CEPC. 
Final states with leptons and photons have smaller QCD background at the LHC and therefore the improvements from the CEPC are limited for these final states. 
\begin{table*}
	\caption{\small\label{tab:summary}The current and projected limits on the Higgs boson exotic decay modes for the (HL-)LHC and the CEPC with an integrated luminosity of $\abih$, based on the results from Ref.~\cite{Liu:2016zki}. The second column shows the current LHC results and the projections for 100 $\fbi$ (in parentheses) and 300 $\fbi$ (in square brackets). The available projections for the HL-LHC are listed in the third column. Pairs of objects in parentheses indicate that they are decay products of intermediate resonances, e.g. $(\bb)+\met$ stands for $X+\met\to (\bb)+\met$ where $X$ is an intermediate resonance.
	}
\begin{center}
\begin{tabular*}{0.8\textwidth}{c@{\extracolsep{\fill}}cccc}
	\toprule
	Decay & \multicolumn{4}{c}{95\% C.L. limit on BR}\\ \cline{2-5}
	Mode & LHC & HL-LHC & CEPC  \\
	\hline\hline
	$\met$              & 0.23 & 0.056   & 0.0030 \\ \hline
	$ (b\bar b)+\met$   & [0.2] & -- & \scidgts{1}{-4}  \\
	$ (jj)+\met$        & -- & -- & \scidgts{4}{-4} \\
	$ (\tau^+\tau^-)+\met$ & [1] &  -- & \scidgts{8}{-5} \\ \hline
	$ b\bar b+\met$        & [0.2] & -- & \scidgts{2}{-4} \\
	$ jj+\met$             & -- & -- & \scidgts{5}{-4} \\
	$ \tau^+\tau^-+\met$   & -- & -- & \scidgts{8}{-5} \\ \hline
	$ (b\bar b)(b\bar b)$  & 1.7 (0.2) & -- & \scidgts{6}{-4} \\
	$(c\bar c)(c\bar c)$   & (0.2) & -- & \scidgts{8}{-4}\\
	$(jj)(jj)$             & [0.1] & -- &  \scidgts{2}{-3} \\
	$ (b\bar b)(\tau^+\tau^-)$  & 0.1 [0.15]& -- & \scidgts{4}{-4} \\
	$ (\tau^+\tau^-)(\tau^+\tau^-)$& 1.2 [0.2$\sim$0.4] & -- & \scidgts{2}{-4}  \\
	$(jj)(\gamma\gamma)$& [0.01] & -- & \scidgts{1}{-4}  \\
	$(\gamma\gamma)(\gamma\gamma)$& \scidgts{7}{-3} & \scidgts{4}{-4} & \scidgts{8}{-5}  \\
	\bottomrule
\end{tabular*}
		
\end{center}
\end{table*}
\begin{figure*}
	\begin{center}
		\centering
		\includegraphics[width=0.8\textwidth]{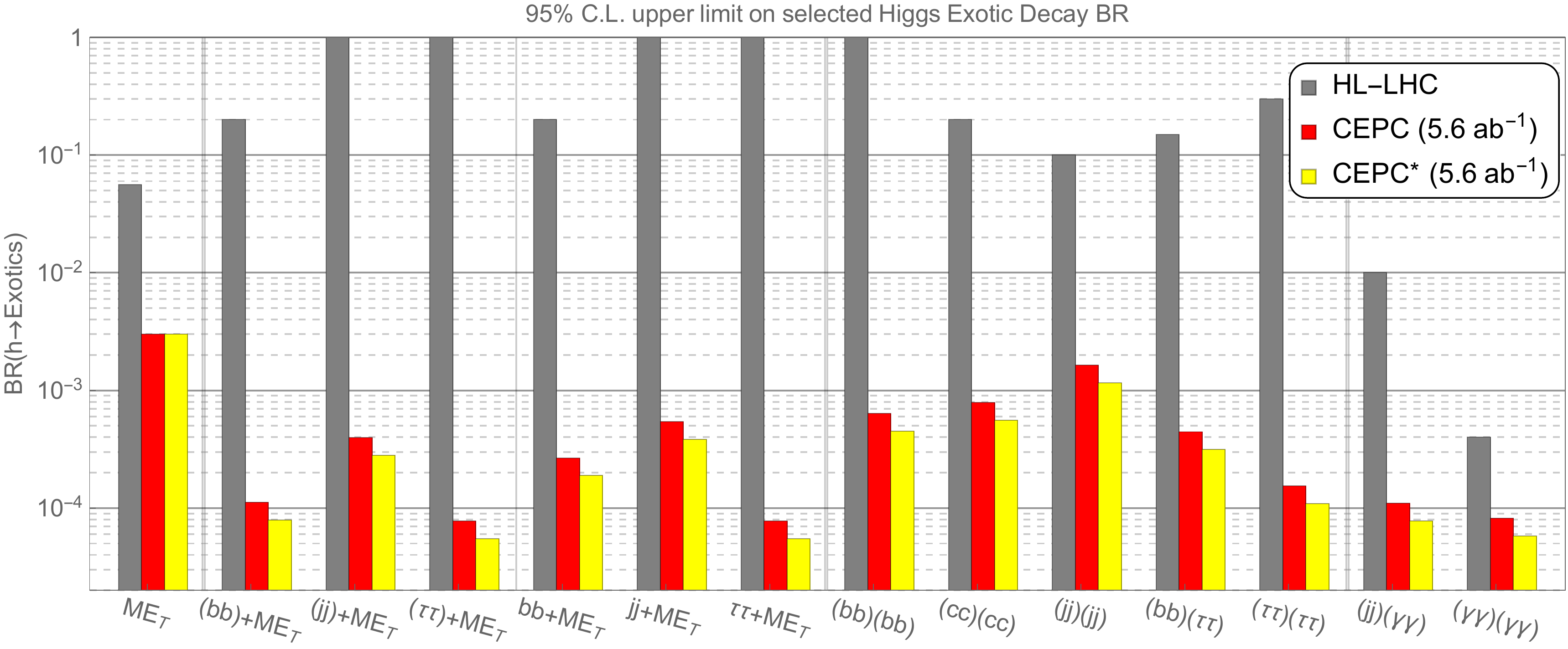}
		\figcaption{\small\label{fig:summary} The 95\% C.L. upper limits on selected Higgs boson exotic decay branching ratios at the HL-LHC and the CEPC, based on Ref.~\cite{Liu:2016zki}. The benchmark parameter choices are the same as in Table~\ref{tab:summary}. The red bars correspond to the results using leptonically decaying spectator $Z$-boson alone. The yellow bars further include extrapolation with the inclusion of the hadronically decaying $Z$-bosons. Several vertical lines are drawn in this figure to divide different types of Higgs boson exotic decays.}
	\end{center}
\end{figure*}

\section{Implications}
\label{sec:implications}

In this section, we briefly discuss the most important physics implications of the Higgs boson measurements at the CEPC. 
The measurements of the Higgs boson properties are essential to the understanding of the nature of electroweak symmetry breaking, which remains to be a central and open question. In the SM, it is parametrized by the so-called ``Mexican Hat'' Higgs potential,

\begin{equation}
V(H) = -\frac{1}{2}\mu^2 |H|^2 + \frac{\lambda}{4} |H|^4 ,  \label{eq:smhpo}
\end{equation}

with the vacuum expectation value (VEV) of the Higgs field spontaneously breaking the $SU(2)_{\rm L} \times U(1)_{\rm Y}$ gauge symmetry down to $U(1)_{\rm em}$, and generating masses for the $W$ and $Z$ bosons.  With the measurements of the Fermi constant (from muon decay) and the Higgs boson mass, the two parameters in Eq.~\ref{eq:smhpo}, $\mu^2$ and $\lambda$, are determined to a very good precision, and thus the SM Higgs potential is fully determined. 
However, we would like to emphasize that this simplicity is somewhat misleading, as our knowledge of the electroweak symmetry breaking is far from complete.
First of all, even though the values of these parameters can be fixed by the experimental measurement, the SM does not contain an explanation of their sizes, and in particular why the electroweak scale appears to be many orders of magnitude smaller than the Planck scale.
Furthermore, the Mexican Hat potential as well as the SM itself are based on assumptions, which need to be explicitly tested by experiments before they are established to be correct.
In this section, we will focus on the potential of using the precision measurements of Higgs boson properties at the CEPC to address these important questions.

\subsection{Naturalness of the electroweak scale}

An important question associated with the electroweak symmetry breaking is naturalness. It arises from the need to explain the presence of the weak scale $\Lambda_{\rm weak} \sim 10^2$ GeV in terms of a more fundamental theory.  
New physics is necessarily involved in such a theory. The SM by itself cannot answer this question, however, there are many new physics models with the potential to provide an answer. 
However, a key question for any model of electroweak symmetry breaking, regardless of the model details, is what the scale of new physics is.
For instance, if the new physics is the quantum gravity scale, $M_{\rm Planck} = 10^{19}$ GeV, then an immediate question is how to explain the 17 orders of magnitude difference between it and the electroweak scale. This is often denoted as the naturalness/hierarchy/fine-tuning problem. 
More generally, the weak scale in any such model can be expressed using dimensional analysis as 
\begin{equation}
\Lambda_{\rm weak}^2 \sim c_1 M_1^2 + c_2 M^2_2 + ...,
\end{equation}
where $M_i \sim M_{\rm NP}$ are the scale of new physics. They are typically the masses of the new physics particles. The $c_i$ are numerical coefficients that depend on the details of the model. 
However, we do note expect them to be very different from order one. 
Therefore, a large and precise cancellation is needed if $M_{\rm NP} \gg \Lambda_{\rm EW}$, 
with the level of tuning proportional to $M_{\rm NP}^2$. 
The discovery of the spin-zero Higgs boson deepens this mystery. 
While it is possible to generate a large cancellation by imposing symmetries instead of tuning  -- one well-known example is the chiral symmetry which protects the masses of the light fermions from receiving large quantum corrections -- there is no obvious symmetry that protects the mass of the Higgs boson if it is an elementary scalar particle.
To avoid an excessive amount of fine tuning in the theory, 
the new physics cannot be too heavy, and should preferably be below the TeV scale. This is the main argument for TeV new physics based on naturalness. 

Searching for new physics which leads to a natural electroweak symmetry breaking has been and will continue to be a main part of the physics program at the LHC. Looking for signals from the direct production of the new physics particles, the LHC will probe the new physics scale up to a few TeV. At the same time, as we will show below, the precision measurements at the CEPC can provide competitive sensitivity reach, and 
has the potential of probing significant higher new physics scales for many scenarios. 
In addition, 
the reach of the LHC searches has a strong dependence on the production and decay modes of the new physics particles.  The measurements at the CEPC thus provides crucial complementary information and can probe scenarios that are difficult at the LHC.
Indeed, the precision measurement of the Higgs boson couplings offers a 
very robust way of probing new physics related to electroweak symmetry breaking. Any such new physics would necessarily contain particles with sizable couplings to the Higgs boson, which leave their imprints in the Higgs boson couplings. 
%
Such a model independent handle is of crucial importance, given the possibility that the new physics 
could simply be missed by the LHC searches designed based on our wrong expectations.  

In the following, we demonstrate the sensitivity potential to new physics in several broad classes of models, which can address the naturalness of the electroweak symmetry breaking. 


\begin{figure*}[t!]
\begin{center}
\subfigure[]{\includegraphics[width=0.4\textwidth]{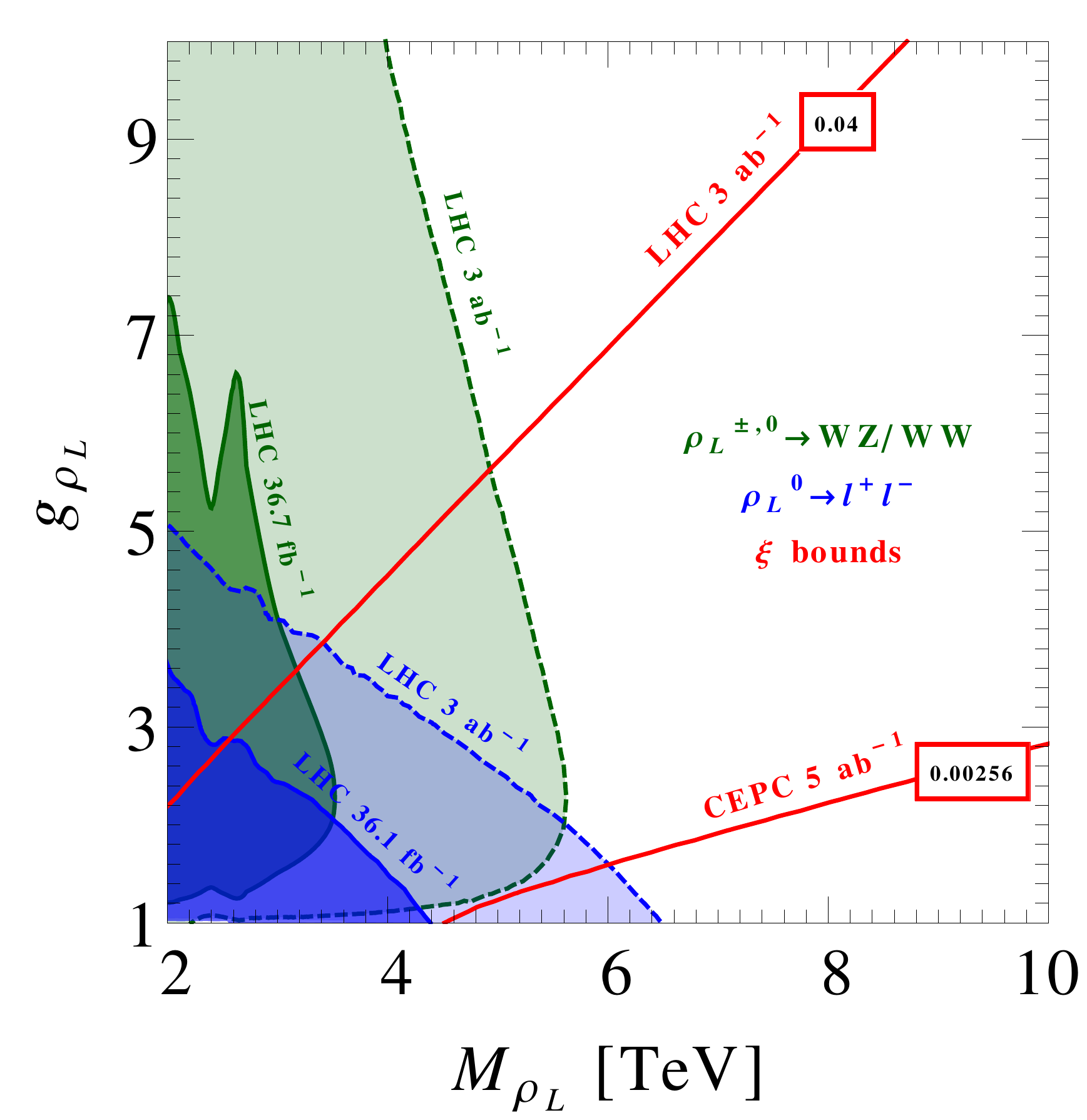}}
\subfigure[]{\includegraphics[width=0.4\textwidth]{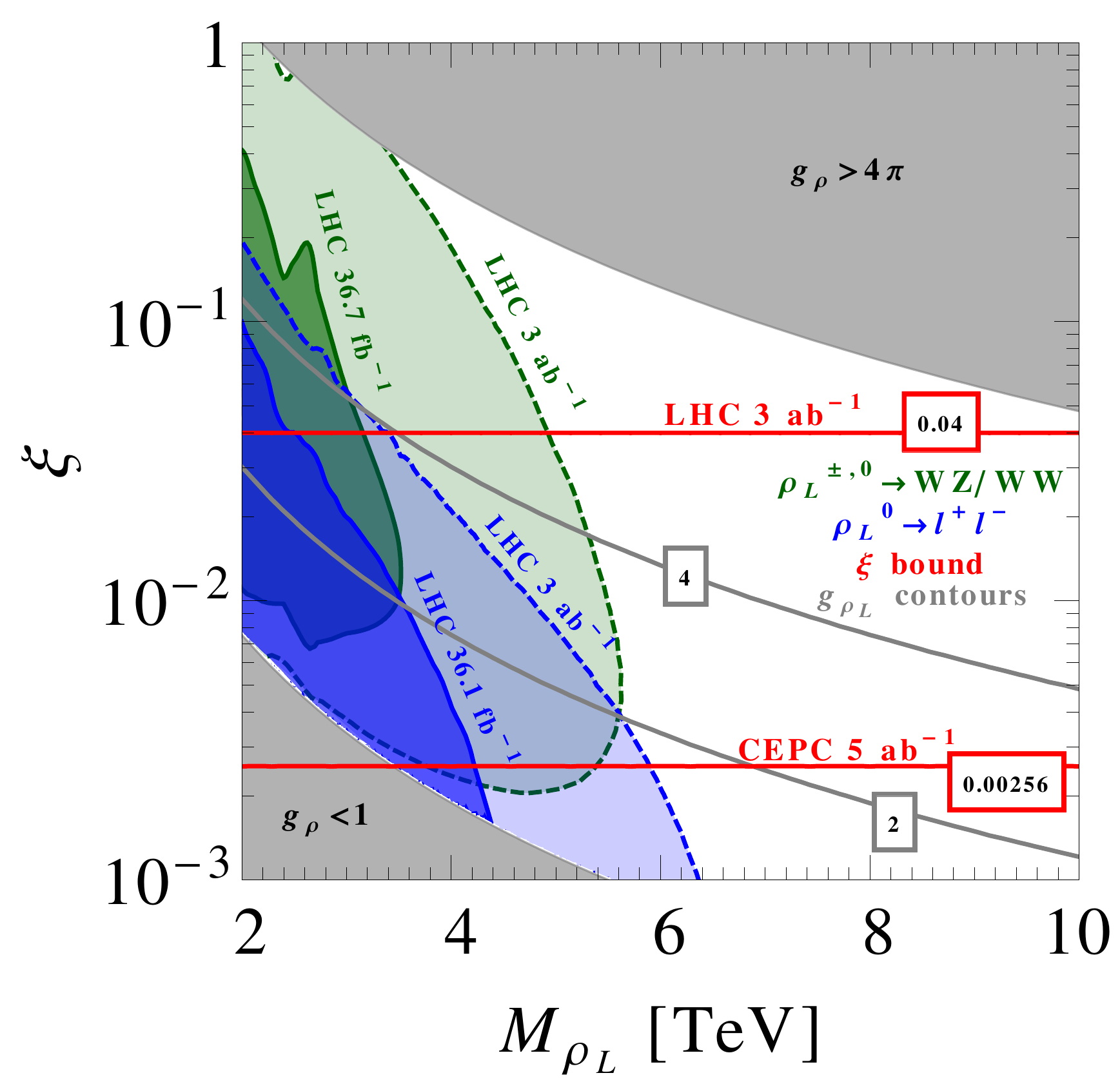}}
\figcaption{\small Limits on the composite Higgs boson model from both direct searches at the LHC and precision measurement at the CEPC. The figures are updated versions of the ones presented in Ref.~\cite{Thamm:2015zwa}.}
\label{fig:compositeH}
\end{center}
\end{figure*}
One obvious idea is that the Higgs boson is a composite particle instead of an elementary one.  After all, many composite light scalars already exist in nature, such as the QCD mesons. 
The composite Higgs boson can thus be regarded as a close analogy of the QCD mesons. A light Higgs boson can be naturally obtained if it is implemented as a pseudo-Nambu-Goldstone boson with new dynamics at scale $f$. Its physics can be described by a chiral Lagrangian similar to that of the low energy QCD. The explicit breaking comes from the couplings which are responsible for the SM fermion masses, and the SM gauge couplings. In this case, 
the Higgs boson would not unitarize the $WW$ scattering amplitude completely, and its coupling to $W$ and $Z$ will be shifted approximately by 
\begin{equation}
\delta \kappa_W, \ \delta \kappa_Z \sim  {\mathcal{O}} \left(\frac{v^2}{f^2} \right).
\end{equation}
Therefore, the measurement of $\kappa_Z$ 
provides a strong and robust constraint on $f$.  Taking the results of the 10-parameter fit in Table~\ref{tab:kappa-fit}, a precision of $0.21\%$ on $\kappa_Z$ implies that values of $f$ below 2.7\,TeV are excluded at 95\%\,CL.  For specific models, an even stronger bound on $f$, up to around $5\,$TeV, can be obtained by exploiting also 
its contributions to other Higgs boson couplings~\cite{Gu:2017ckc}.  The masses of the composite resonances are given by $m_\rho \sim g_\rho f$, where $g_\rho$ is the coupling of the new strong interaction, with a size typically much larger than one.
%
This indicates that the CEPC has the potential to probe composite resonance scales 
much above 10\,TeV, which is far beyond the reach of the LHC direct searches. 
The Higgs boson measurements at the CEPC thus provides a strong and robust test of the idea of naturalness in the composite Higgs boson models. The detailed exclusion regions from the CEPC and the LHC are shown in Fig.~\ref{fig:compositeH}, in terms of resonance mass $ m_\rho $, coupling parameter $ g_{\rho L}$ and mixing parameter $ \xi \equiv v^2/f^2 $.

Due to the large Higgs boson coupling to the top quark, arguably the most important particle in addressing the naturalness problem is the top-quark partner. For example, in supersymmetric models (SUSY), the particle mainly responsible for 
 stabilizing the electroweak scale is the scalar top, $\tilde{t}$ (stop).  The presence of stop will modify the Higgs boson couplings via a loop contribution, which is most notable for the $Hgg$ and $H \gamma \gamma$ couplings since they are also generated at the one-loop level in the SM. The dominant effect is on the $Hgg$ coupling,
\begin{equation}
\kappa_g -1 \simeq \frac{m_t^2}{4 m_{\tilde{t}}^2}.
\end{equation}
The measurement of $\kappa_g$ at the CEPC, up to $1 \% $ accuracy,  will allow us to probe stop mass up to 900 GeV~\cite{Fan:2014axa, Essig:2017zwe}. 
The situation is also very similar for non-SUSY models with fermionic top-quark partners, with the bounds on the top-quark partner mass being even stronger than the stop one~\cite{Essig:2017zwe}.  The more detailed exclusion region in the top-quark partner parameter space is presented in Fig.~\ref{fig:TopPartner} for both scenarios. 
\begin{figure*}[t!]
\begin{center}
\subfigure[]{\includegraphics[width=0.4\textwidth]{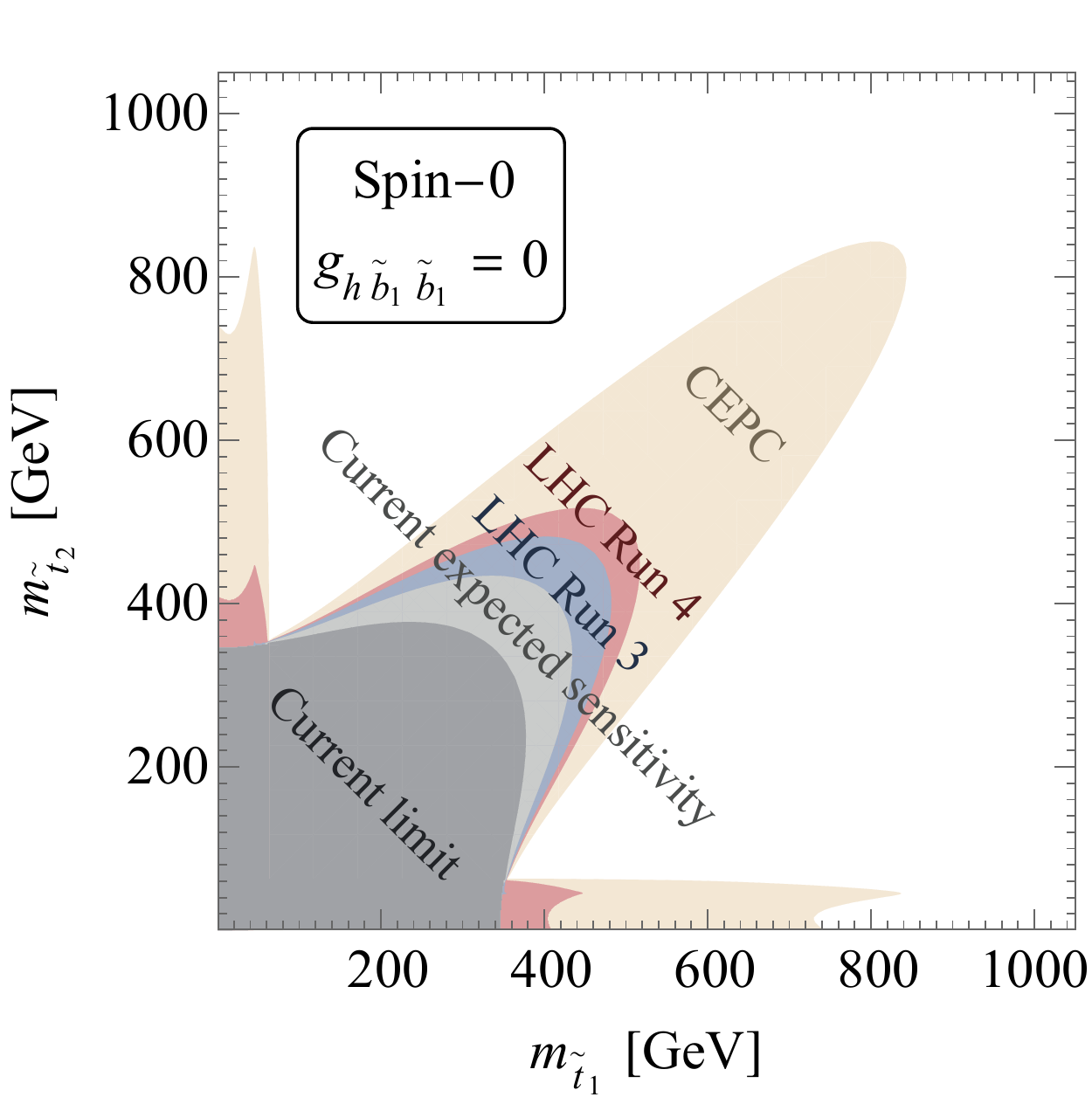}}
\subfigure[]{\includegraphics[width=0.4\textwidth]{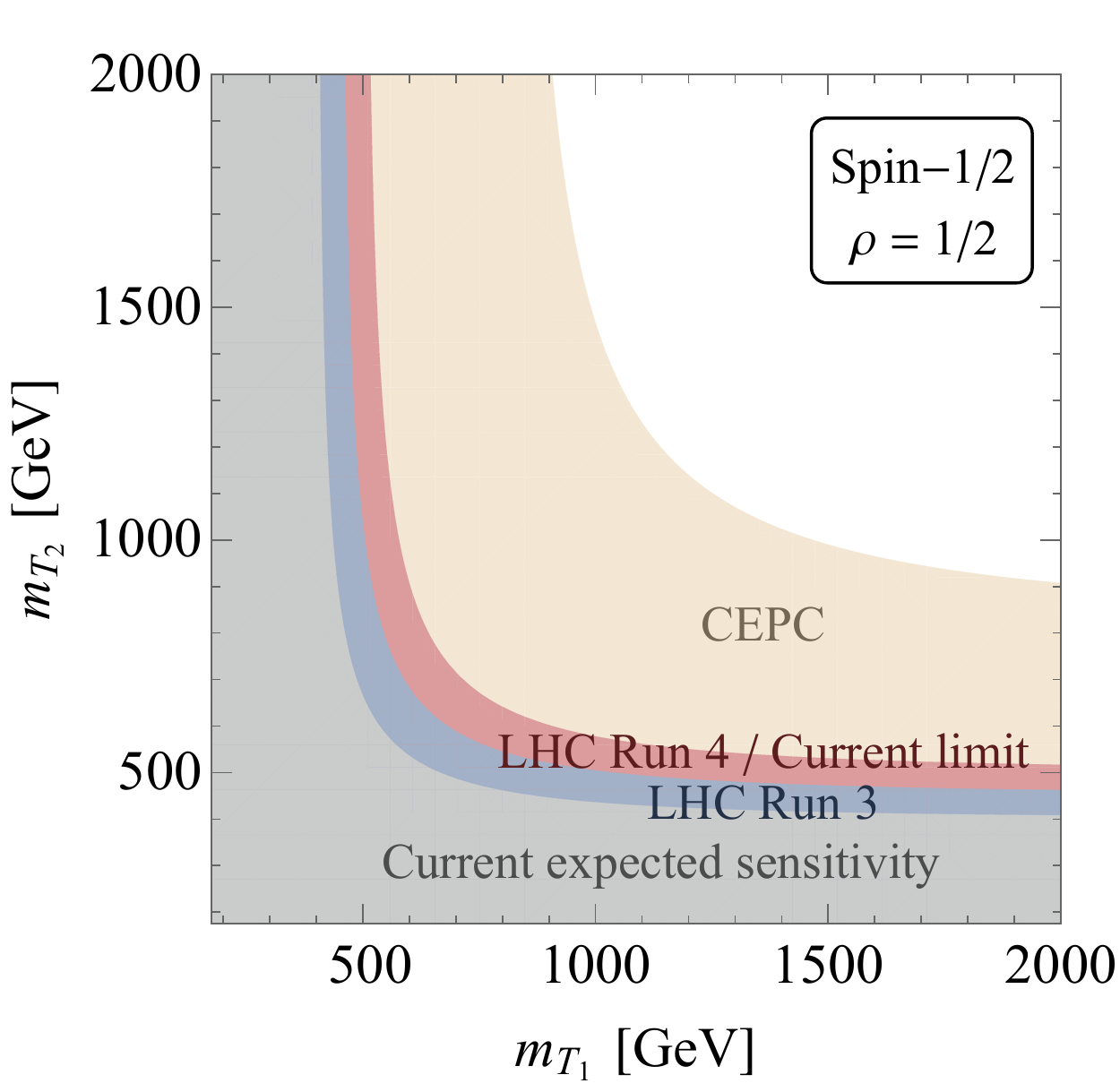}}
\figcaption{\small 95\%\,CL Limits on the stop (a) and fermionic top-quark partner (b) from Higgs boson coupling measurements at various current and future collider scenarios, including the CEPC.  This figure is reproduced from Ref.~\cite{Essig:2017zwe}. }
\label{fig:TopPartner}
\end{center}
\end{figure*}

This gives us another 
important handle to test the idea of naturalness. We note that, in favorable cases, the search of stop at the LHC run 2 can set a stronger limit on the stop mass. However, this limit depends strongly on the assumption of the mass spectrum of the other superpartners, as well as the relevant decay modes of the stop.  As a result, 
there will still be significant gaps remaining in the parameter space after the upcoming runs of the LHC, and even very light stops cannot be completely excluded. On the other hand, the measurement of the $Hgg$ coupling offers a complementary way of probing the stop that is independent of the decay modes of the stop.

It is also possible that the top-quark partner does not have the same SM gauge quantum numbers as the top quark. 
A particularly interesting possibility is that the top-quark partner is a SM singlet.  In such scenarios, it is very difficult to search for the top-quark partner at the LHC.  It is nontrivial to construct models with SM-singlet top-quark partners that resolve the fine-tuning problem of the electroweak scale~\cite{Chacko:2005pe, Burdman:2006tz}.  Nevertheless, they offer an extreme example that new physics with a scale of a few hundred GeVs could still be alive after the current and future LHC runs.
However, as mentioned earlier, any model that addresses the electroweak naturalness problem would inevitably contain sizable couplings to the Higgs boson.  The Higgs boson coupling measurements at the CEPC thus offer an ideal way of testing this type of models, which is very important for making robust arguments on the naturalness problem.  
As an example, we consider a scalar top-quark partner $\phi_t$ with its only interaction to the SM fields given by $H^\dagger H \phi_t^\dagger \phi_t$~\cite{Craig:2013xia,Craig:2014una}.   
This interaction contributes to the Higgs propagator at one-loop level, and induces a universal shift to all Higgs boson couplings.  The precise measurement of the inclusive $ZH$ cross section imposes a strong constraint on $\kappa_Z$ and provides the best constraint on the mass of the top-quark partner, $m_\phi$.  
As we can see from the left panel of Fig.~\ref{fig:HiddenTopPartner}, the CEPC will be able to probe 
$m_\phi$ up to around 700\,GeV, giving an non-trivial test of naturalness even in this very difficult scenario. 
A more concrete model is the so-called ``folded SUSY''~\cite{Burdman:2006tz}, in which the top-quark partners are scalars analogous to the stops in SUSY.  The projected constraints in the folded stop mass plane is shown on in the right panel of Fig.~\ref{fig:HiddenTopPartner}, which are at least around $350\,$GeV for both stops.
\begin{figure*}[t!]
	\begin{center}
		\subfigure[]{\includegraphics[width=0.4\textwidth]{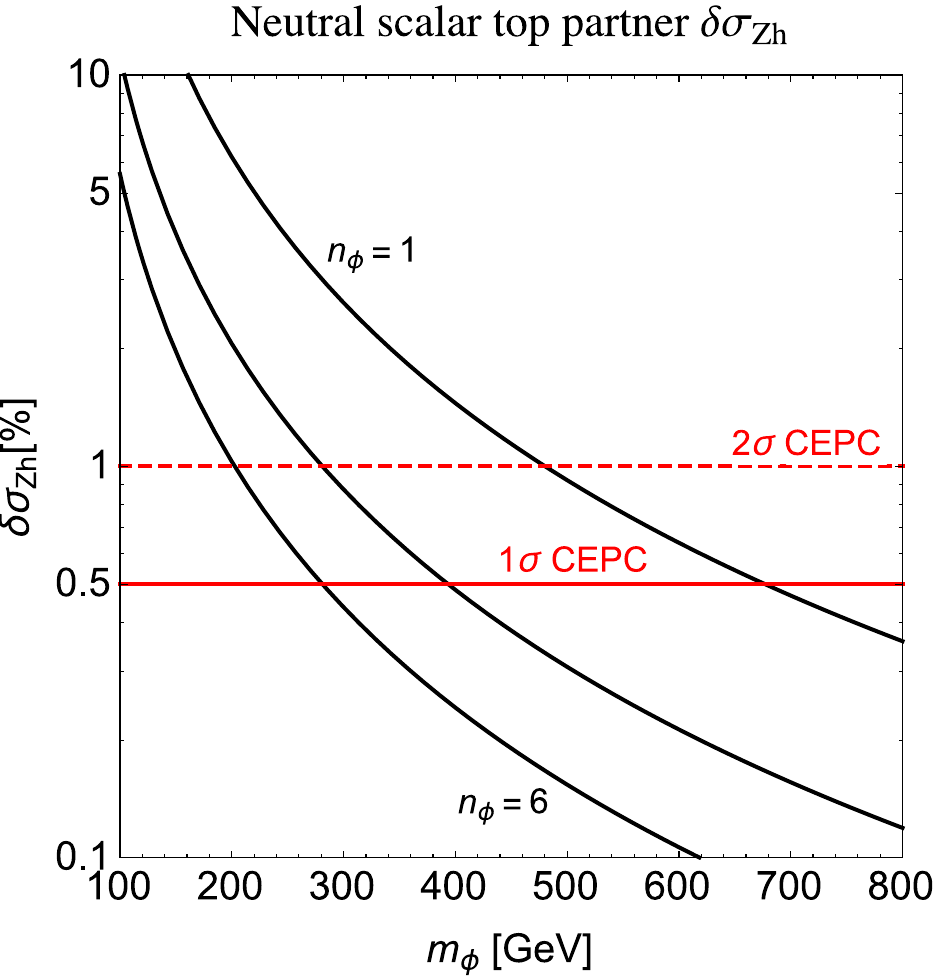}}
		\subfigure[]{\includegraphics[width=0.4\textwidth]{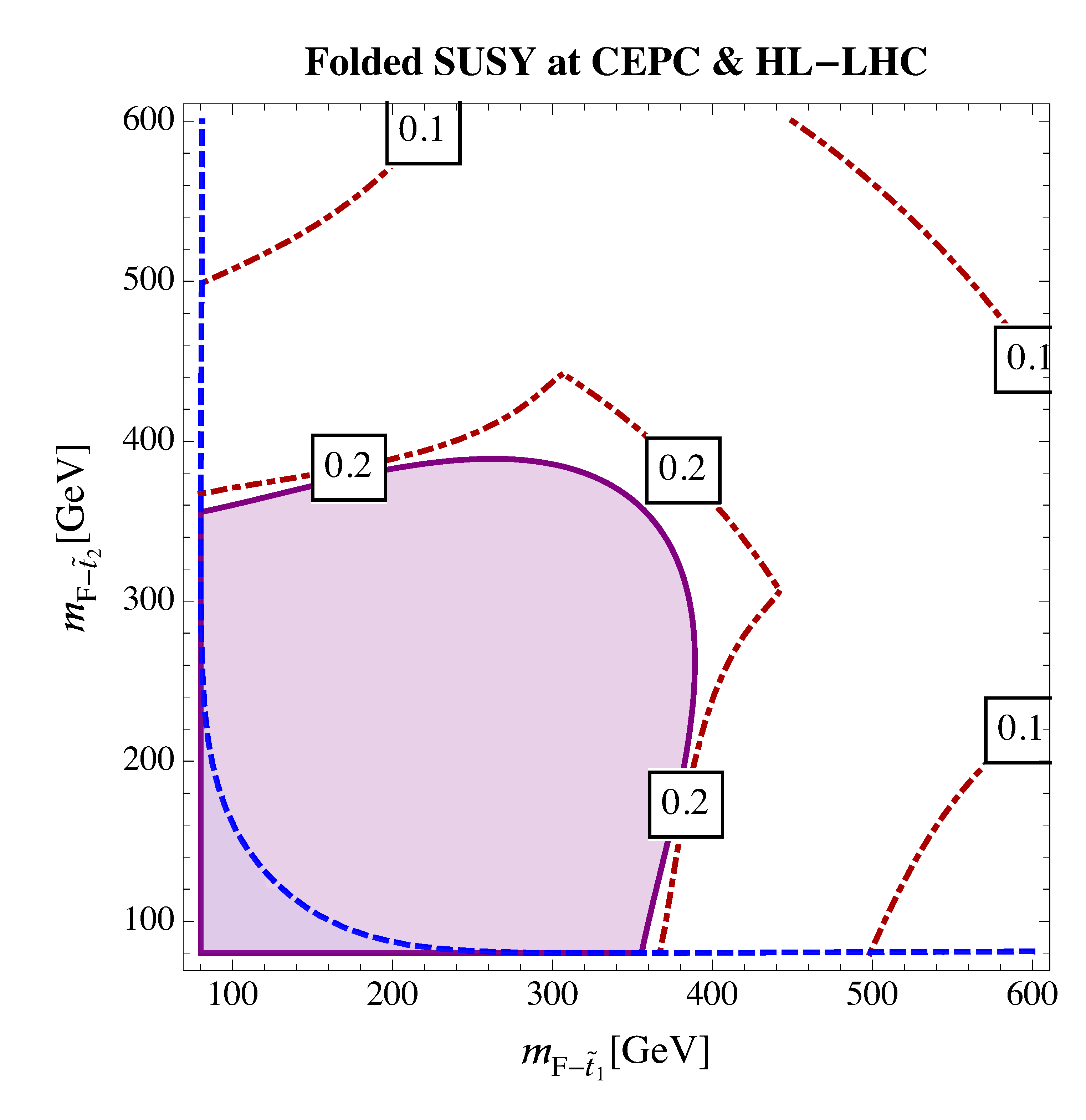}}
		\figcaption{\small (a) The fractional deviation of $\sigma_{ZH}$  at the Higgs factory in the scalar singlet top-quark partner model with the $H^\dagger H \phi^\dagger_t \phi_t $ interaction, reproduced from Ref.~\cite{Craig:2013xia}. (b) Projected constraints in the folded stop mass plane from the $H\gamma\gamma$ coupling measurements at HL-LHC and CEPC, reproduced from Ref.~\cite{Fan:2014axa}. The dot-dashed red contours indicates the fine-tuning in the Higgs boson mass from the quadratic sensitivity to stop soft terms.  }    
		\label{fig:HiddenTopPartner}
	\end{center}
\end{figure*}

\subsection{Electroweak phase transition}

 The measurement of the properties of the Higgs boson at the LHC has been consistent with the SM so far. At the same time, the nature of the electroweak phase transition remains unknown.
 While we have a very good knowledge of the sizes of the electroweak VEV and the Higgs boson mass, 
they only allow to probe a small region of the Higgs potential near the minimum, whereas the global picture of the potential is largely undetermined.
 This is shown schematically in Fig.~\ref{fig:ewpt_schematic}.
\begin{figure*}
\begin{center}
  \includegraphics[width=0.8\textwidth]{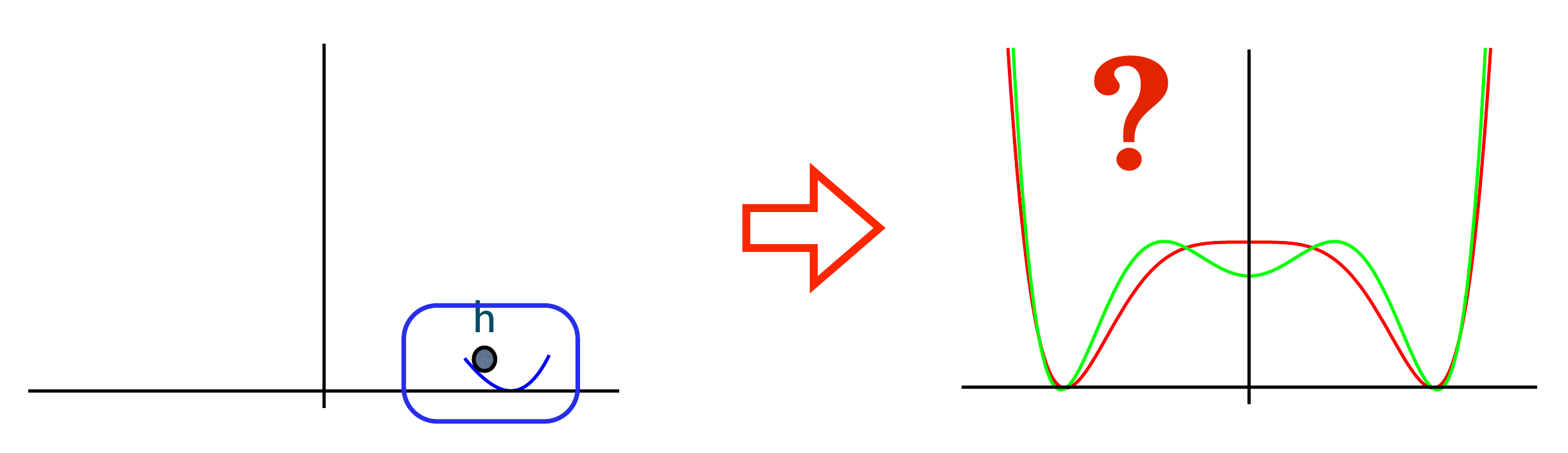}
\end{center}\vspace*{-0.4cm}
\caption{\small  A schematic drawing illustrating the question of the nature of the electroweak phase transition. 
Left:  Our current knowledge of the Higgs potential.  Right: Based on our current knowledge, we could not distinguish the SM Mexican Hat potential from an alternative one with more wiggles.  
}
\label{fig:ewpt_schematic}
\end{figure*}

%
The remaining region of the Higgs potential is difficult to probe, even with an upgraded LHC.  Meanwhile, it has important consequences on the early universe cosmology and 
the understanding of our observable world.   
For example, it is crucial in determining whether the electroweak phase transition is of first or second order. 
The nature of the electroweak phase transition can also be relevant for the matter anti-matter asymmetry in the Universe, as a large class of models of baryogenesis rely on a first order electroweak phase transition. 
The CEPC has the capability of probing many of these models and potentially revealing the nature of the electroweak phase transition and the origin of baryogenesis.

It is well known that with a minimal Higgs potential and the SM Higgs sector, the electroweak phase transition is of second order~\cite{Kajantie:1996mn}.  New physics with sizable couplings to the Higgs boson are needed to make the phase transition a first order one. 
The measurement of the triple Higgs boson coupling offers an ideal testing ground for these new physics models.  Being the third derivative, it carries more information about the global shape of the Higgs potential than the mass.  It can also be determined to a reasonable precision at the future colliders, unlike the quartic Higgs boson coupling.  Indeed, most models with first order electroweak phase transition predict a triple  Higgs boson coupling with large deviations from the SM prediction. 
This is demonstrated with a simple example in Fig.~\ref{fig:self_coupling_PT}, which shows the deviation in the triple Higgs boson coupling for a generic singlet model.  For the model points that produces a first order phase transition, the value of triple Higgs boson coupling indeed covers a wide range and can be different from the SM prediction by up to $100\%$.
\begin{center}
  \includegraphics[width=0.45\textwidth]{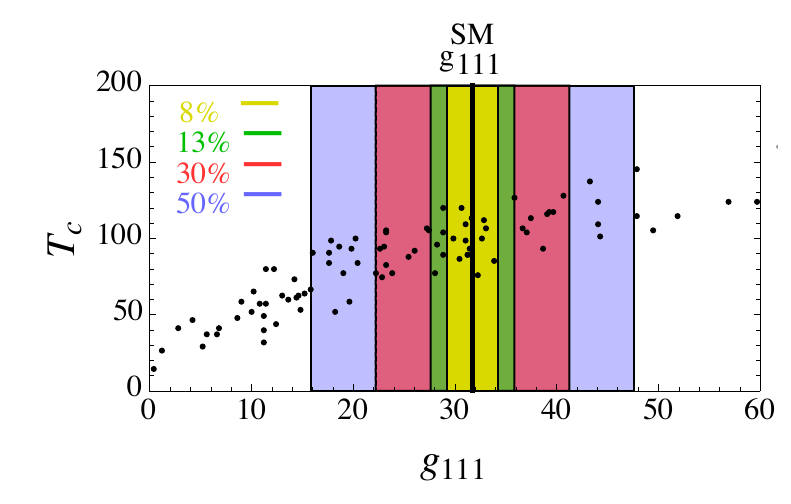}
  \figcaption{\small The deviation in the triple Higgs boson coupling in a generic singlet model that could produce first order electroweak phase transition, reproduced from Ref.~\cite{Profumo:2014opa}.  Black dots are points where the phase transition is of first order. The parameter $g_{111}$ is the triple Higgs boson coupling.  
	}
\label{fig:self_coupling_PT}
\end{center}

The CEPC could probe the triple Higgs boson coupling via its loop contributions to single Higgs boson processes.  As pointed out in Section~\ref{subsec:hhh}, it will have a limited reach to the most general scenario in which all Higgs boson couplings are allowed to deviate from their SM values. 
An additional run at $350\,$GeV will help to  improve the sensitivity, while a direct measurement using di-Higgs production would have to wait for a future proton-proton collider, or a lepton collider running at much higher energies.  
%
However, it should be noted that the model independent approach in Section~\ref{subsec:hhh} makes no assumption on any possible connection between the triple Higgs boson coupling and other couplings.
In practice, to induce large deviation in triple Higgs boson coupling requires the new physics to be close to the weak scale, while the presence of such new physics will most likely induce deviations in other Higgs boson couplings as well, such as the couplings to the electroweak gauge bosons. 
Without some symmetry or fine tuning, both deviations are expected to come in at the order of $v^2/M_{\rm NP}^2$.  Such deviations can be probed very well at lepton colliders. 

\begin{figure*}[t!]
\begin{center}
\subfigure[]{\includegraphics[width=0.5\textwidth]{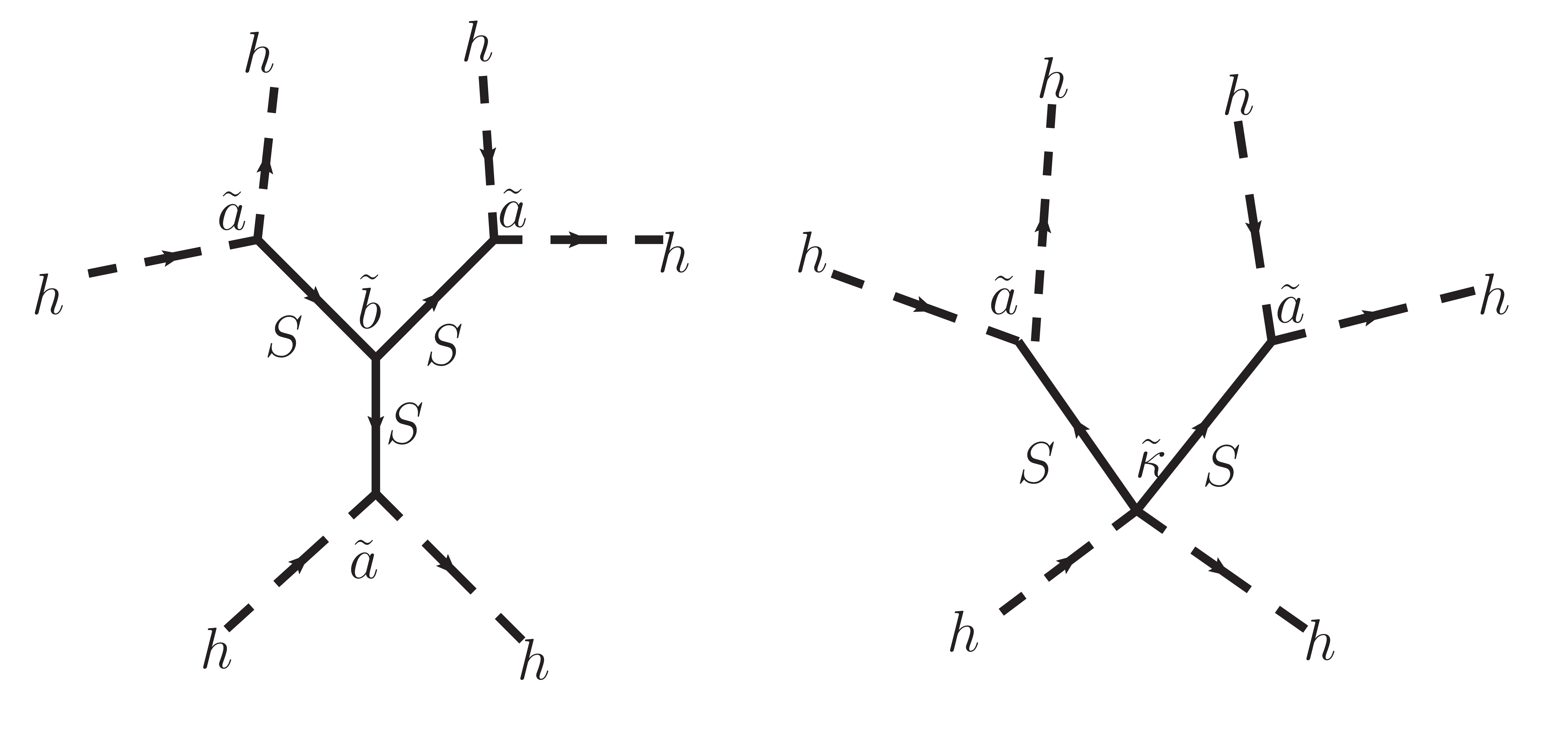}}
\subfigure[]{\includegraphics[width=0.4\textwidth]{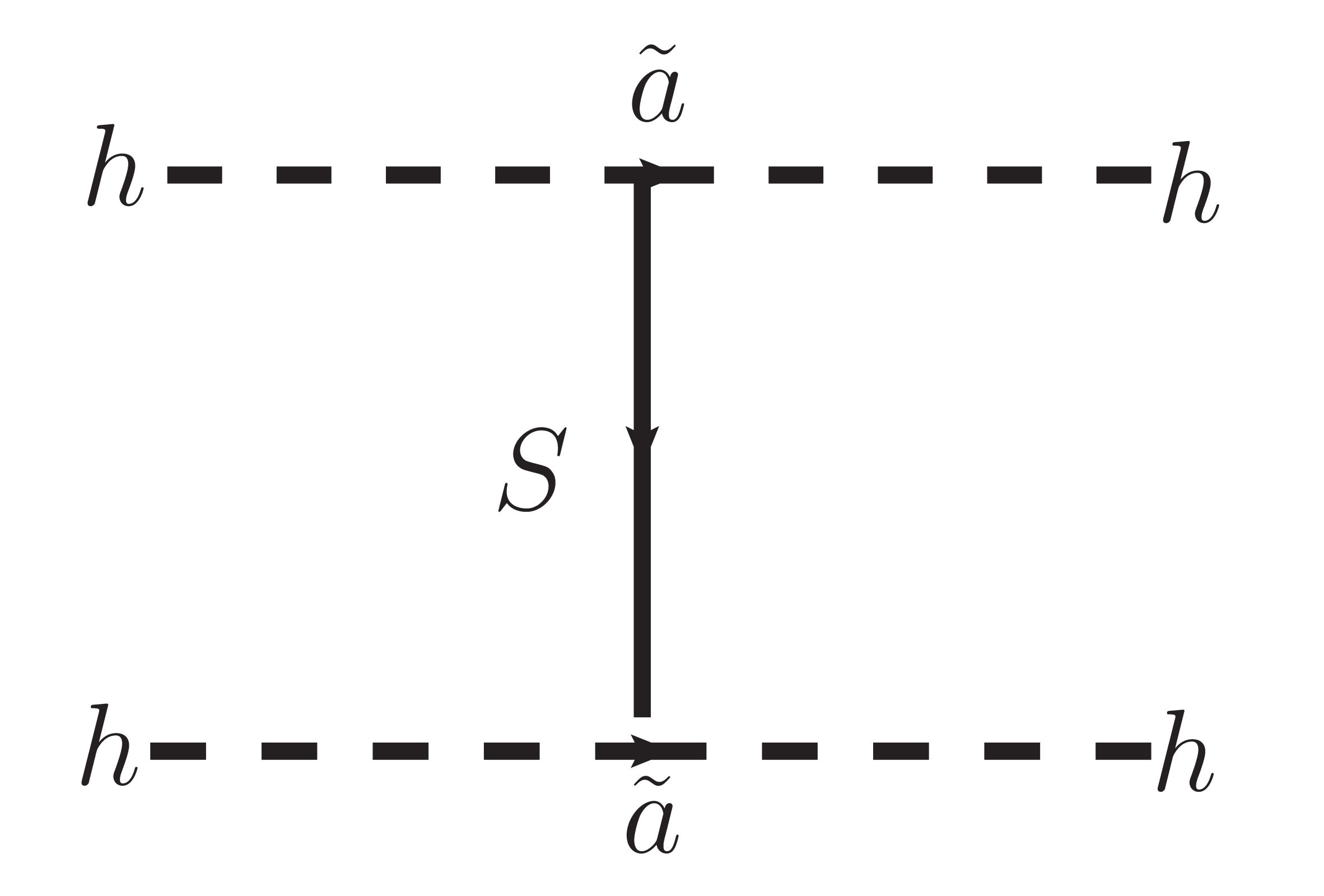}}
\figcaption{\small  (a) Induced $|H|^6$ couplings after integrating out the singlet. (b) Induced wave function renormalization of the Higgs, $|H^\dagger \partial H|^2 $. }
\label{fig:singlet_Higgs_coupling}
\end{center}
\end{figure*}
We will now demonstrate this in the context of  models.  
Instead of a comprehensive survey, we will focus here on some of the simplest possibilities which are also difficult to probe. The minimal model that has been well studied  in this class introduces an additional singlet scalar which couples to the Higgs 
boson~\cite{Profumo:2014opa,Katz:2014bha,Noble:2007kk,Henning:2014gca,Profumo:2007wc,Curtin:2014jma}.  The general 
potential of the Higgs boson and the new scalar $S$ is 
\begin{eqnarray}
V(H,S) =  \frac{1}{2}  \mu^2 |H|^2 + \frac{\lambda}{4} |H|^4 +  m_S^2 S^2 + \nonumber\\
 \tilde{a} S |H|^2 + \tilde{\kappa} S^2 |H|^2  + \tilde{b} S^3 + \lambda_S S^4. \label{eq:realsing}
\end{eqnarray}
After integrating out the singlet, it will generate an $|H|^6$ interaction (shown in panel (a) in Fig.~\ref{fig:singlet_Higgs_coupling}), which, after electroweak symmetry breaking, leads to a modification of the triple Higgs boson coupling on the order of $v^2/m_S^2$. At the same time, it will also generate the operator $|H^\dagger \partial H|^2 $. This leads to a wave function renormalization, which gives rises to universal shift of the Higgs boson couplings. In particular, the modification of the $HZZ$ coupling is also of order $ \sim v^2 / m_{S}^2$.  
We thus expect $\kappa_Z$, which is constrained within $0.25\%$ even with the inclusive $ZH$ measurement alone, to provide the best constraining power on this model.  
%
This is explicitly verified with a scan in the model parameter space, shown in Fig.~\ref{fig:singlet_dZ_triple}.  The model points with a first order phase transition are projected on the plane of the $HZZ$ and triple Higgs boson couplings.  Indeed, for model points with a large deviation in the triple Higgs boson coupling, a sizable deviation in the $HZZ$ coupling is also present. In this model, constraining power of the $HZZ$ coupling measurement at CEPC is almost the same as the  triple Higgs boson coupling measurement at a future $100\,$TeV hadron collider. 
A more detailed view of the parameter space of the real singlet model is presented in Fig.~\ref{fig:singletps}.  In addition to the deviations in $\sigma(ZH)$ at CEPC, the sensitivities of the current and future electroweak precision tests are also presented~\cite{Cao:2017oez}.  The  $\sigma(ZH)$ measurement, with a projected precision of $0.5\%$, indeed provides the best sensitivity in this scenario.
We thus conclude that CEPC has an excellent coverage in the full model space that gives a first order electroweak phase transition. 
\begin{center} 
\includegraphics[width=0.4\textwidth]{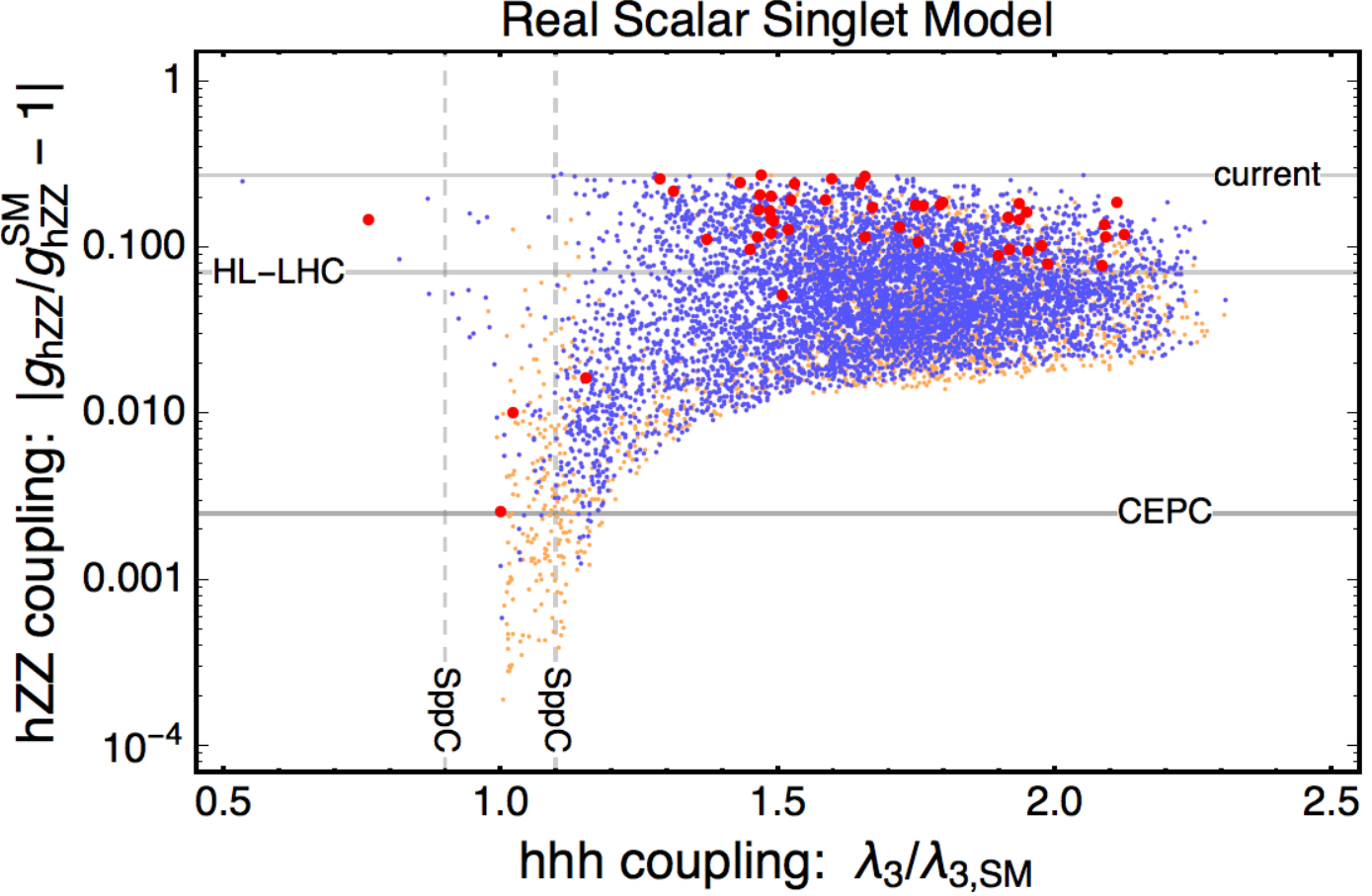}
\figcaption{\small  
The $HZZ$ and $HHH$ couplings in the real scalar singlet model of Eq.~\ref{eq:realsing}. The points in this figure represent models with a first order electroweak phase transition, and are obtained by scanning over the theory space. Points 
with a first order phase transition are shown in orange, points with a strongly first order phase transition are shown in blue, and points with a strongly first order phase transition that also produces detectable gravitational waves are shown in red. This figure is reproduced from Ref.~\cite{Huang:2016cjm}. }
\label{fig:singlet_dZ_triple}
\end{center}
%
%
\begin{figure*}[t!]
\begin{center} 
\subfigure[]{\includegraphics[width=0.3\textwidth]{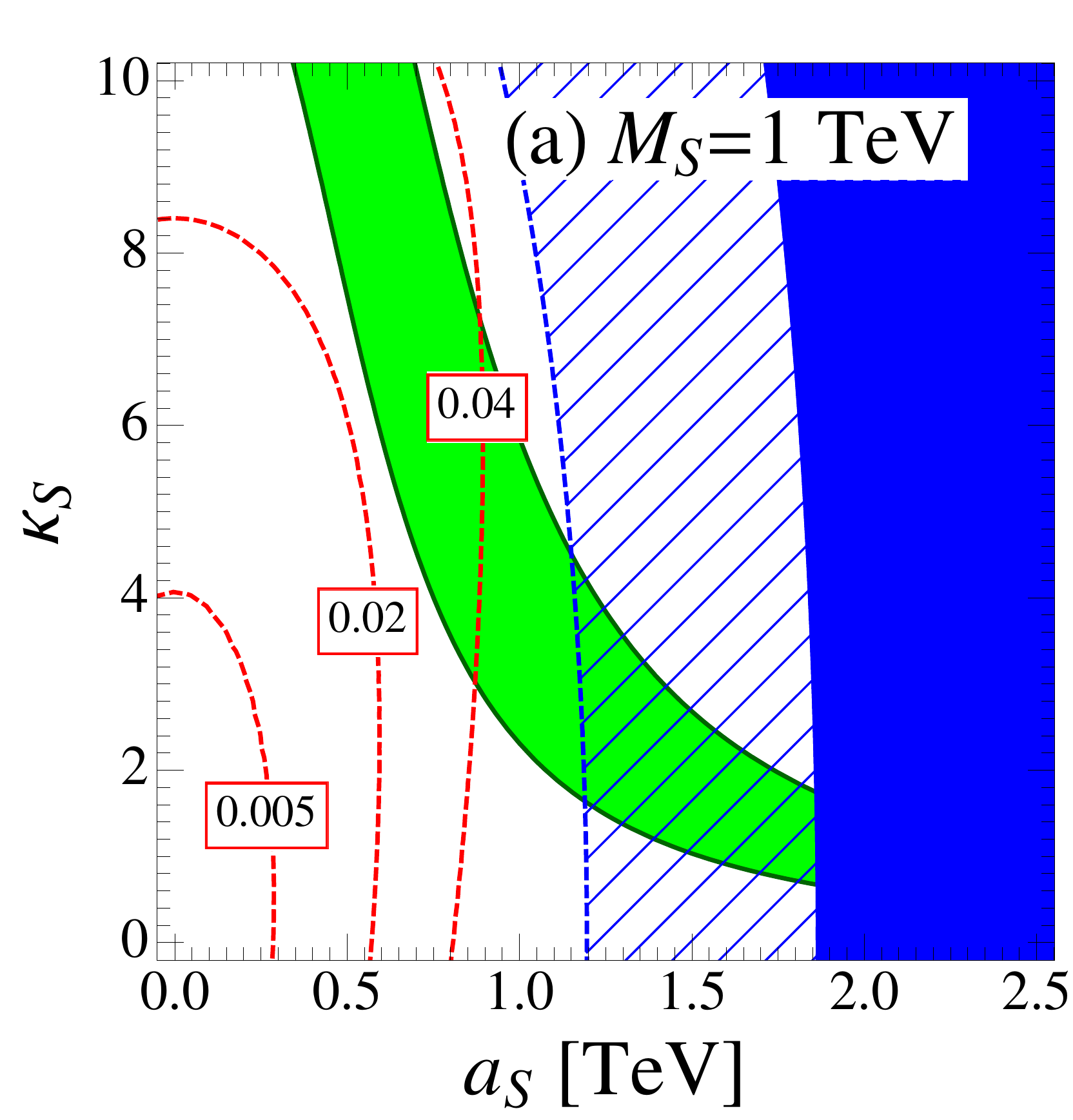}}
\subfigure[]{\includegraphics[width=0.3\textwidth]{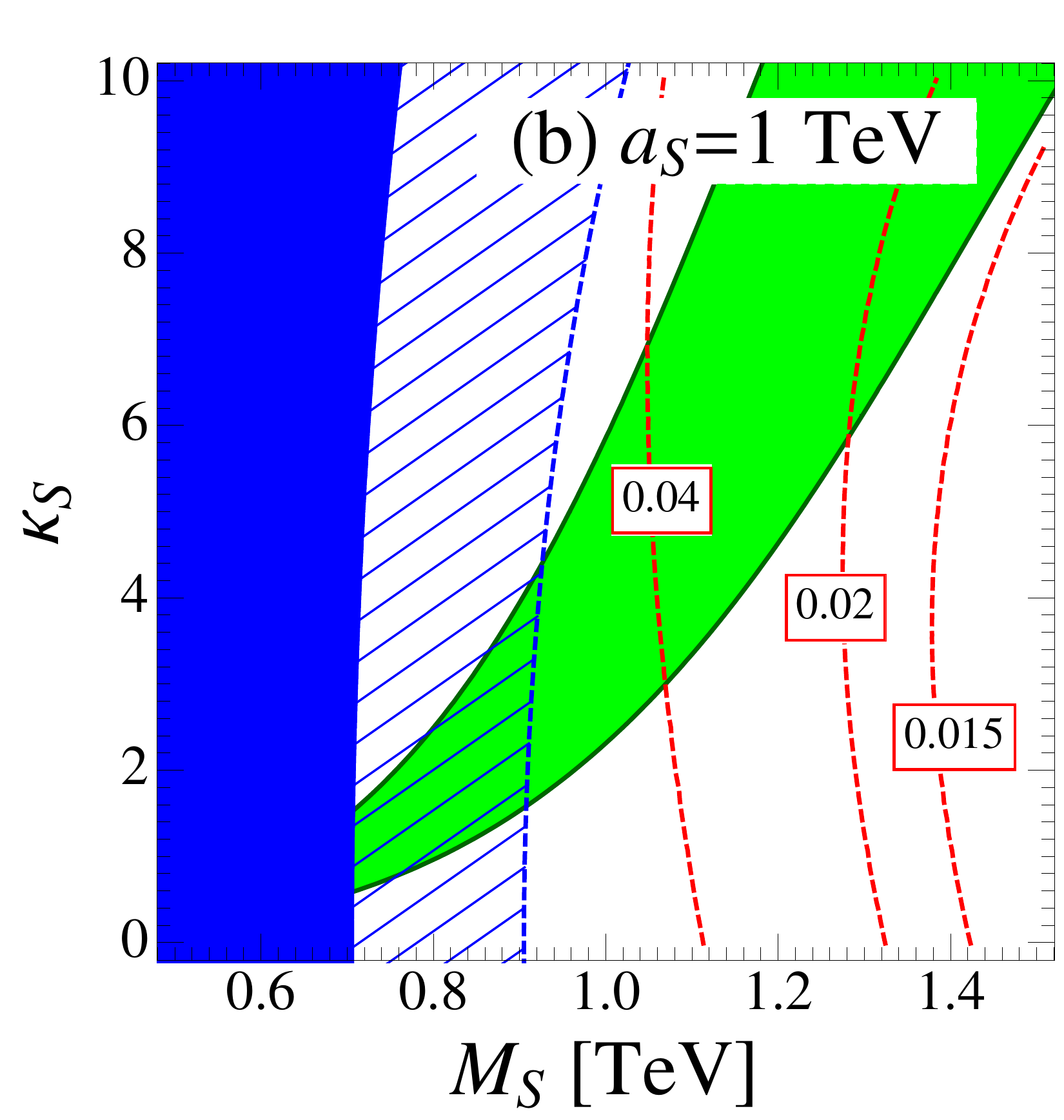}}
\subfigure[]{\includegraphics[width=0.3\textwidth]{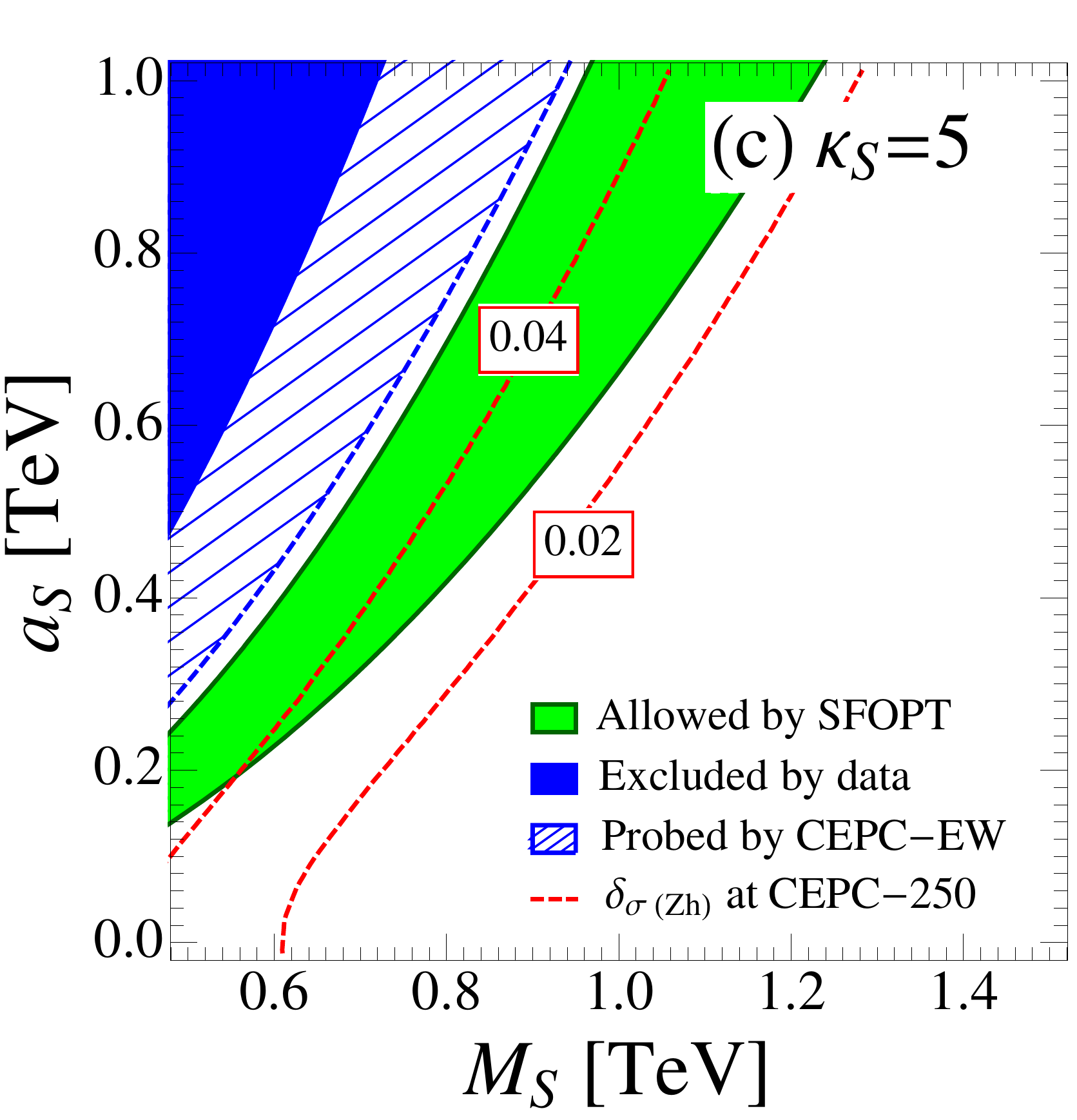}}
\figcaption{\small    The parameter space compatible with a strong first order phase transition (green region) and the deviations in $\sigma(ZH)$ (dashed red contours) in the real singlet scalar model, reproduced from Ref.~\cite{Cao:2017oez}.  The solid blue region is excluded by current EW and Higgs boson data, and the region with dashed blue lines can be probed by the CEPC $Z$-pole run.}
\label{fig:singletps}
\end{center}
\end{figure*}

A more restricted scenario, in which a discrete $Z_2$ symmetry is imposed on the singlet, has also been considered \cite{Katz:2014bha,Curtin:2014jma}. It is significantly more difficult to achieve a first order electroweak phase transition in this scenario, since the singlet could only modify the Higgs potential at loop levels.  To produce the same level of deviation in the Higgs potential, a much stronger coupling between the Higgs boson and the singlet is required, which often exceeds the limits imposed by the requirement of perturbativity.  
For the same reason, the expected loop induced deviation in the triple Higgs boson coupling is also generically smaller in this case, and is about $10-15\%$, as shown in Fig.~\ref{subfig:Z2a}.  
Even in this difficult case, we see in Fig.~\ref{subfig:Z2b} that the expected deviation of the cross section $\sigma(ZH)$ 
is about $0.6 \%$.
Therefore, the CEPC will see the first evidence of new physics even in this very difficult case. In the more general classes of models, the new physics which modifies the Higgs boson coupling could carry other SM gauge quantum numbers, such as electric charge and/or color. In such cases, there will be significant modifications to the $Hgg$ and $H \gamma \gamma$ couplings.  One such example is shown in  Fig.~\ref{subfig:Z2c}, with a $6 \%$ deviation in the $H\gamma\gamma$ coupling expected in order to obtain a first order phase transition.  As shown in Table~\ref{tab:kappa-fit}, 
the combination of CEPC and HL-LHC measurements could constrain $\kappa_\gamma$ to a precision of $1.7\%$, and would test this scenario with a sensitivity of more than three standard deviations.

%
\begin{figure*}[t!]
\begin{center} 
\subfigure[triple Higgs boson coupling]{\includegraphics[width=0.3\textwidth]{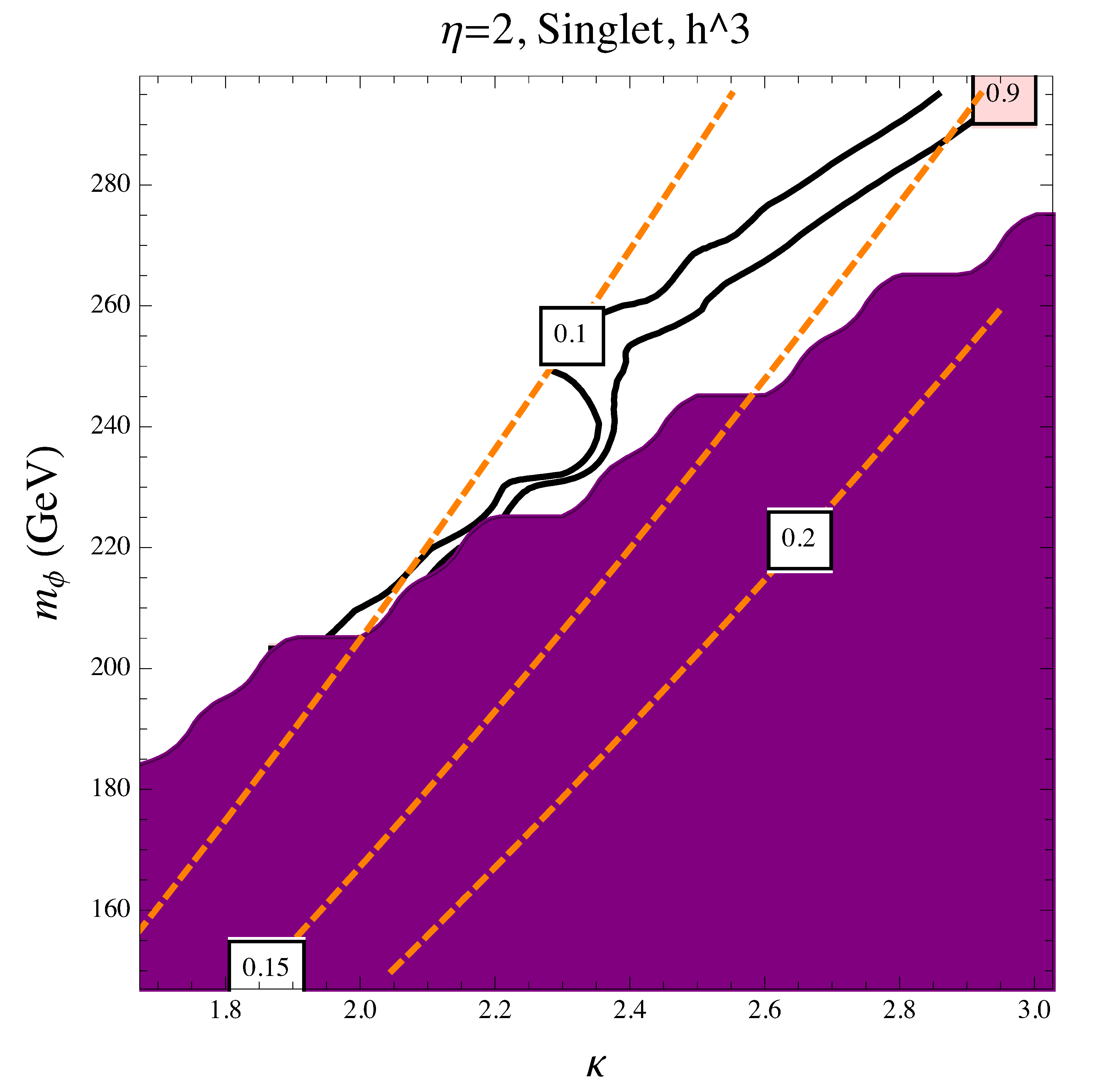} \label{subfig:Z2a}}
\subfigure[$\sigma(ZH)$]{\includegraphics[width=0.3\textwidth]{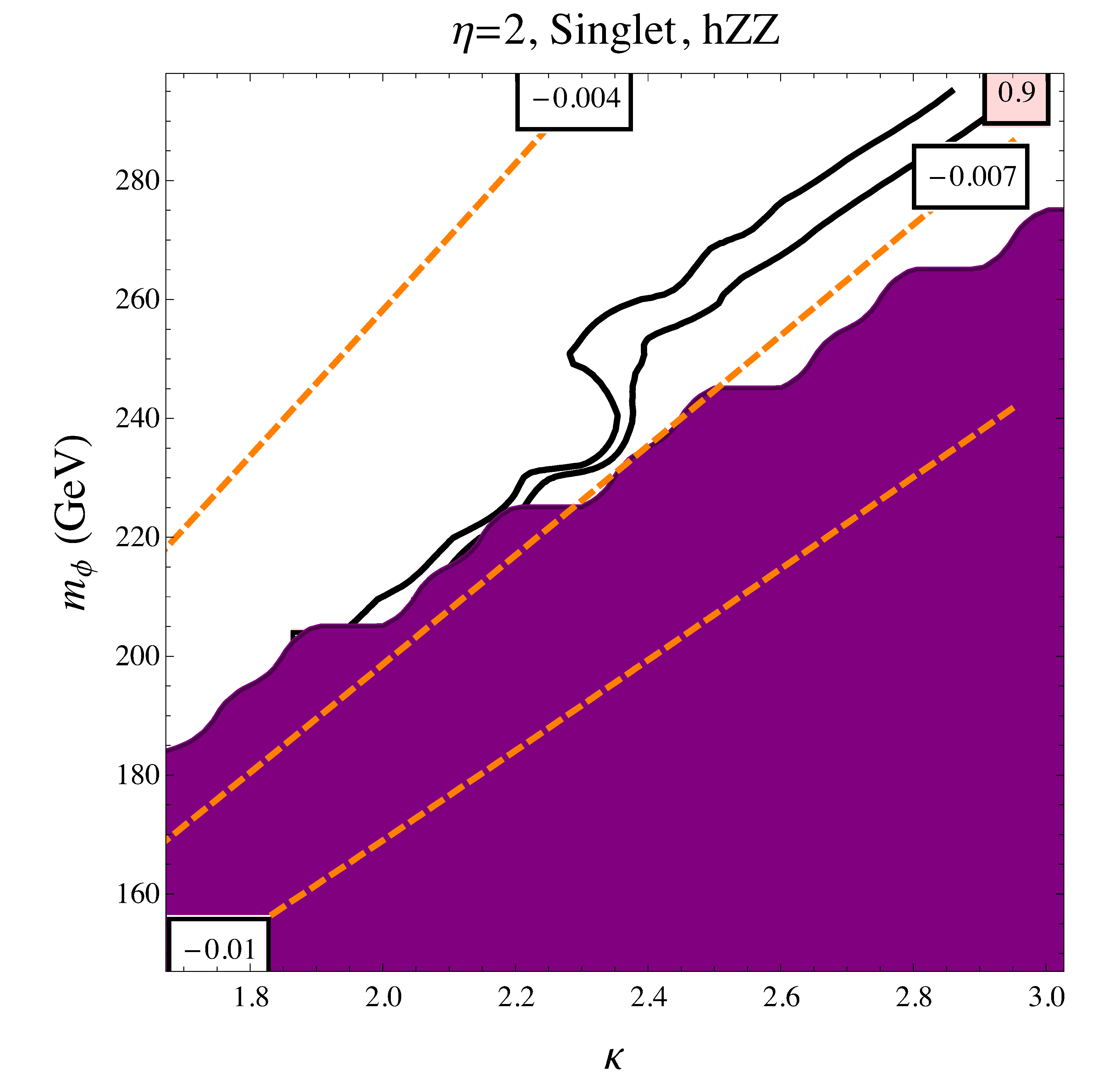} \label{subfig:Z2b}}
\subfigure[$H\gamma\gamma$ coupling]{\includegraphics[width=0.3\textwidth]{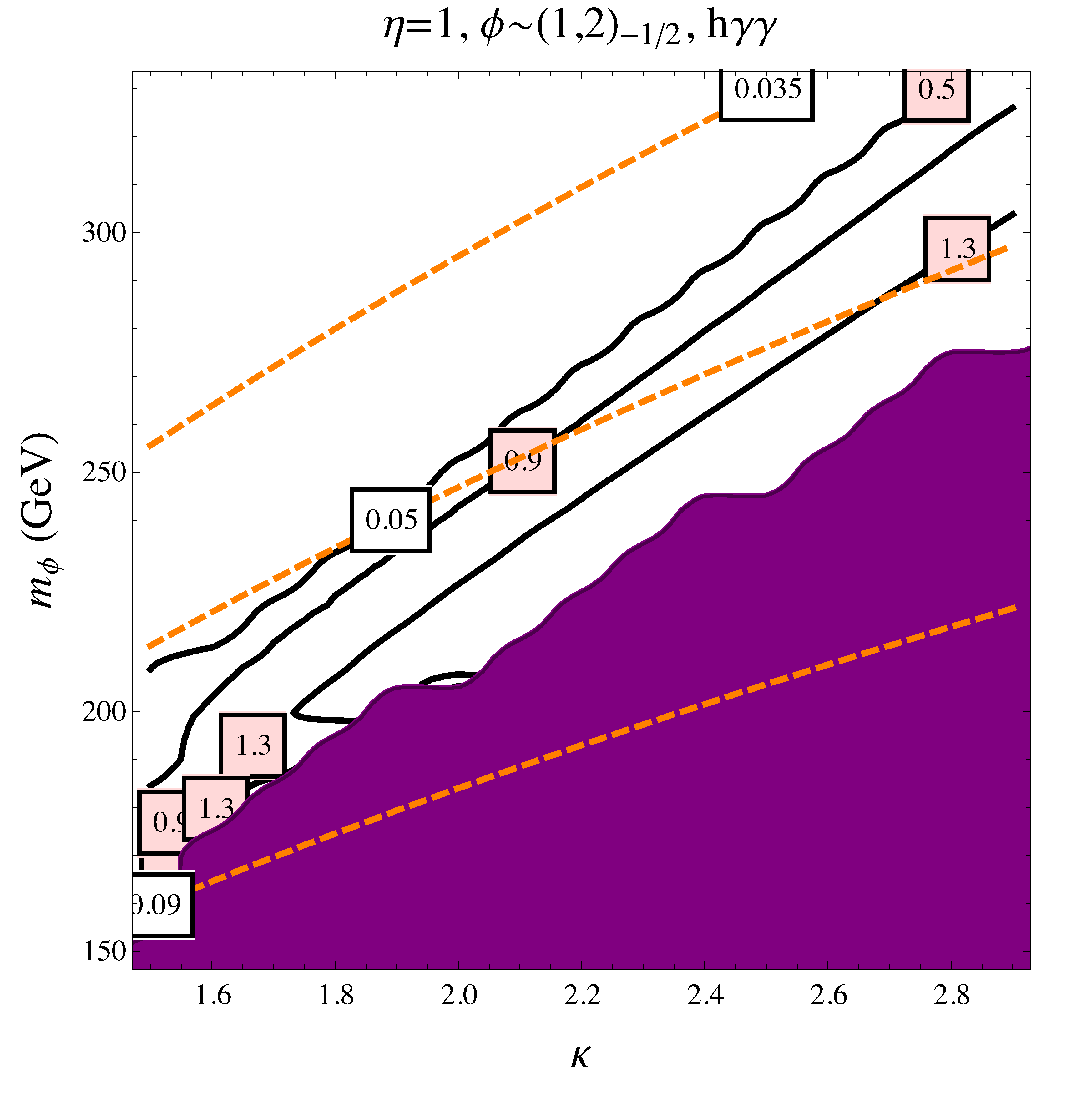} \label{subfig:Z2c}}
\figcaption{\small  Deviations in the triple Higgs, $\sigma(ZH)$ and $H\gamma\gamma$ couplings in models with $Z_2$ symmetry.  In each plot, the dashed orange lines are contours of constant deviations in the corresponding quantity, the solid black lines are contours of constant electroweak phase transition strength parameter $\xi = v(T_c)/T_c$, where $v(T_c)$ is the Higgs VEV at temperature $T_c$.  The shaded region is excluded for producing a color-breaking vacuum.  
}
\label{fig:Z2}
\end{center}
\end{figure*}
%

%
%
%


In general, the newly discovered Higgs particle could serve as a gateway to new physics. One generic form of the Higgs boson coupling to new physics is the so-called Higgs portal, $H^\dagger H {\mathcal{O}}_{\rm NP}$, where ${\mathcal{O}}_{\rm NP}$ is an operator composed out of new physics fields.    Since $H^\dagger H$ is the lowest dimensional operator that is consistent with all the symmetries in the SM, it is easy to construct scenarios in which such Higgs portal couplings are the most relevant ones for the low energy phenomenology of new physics. The singlet extended Higgs sector and the scalar top-quark partner, discussed earlier, are special examples of this scenario. In general, the Higgs portal interactions will shift the Higgs boson couplings, and can be thoroughly tested at the CEPC. Moreover, if the new physics is lighter than $m_H /2$, the Higgs portal coupling will lead to new Higgs boson decay channels.  We have already seen in Section~\ref{subsec:exotic} that the CEPC has an excellent capability of probing such exotic decays, and could cover a vast range of decay signals.




\section{Conclusion}
\label{sec:conclusion}

The Higgs boson is responsible for the electroweak symmetry breaking. It is the only fundamental scalar particle in the Standard Model observed so far. The discovery of such a particle at the LHC is a major breakthrough on both theoretical and experimental fronts. However, the Standard Model is likely only an effective theory at the electroweak scale. To explore potential new physics at the electroweak scale and beyond, complementary approaches of direct searches at the energy frontier as well as precision measurements will be needed. The current LHC and the planned HL-LHC have the potential to significantly extend its new physics reach and to measure many of the Higgs boson couplings with precision of a few percent. 

However, many new physics models predict Higgs boson coupling deviations at the sub-percent level, beyond those achievable at the LHC. The CEPC complements the LHC and will be able to study the properties of the  Higgs boson in great detail with unprecedented precision.   Therefore it is capable of unveiling the true nature of this particle. At the CEPC, most Higgs boson couplings can be measured with precision at a sub-percent level. More importantly, the CEPC will able to measure many of the key Higgs boson properties such as the total width and decay branching ratios in a model-independent way, greatly enhancing the coverage of new physics searches. Furthermore, the clean event environment of the CEPC will allow the detailed study of known decay modes and the identification of potential unknown decay modes that are impractical to test at the LHC.  

This paper provides a snapshot of the current studies, many  of which are still ongoing. More analyses are needed to   fully understand the physics potential of the CEPC. Nevertheless, the results presented here have already built a strong case for the CEPC as a Higgs factory. The CEPC has the potential to characterize the Higgs boson in the same way LEP did with the $Z$ boson, and potentially shed light on new physics.


\end{multicols}
\clearpage
\begin{multicols}{2}
{
\bibliographystyle{style/atlasnote}
\bibliography{CEPCHiggsCPC}

\providecommand{\href}[2]{#2}\begingroup\raggedright\begin{thebibliography}{100}

\bibitem{atlas:2012obs}
{ATLAS} Collaboration, {\em {Observation of a new particle in the search for
  the Standard Model Higgs boson with the ATLAS detector at the LHC}\/},
  \href{http://dx.doi.org/10.1016/j.physletb.2012.08.020}{Phys. Lett. {\bf
  B716} (2012)  1--29},
\href{http://arxiv.org/abs/1207.7214}{{\tt arXiv:1207.7214 [hep-ex]}}.

\bibitem{cms:2012obs}
{CMS} Collaboration, {\em {Observation of a new boson at a mass of 125 GeV with
  the CMS experiment at the LHC}\/},
  \href{http://dx.doi.org/10.1016/j.physletb.2012.08.021}{Phys. Lett. {\bf
  B716} (2012)  30--61},
\href{http://arxiv.org/abs/1207.7235}{{\tt arXiv:1207.7235 [hep-ex]}}.

\bibitem{Aad:2013wqa}
{ATLAS} Collaboration, {\em {Measurements of Higgs boson production and
  couplings in diboson final states with the ATLAS detector at the LHC}\/},
  \href{http://dx.doi.org/10.1016/j.physletb.2014.05.011,
  10.1016/j.physletb.2013.08.010}{Phys. Lett. {\bf B726} (2013)  88--119},
  \href{http://arxiv.org/abs/1307.1427}{{\tt arXiv:1307.1427 [hep-ex]}}.
[Erratum: Phys. Lett.B734,406(2014)].

\bibitem{Aad:2013xqa}
{ATLAS} Collaboration, {\em {Evidence for the spin-0 nature of the Higgs boson
  using ATLAS data}\/},
  \href{http://dx.doi.org/10.1016/j.physletb.2013.08.026}{Phys. Lett. {\bf
  B726} (2013)  120--144},
\href{http://arxiv.org/abs/1307.1432}{{\tt arXiv:1307.1432 [hep-ex]}}.

\bibitem{Chatrchyan:2013lba}
{CMS} Collaboration, {\em {Observation of a new boson with mass near 125 GeV in
  pp collisions at $\sqrt{s}$ = 7 and 8 TeV}\/},
  \href{http://dx.doi.org/10.1007/JHEP06(2013)081}{JHEP {\bf 1306} (2013)
  081},
\href{http://arxiv.org/abs/1303.4571}{{\tt arXiv:1303.4571 [hep-ex]}}.

\bibitem{Chatrchyan:2014vua}
{CMS} Collaboration, {\em {Evidence for the direct decay of the 125 GeV Higgs
  boson to fermions}\/},  \href{http://dx.doi.org/10.1038/nphys3005}{Nature
  Phys. {\bf 10} (2014)  },
\href{http://arxiv.org/abs/1401.6527}{{\tt arXiv:1401.6527 [hep-ex]}}.

\bibitem{Khachatryan:2014kca}
{CMS} Collaboration, {\em {Constraints on the spin-parity and anomalous HVV
  couplings of the Higgs boson in proton collisions at 7 and 8 TeV}\/},
  \href{http://dx.doi.org/10.1103/PhysRevD.92.012004}{Phys. Rev. {\bf D92}
  (2015)  012004},
\href{http://arxiv.org/abs/1411.3441}{{\tt arXiv:1411.3441 [hep-ex]}}.

\bibitem{Khachatryan:2016vau}
{ATLAS and CMS Collaborations}, {\em {Measurements of the Higgs boson
  production and decay rates and constraints on its couplings from a combined
  ATLAS and CMS analysis of the LHC pp collision data at $ \sqrt{s}=7 $ and 8
  TeV}\/},  \href{http://dx.doi.org/10.1007/JHEP08(2016)045}{JHEP {\bf 08}
  (2016)  045},
\href{http://arxiv.org/abs/1606.02266}{{\tt arXiv:1606.02266 [hep-ex]}}.

\bibitem{Aad:2015zhl}
{ATLAS and CMS Collaborations}, {\em {Combined Measurement of the Higgs Boson
  Mass in $pp$ Collisions at $\sqrt{s}=7$ and 8 TeV with the ATLAS and CMS
  Experiments}\/},
  \href{http://dx.doi.org/10.1103/PhysRevLett.114.191803}{Phys. Rev. Lett. {\bf
  114} (2015)  191803},
\href{http://arxiv.org/abs/1503.07589}{{\tt arXiv:1503.07589 [hep-ex]}}.

\bibitem{ATL-PHYS-PUB-2014-016}
{ATLAS Collaboration}, {\em {Projections for measurements of Higgs boson signal
  strengths and coupling parameters with the ATLAS detector at a HL-LHC}\/},
  ATL-PHYS-PUB-2014-016 (2014)  . \url{http://cds.cern.ch/record/1956710}.

\bibitem{CMS:2013xfa}
{CMS} Collaboration, {\em {Projected Performance of an Upgraded CMS Detector at
  the LHC and HL-LHC: Contribution to the Snowmass Process}\/},
  \href{http://arxiv.org/abs/1307.7135}{{\tt arXiv:1307.7135 [hep-ex]}}.
\url{http://www.slac.stanford.edu/econf/C1307292/docs/submittedArxivFiles/1307.7135.pdf}.

\bibitem{CEPCStudyGroup:2018rmc}
{CEPC Study Group}, {\em {CEPC Conceptual Design Report}\/},
\href{http://arxiv.org/abs/1809.00285}{{\tt arXiv:1809.00285
  [physics.acc-ph]}}.

\bibitem{MoradeFreitas:2002kj}
P.~Mora~de Freitas and H.~Videau, {\em {Detector simulation with MOKKA /
  GEANT4: Present and future}\/},
\newblock {Presented at the International Workshop on physics and experiments
  with future electron-positron linear colliders, Jeju Island, Korea (2002)}.
\newblock
\url{http://inspirehep.net/record/609687}.
\newblock

\bibitem{geant4}
S.~Agostinelli et al., {\em {GEANT4: A Simulation toolkit}\/},
\href{http://dx.doi.org/10.1016/S0168-9002(03)01368-8}{Nucl. Instrum. Meth.
  {\bf A506} (2003)  250}.

\bibitem{Abe:2010aa}
{ILD Concept Group - Linear Collider Collaboration}, {\em {The International
  Large Detector: Letter of Intent}\/},
\href{http://arxiv.org/abs/1006.3396}{{\tt arXiv:1006.3396 [hep-ex]}}.

\bibitem{Behnke:2013lya}
T.~Behnke, J.~E. Brau, P.~N. Burrows, J.~Fuster, M.~Peskin, et al., {\em {The
  International Linear Collider Technical Design Report - Volume 4:
  Detectors}\/},
\href{http://arxiv.org/abs/1306.6329}{{\tt arXiv:1306.6329 [physics.ins-det]}}.

\bibitem{CEPC-SPPCStudyGroup:2015csa}
{CEPC-SPPC Study Group, } {CEPC-SPPC Preliminary Conceptual Design Report. 1.
  Physics and Detector} (2015)  .
\url{http://cepc.ihep.ac.cn/preCDR/volume.html}.

\bibitem{PatrickALEPHEnergyFlow}
P.~Janot, {\em {Particle Flow Event Reconstruction from LEP to LHC}\/},
\newblock Presented at Excellence in Detectors and Instrumentation Technologies
  workshop, CERN (2011).
\newblock \url{https://indico.cern.ch/event/96989/contribution/15}.

\bibitem{ALEPHEnergyFlow}
M.~Minard, {\em {Jet energy measurement with the ALEPH detector at LEP2}\/},
\newblock Presented at CALOR2002 Conference, Pasadena, California, USA (2002).
\newblock
\url{http://inspirehep.net/record/608013}.
\newblock

\bibitem{PandoraPFA}
M.~Thomson, {\em {Particle Flow Calorimetry and the PandoraPFA Algorithm}\/},
  \href{http://dx.doi.org/10.1016/j.nima.2009.09.009}{Nucl. Instrum. Meth. {\bf
  A611} (2009)  25},
\href{http://arxiv.org/abs/0907.3577}{{\tt arXiv:0907.3577 [physics.ins-det]}}.

\bibitem{ArborPFA}
M.~Ruan and H.~Videau, {\em {Arbor, a new approach of the Particle Flow
  Algorithm}\/},
\href{http://arxiv.org/abs/1403.4784}{{\tt arXiv:1403.4784 [physics.ins-det]}}.

\bibitem{CMS:2009nxa}
{CMS} Collaboration, {\em {Particle-Flow Event Reconstruction in CMS and
  Performance for Jets, Taus, and MET}\/},  CMS-PAS-PFT-09-001 (2009)  .
\url{http://cds.cern.ch/record/1194487}.

\bibitem{Beaudette:2014cea}
F.~Beaudette, {\em {The CMS Particle Flow Algorithm}\/},  pp.~295--304.
\newblock {Proceedings, International Conference on Calorimetry for the High
  Energy Frontier (CHEF 2013): Paris, France, April 22-25, 2013}.
\newblock
\href{http://arxiv.org/abs/1401.8155}{{\tt arXiv:1401.8155 [hep-ex]}}.
\newblock

\bibitem{Ruan:2018yrh}
M.~Ruan et al., {\em {Reconstruction of physics objects at the Circular
  Electron Positron Collider with Arbor}\/},
\href{http://dx.doi.org/10.1140/epjc/s10052-018-5876-z}{Eur. Phys. J. {\bf C78}
  (2018)  426}.

\bibitem{Zhao:2018jiq}
H.~Zhao, Y.-F. Zhu, C.-D. Fu, D.~Yu, and M.-Q. Ruan, {\em {The Higgs Signatures
  at the CEPC CDR Baseline}\/},
\href{http://arxiv.org/abs/1806.04992}{{\tt arXiv:1806.04992 [hep-ex]}}.

\bibitem{Yu:2017mpx}
D.~Yu, M.~Ruan, V.~Boudry, and H.~Videau, {\em {Lepton identification at
  particle flow oriented detector for the future $e^{+}e^{-}$ Higgs
  factories}\/},  \href{http://dx.doi.org/10.1140/epjc/s10052-017-5146-5}{Eur.
  Phys. J. {\bf C77} (2017)  591},
\href{http://arxiv.org/abs/1701.07542}{{\tt arXiv:1701.07542
  [physics.ins-det]}}.

\bibitem{Catani:1991hj}
S.~Catani, Y.~L. Dokshitzer, M.~Olsson, G.~Turnock, and B.~Webber, {\em {New
  clustering algorithm for multi-jet cross-sections in $e^+ e^-$
  annihilation}\/},
\href{http://dx.doi.org/10.1016/0370-2693(91)90196-W}{Phys. Lett. {\bf B269}
  (1991)  432}.

\bibitem{Aad:2012ag}
{ATLAS} Collaboration, {\em {Jet energy resolution in proton-proton collisions
  at $\sqrt{s}=7$ TeV recorded in 2010 with the ATLAS detector}\/},
  \href{http://dx.doi.org/10.1140/epjc/s10052-013-2306-0}{Eur. Phys. J. {\bf
  C73} (2013)  2306},
\href{http://arxiv.org/abs/1210.6210}{{\tt arXiv:1210.6210 [hep-ex]}}.

\bibitem{Khachatryan:2016kdb}
{CMS} Collaboration, {\em {Jet energy scale and resolution in the CMS
  experiment in pp collisions at 8 TeV}\/},
  \href{http://dx.doi.org/10.1088/1748-0221/12/02/P02014}{JINST {\bf 12} (2017)
   P02014},
\href{http://arxiv.org/abs/1607.03663}{{\tt arXiv:1607.03663 [hep-ex]}}.

\bibitem{LCFIPlus}
T.~Tanabe and T.~Suehara, {\em {LCFIPlus}\/},
\newblock Presented at ILD workshop at Kyushu University (2012).
\newblock
  \url{https://agenda.linearcollider.org/event/5496/session/1/contribution/16}.

\bibitem{LHCCrossSectionPaper2011}
{LHC Higgs Cross Section Working Group}, {\em Handbook of LHC Higgs Cross
  Sections: 1. Inclusive Observables\/},
  \href{http://arxiv.org/abs/1101.0593}{{\tt arXiv:1101.0593 [hep-ph]}}.

\bibitem{LHCCrossSectionPaper2012}
{LHC Higgs Cross Section Working Group}, {\em Handbook of LHC Higgs Cross
  Sections: 2. Differential Distributions\/},
  \href{http://arxiv.org/abs/1201.3084}{{\tt arXiv:1201.3084 [hep-ph]}}.

\bibitem{Heinemeyer:2013tqa}
{LHC Higgs Cross Section Working Group}, {\em {Handbook of LHC Higgs Cross
  Sections: 3. Higgs Properties}\/},
\href{http://arxiv.org/abs/1307.1347}{{\tt arXiv:1307.1347 [hep-ph]}}.

\bibitem{Kilian:2007gr}
W.~Kilian, T.~Ohl, and J.~Reuter, {\em {WHIZARD: Simulating Multi-Particle
  Processes at LHC and ILC}\/},
  \href{http://dx.doi.org/10.1140/epjc/s10052-011-1742-y}{Eur. Phys. J. {\bf
  C71} (2011)  1742},
\href{http://arxiv.org/abs/0708.4233}{{\tt arXiv:0708.4233 [hep-ph]}}.

\bibitem{CarloniCalame:2003yt}
C.~M. Carloni~Calame, G.~Montagna, O.~Nicrosini, and F.~Piccinini, {\em {The
  BABAYAGA event generator}\/},
  \href{http://dx.doi.org/10.1016/j.nuclphysbps.2004.02.008}{Nucl. Phys. Proc.
  Suppl. {\bf 131} (2004)  48--55},
  \href{http://arxiv.org/abs/hep-ph/0312014}{{\tt arXiv:hep-ph/0312014
  [hep-ph]}}.
[,48(2003)].

\bibitem{ilcsoft}
S.~Aplin, J.~Engels, and F.~Gaede, {\em {A production system for massive data
  processing in ILCSoft}\/}, .
\url{http://inspirehep.net/record/889841}.

\bibitem{Asner:2013psa}
D.~Asner, T.~Barklow, C.~Calancha, K.~Fujii, N.~Graf, et al., {\em {ILC Higgs
  White Paper}\/},
\href{http://arxiv.org/abs/1310.0763}{{\tt arXiv:1310.0763 [hep-ph]}}.

\bibitem{Haddad:2014fma}
Y.~Haddad, {\em {Feasibility of a minimum bias analysis of $e^+e^-\to ZH \to
  q\bar{q}+X$ at a 250 GeV ILC}\/},
\href{http://arxiv.org/abs/1404.3164}{{\tt arXiv:1404.3164 [hep-ph]}}.

\bibitem{Oreglia:1980cs}
M.~Oreglia, {\em {A Study of the Reactions $\psi^\prime \to \gamma
  \gamma\psi$}\/},  SLAC-R-0236 (1980)  .
\url{http://www.slac.stanford.edu/cgi-wrap/getdoc/slac-r-23\ 6.pdf}.

\bibitem{Aaboud:2018zhk}
{ATLAS} Collaboration, {\em {Observation of $H \rightarrow b\bar{b}$ decays and
  $VH$ production with the ATLAS detector}\/},  Submitted to: Phys. Lett. B
  (2018)  ,
\href{http://arxiv.org/abs/1808.08238}{{\tt arXiv:1808.08238 [hep-ex]}}.

\bibitem{Sirunyan:2018kst}
{CMS} Collaboration, {\em {Observation of Higgs boson decay to bottom
  quarks}\/},  Submitted to: Phys. Rev. Lett. (2018)  ,
\href{http://arxiv.org/abs/1808.08242}{{\tt arXiv:1808.08242 [hep-ex]}}.

\bibitem{Shrock:1982kd}
R.~E. Shrock and M.~Suzuki, {\em {Invisible Decays of Higgs Bosons}\/},
\href{http://dx.doi.org/10.1016/0370-2693(82)91247-3}{Phys. Lett. {\bf 110B}
  (1982)  250}.

\bibitem{Griest:1987qv}
K.~Griest and H.~E. Haber, {\em {Invisible Decays of Higgs Bosons in
  Supersymmetric Models}\/},
\href{http://dx.doi.org/10.1103/PhysRevD.37.719}{Phys. Rev. {\bf D37} (1988)
  719}.

\bibitem{Englert:2011yb}
C.~Englert, T.~Plehn, D.~Zerwas, and P.~M. Zerwas, {\em {Exploring the Higgs
  portal}\/},  \href{http://dx.doi.org/10.1016/j.physletb.2011.08.002}{Phys.
  Lett. {\bf B703} (2011)  298--305},
\href{http://arxiv.org/abs/1106.3097}{{\tt arXiv:1106.3097 [hep-ph]}}.

\bibitem{Bonilla:2015uwa}
C.~Bonilla, J.~W.~F. Valle, and J.~C. Romão, {\em {Neutrino mass and invisible
  Higgs decays at the LHC}\/},
  \href{http://dx.doi.org/10.1103/PhysRevD.91.113015}{Phys. Rev. {\bf D91}
  (2015)  113015},
\href{http://arxiv.org/abs/1502.01649}{{\tt arXiv:1502.01649 [hep-ph]}}.

\bibitem{Schael:2013ita}
{ALEPH, DELPHI, L3, OPAL, LEP Electroweak} Collaboration, {\em {Electroweak
  Measurements in Electron-Positron Collisions at W-Boson-Pair Energies at
  LEP}\/},  \href{http://dx.doi.org/10.1016/j.physrep.2013.07.004}{Phys. Rept.
  {\bf 532} (2013)  119},
\href{http://arxiv.org/abs/1302.3415}{{\tt arXiv:1302.3415 [hep-ex]}}.

\bibitem{Dawson:2013bba}
S.~Dawson, A.~Gritsan, H.~Logan, J.~Qian, C.~Tully, et al., {\em {Working Group
  Report: Higgs Boson}\/},
\href{http://arxiv.org/abs/1310.8361}{{\tt arXiv:1310.8361 [hep-ex]}}.

\bibitem{deFlorian:2016spz}
{LHC Higgs Cross Section Working Group}, {\em {Handbook of LHC Higgs Cross
  Sections: 4. Deciphering the Nature of the Higgs Sector}\/},
\href{http://arxiv.org/abs/1610.07922}{{\tt arXiv:1610.07922 [hep-ph]}}.

\bibitem{Han:2013kya}
T.~Han, Z.~Liu, and J.~Sayre, {\em {Potential Precision on Higgs Couplings and
  Total Width at the ILC}\/},
  \href{http://dx.doi.org/10.1103/PhysRevD.89.113006}{Phys. Rev. {\bf D89}
  (2014)  113006},
\href{http://arxiv.org/abs/1311.7155}{{\tt arXiv:1311.7155 [hep-ph]}}.

\bibitem{Peskin:2013xra}
M.~E. Peskin, {\em {Estimation of LHC and ILC Capabilities for Precision Higgs
  Boson Coupling Measurements}\/},
\href{http://arxiv.org/abs/1312.4974}{{\tt arXiv:1312.4974 [hep-ph]}}.

\bibitem{Giardino:2013bma}
P.~P. Giardino, K.~Kannike, I.~Masina, M.~Raidal, and A.~Strumia, {\em {The
  universal Higgs fit}\/},
  \href{http://dx.doi.org/10.1007/JHEP05(2014)046}{JHEP {\bf 05} (2014)  046},
\href{http://arxiv.org/abs/1303.3570}{{\tt arXiv:1303.3570 [hep-ph]}}.

\bibitem{Belanger:2013xza}
G.~Belanger, B.~Dumont, U.~Ellwanger, J.~F. Gunion, and S.~Kraml, {\em {Global
  fit to Higgs signal strengths and couplings and implications for extended
  Higgs sectors}\/},  \href{http://dx.doi.org/10.1103/PhysRevD.88.075008}{Phys.
  Rev. {\bf D88} (2013)  075008},
\href{http://arxiv.org/abs/1306.2941}{{\tt arXiv:1306.2941 [hep-ph]}}.

\bibitem{Bechtle:2014ewa}
P.~Bechtle, S.~Heinemeyer, O.~Stal, T.~Stefaniak, and G.~Weiglein, {\em
  {Probing the Standard Model with Higgs signal rates from the Tevatron, the
  LHC and a future ILC}\/},
  \href{http://dx.doi.org/10.1007/JHEP11(2014)039}{JHEP {\bf 11} (2014)  039},
\href{http://arxiv.org/abs/1403.1582}{{\tt arXiv:1403.1582 [hep-ph]}}.

\bibitem{Cheung:2014noa}
K.~Cheung, J.~S. Lee, and P.-Y. Tseng, {\em {Higgs precision analysis updates
  2014}\/},  \href{http://dx.doi.org/10.1103/PhysRevD.90.095009}{Phys. Rev.
  {\bf D90} (2014)  095009},
\href{http://arxiv.org/abs/1407.8236}{{\tt arXiv:1407.8236 [hep-ph]}}.

\bibitem{Fichet:2015xla}
S.~Fichet and G.~Moreau, {\em {Anatomy of the Higgs fits: a first guide to
  statistical treatments of the theoretical uncertainties}\/},
  \href{http://dx.doi.org/10.1016/j.nuclphysb.2016.02.019}{Nucl. Phys. {\bf
  B905} (2016)  391--446},
\href{http://arxiv.org/abs/1509.00472}{{\tt arXiv:1509.00472 [hep-ph]}}.

\bibitem{Lafaye:2017kgf}
R.~Lafaye, T.~Plehn, M.~Rauch, and D.~Zerwas, {\em {Higgs factories:
  Higgsstrahlung versus $W$ fusion}\/},
  \href{http://dx.doi.org/10.1103/PhysRevD.96.075044}{Phys. Rev. {\bf D96}
  (2017) no.~7, 075044},
\href{http://arxiv.org/abs/1706.02174}{{\tt arXiv:1706.02174 [hep-ph]}}.

\bibitem{Barger:2003rs}
V.~Barger, T.~Han, P.~Langacker, B.~McElrath, and P.~Zerwas, {\em {Effects of
  genuine dimension-six Higgs operators}\/},
  \href{http://dx.doi.org/10.1103/PhysRevD.67.115001}{Phys. Rev. {\bf D67}
  (2003)  115001},
\href{http://arxiv.org/abs/hep-ph/0301097}{{\tt arXiv:hep-ph/0301097
  [hep-ph]}}.

\bibitem{Corbett:2012ja}
T.~Corbett, O.~J.~P. Eboli, J.~Gonzalez-Fraile, and M.~C. Gonzalez-Garcia, {\em
  {Robust Determination of the Higgs Couplings: Power to the Data}\/},
  \href{http://dx.doi.org/10.1103/PhysRevD.87.015022}{Phys. Rev. {\bf D87}
  (2013)  015022},
\href{http://arxiv.org/abs/1211.4580}{{\tt arXiv:1211.4580 [hep-ph]}}.

\bibitem{Elias-Miro:2013mua}
J.~Elias-Miro, J.~R. Espinosa, E.~Masso, and A.~Pomarol, {\em {Higgs windows to
  new physics through d=6 operators: constraints and one-loop anomalous
  dimensions}\/},  \href{http://dx.doi.org/10.1007/JHEP11(2013)066}{JHEP {\bf
  11} (2013)  066},
\href{http://arxiv.org/abs/1308.1879}{{\tt arXiv:1308.1879 [hep-ph]}}.

\bibitem{Pomarol:2013zra}
A.~Pomarol and F.~Riva, {\em {Towards the Ultimate SM Fit to Close in on Higgs
  Physics}\/},  \href{http://dx.doi.org/10.1007/JHEP01(2014)151}{JHEP {\bf 01}
  (2014)  151},
\href{http://arxiv.org/abs/1308.2803}{{\tt arXiv:1308.2803 [hep-ph]}}.

\bibitem{Amar:2014fpa}
G.~Amar, S.~Banerjee, S.~von Buddenbrock, A.~S. Cornell, T.~Mandal, B.~Mellado,
  and B.~Mukhopadhyaya, {\em {Exploration of the tensor structure of the Higgs
  boson coupling to weak bosons in $e^{+} e^{-}$ collisions}\/},
  \href{http://dx.doi.org/10.1007/JHEP02(2015)128}{JHEP {\bf 02} (2015)  128},
\href{http://arxiv.org/abs/1405.3957}{{\tt arXiv:1405.3957 [hep-ph]}}.

\bibitem{Ellis:2014jta}
J.~Ellis, V.~Sanz, and T.~You, {\em {The Effective Standard Model after LHC Run
  I}\/},  \href{http://dx.doi.org/10.1007/JHEP03(2015)157}{JHEP {\bf 03} (2015)
   157},
\href{http://arxiv.org/abs/1410.7703}{{\tt arXiv:1410.7703 [hep-ph]}}.

\bibitem{Falkowski:2015fla}
A.~Falkowski, {\em {Effective field theory approach to LHC Higgs data}\/},
  \href{http://dx.doi.org/10.1007/s12043-016-1251-5}{Pramana {\bf 87} (2016)
  39},
\href{http://arxiv.org/abs/1505.00046}{{\tt arXiv:1505.00046 [hep-ph]}}.

\bibitem{Butter:2016cvz}
A.~Butter, O.~J.~P. \'Eboli, J.~Gonzalez-Fraile, M.~C. Gonzalez-Garcia,
  T.~Plehn, and M.~Rauch, {\em {The Gauge-Higgs Legacy of the LHC Run I}\/},
  \href{http://dx.doi.org/10.1007/JHEP07(2016)152}{JHEP {\bf 07} (2016)  152},
\href{http://arxiv.org/abs/1604.03105}{{\tt arXiv:1604.03105 [hep-ph]}}.

\bibitem{Craig:2014una}
N.~Craig, M.~Farina, M.~McCullough, and M.~Perelstein, {\em {Precision
  Higgsstrahlung as a Probe of New Physics}\/},
  \href{http://dx.doi.org/10.1007/JHEP03(2015)146}{JHEP {\bf 03} (2015)  146},
\href{http://arxiv.org/abs/1411.0676}{{\tt arXiv:1411.0676 [hep-ph]}}.

\bibitem{Beneke:2014sba}
M.~Beneke, D.~Boito, and Y.-M. Wang, {\em {Anomalous Higgs couplings in angular
  asymmetries of $H \to Z\ell^{+} \ell^{-}$ and $e^+ e^- \to HZ$}\/},
  \href{http://dx.doi.org/10.1007/JHEP11(2014)028}{JHEP {\bf 11} (2014)  028},
\href{http://arxiv.org/abs/1406.1361}{{\tt arXiv:1406.1361 [hep-ph]}}.

\bibitem{Craig:2015wwr}
N.~Craig, J.~Gu, Z.~Liu, and K.~Wang, {\em {Beyond Higgs Couplings: Probing the
  Higgs with Angular Observables at Future e$^{+}$ e$^{-}$ Colliders}\/},
  \href{http://dx.doi.org/10.1007/JHEP03(2016)050}{JHEP {\bf 03} (2016)  050},
\href{http://arxiv.org/abs/1512.06877}{{\tt arXiv:1512.06877 [hep-ph]}}.

\bibitem{Ellis:2015sca}
J.~Ellis and T.~You, {\em {Sensitivities of Prospective Future e+e- Colliders
  to Decoupled New Physics}\/},
  \href{http://dx.doi.org/10.1007/JHEP03(2016)089}{JHEP {\bf 03} (2016)  089},
\href{http://arxiv.org/abs/1510.04561}{{\tt arXiv:1510.04561 [hep-ph]}}.

\bibitem{Ge:2016zro}
S.-F. Ge, H.-J. He, and R.-Q. Xiao, {\em {Probing new physics scales from Higgs
  and electroweak observables at $e^+e^-$ Higgs factory}\/},
  \href{http://dx.doi.org/10.1007/JHEP10(2016)007}{JHEP {\bf 10} (2016)  007},
\href{http://arxiv.org/abs/1603.03385}{{\tt arXiv:1603.03385 [hep-ph]}}.

\bibitem{Ellis:2017kfi}
J.~Ellis, P.~Roloff, V.~Sanz, and T.~You, {\em {Dimension-6 Operator Analysis
  of the CLIC Sensitivity to New Physics}\/},
\href{http://arxiv.org/abs/1701.04804}{{\tt arXiv:1701.04804 [hep-ph]}}.

\bibitem{Durieux:2017rsg}
G.~Durieux, C.~Grojean, J.~Gu, and K.~Wang, {\em {The leptonic future of the
  Higgs}\/},  \href{http://dx.doi.org/10.1007/JHEP09(2017)014}{JHEP {\bf 09}
  (2017)  014},
\href{http://arxiv.org/abs/1704.02333}{{\tt arXiv:1704.02333 [hep-ph]}}.

\bibitem{Barklow:2017suo}
T.~Barklow, K.~Fujii, S.~Jung, R.~Karl, J.~List, T.~Ogawa, M.~E. Peskin, and
  J.~Tian, {\em {Improved Formalism for Precision Higgs Coupling Fits}\/},
  \href{http://dx.doi.org/10.1103/PhysRevD.97.053003}{Phys. Rev. {\bf D97}
  (2018) no.~5, 053003},
\href{http://arxiv.org/abs/1708.08912}{{\tt arXiv:1708.08912 [hep-ph]}}.

\bibitem{Barklow:2017awn}
T.~Barklow, K.~Fujii, S.~Jung, M.~E. Peskin, and J.~Tian, {\em
  {Model-Independent Determination of the Triple Higgs Coupling at e+e-
  Colliders}\/},  \href{http://dx.doi.org/10.1103/PhysRevD.97.053004}{Phys.
  Rev. {\bf D97} (2018) no.~5, 053004},
\href{http://arxiv.org/abs/1708.09079}{{\tt arXiv:1708.09079 [hep-ph]}}.

\bibitem{DiVita:2017vrr}
S.~Di~Vita, G.~Durieux, C.~Grojean, J.~Gu, Z.~Liu, G.~Panico, M.~Riembau, and
  T.~Vantalon, {\em {A global view on the Higgs self-coupling at lepton
  colliders}\/},  \href{http://dx.doi.org/10.1007/JHEP02(2018)178}{JHEP {\bf
  02} (2018)  178},
\href{http://arxiv.org/abs/1711.03978}{{\tt arXiv:1711.03978 [hep-ph]}}.

\bibitem{Chiu:2017yrx}
W.~H. Chiu, S.~C. Leung, T.~Liu, K.-F. Lyu, and L.-T. Wang, {\em {Probing 6D
  operators at future e$^{-}$e$^{+}$ colliders}\/},
  \href{http://dx.doi.org/10.1007/JHEP05(2018)081}{JHEP {\bf 05} (2018)  081},
\href{http://arxiv.org/abs/1711.04046}{{\tt arXiv:1711.04046 [hep-ph]}}.

\bibitem{Ellis:2018gqa}
J.~Ellis, C.~W. Murphy, V.~Sanz, and T.~You, {\em {Updated Global SMEFT Fit to
  Higgs, Diboson and Electroweak Data}\/},
  \href{http://dx.doi.org/10.1007/JHEP06(2018)146}{JHEP {\bf 06} (2018)  146},
\href{http://arxiv.org/abs/1803.03252}{{\tt arXiv:1803.03252 [hep-ph]}}.

\bibitem{Durieux:2018ggn}
G.~Durieux, J.~Gu, E.~Vryonidou, and C.~Zhang, {\em {Probing top-quark
  couplings indirectly at Higgs factories}\/},
  \href{http://dx.doi.org/10.1088/1674-1137/42/12/123107}{Chin. Phys. {\bf C42}
  (2018) no.~12, 123107},
\href{http://arxiv.org/abs/1809.03520}{{\tt arXiv:1809.03520 [hep-ph]}}.

\bibitem{Gao:2016jcm}
J.~Gao, {\em {Probing light-quark Yukawa couplings via hadronic event shapes at
  lepton colliders}\/},  \href{http://dx.doi.org/10.1007/JHEP01(2018)038}{JHEP
  {\bf 01} (2018)  038},
\href{http://arxiv.org/abs/1608.01746}{{\tt arXiv:1608.01746 [hep-ph]}}.

\bibitem{LHCHiggsCrossSectionWorkingGroup:2012nn}
{LHC Higgs Cross Section Working Group}, {\em {LHC HXSWG interim
  recommendations to explore the coupling structure of a Higgs-like
  particle}\/},
\href{http://arxiv.org/abs/1209.0040}{{\tt arXiv:1209.0040 [hep-ph]}}.

\bibitem{Sirunyan:2018hoz}
{CMS} Collaboration, {\em {Observation of $\mathrm{t\overline{t}}$H
  production}\/},  \href{http://dx.doi.org/10.1103/PhysRevLett.120.231801,
  10.1130/PhysRevLett.120.231801}{Phys. Rev. Lett. {\bf 120} (2018)  231801},
\href{http://arxiv.org/abs/1804.02610}{{\tt arXiv:1804.02610 [hep-ex]}}.

\bibitem{Aaboud:2018urx}
{ATLAS} Collaboration, {\em {Observation of Higgs boson production in
  association with a top quark pair at the LHC with the ATLAS detector}\/},
  \href{http://dx.doi.org/10.1016/j.physletb.2018.07.035}{Phys. Lett. {\bf
  B784} (2018)  173--191},
\href{http://arxiv.org/abs/1806.00425}{{\tt arXiv:1806.00425 [hep-ex]}}.

\bibitem{Banfi:2013yoa}
A.~Banfi, A.~Martin, and V.~Sanz, {\em {Probing top-partners in Higgs+jets}\/},
   \href{http://dx.doi.org/10.1007/JHEP08(2014)053}{JHEP {\bf 1408} (2014)
  053},
\href{http://arxiv.org/abs/1308.4771}{{\tt arXiv:1308.4771 [hep-ph]}}.

\bibitem{Azatov:2013xha}
A.~Azatov and A.~Paul, {\em {Probing Higgs couplings with high $p_T$ Higgs
  production}\/},  \href{http://dx.doi.org/10.1007/JHEP01(2014)014}{JHEP {\bf
  1401} (2014)  014},
\href{http://arxiv.org/abs/1309.5273}{{\tt arXiv:1309.5273 [hep-ph]}}.

\bibitem{Grojean:2013nya}
C.~Grojean, E.~Salvioni, M.~Schlaffer, and A.~Weiler, {\em {Very boosted Higgs
  in gluon fusion}\/},  \href{http://dx.doi.org/10.1007/JHEP05(2014)022}{JHEP
  {\bf 1405} (2014)  022},
\href{http://arxiv.org/abs/1312.3317}{{\tt arXiv:1312.3317 [hep-ph]}}.

\bibitem{Buschmann:2014twa}
M.~Buschmann, C.~Englert, D.~Goncalves, T.~Plehn, and M.~Spannowsky, {\em
  {Resolving the Higgs-Gluon Coupling with Jets}\/},
  \href{http://dx.doi.org/10.1103/PhysRevD.90.013010}{Phys. Rev. {\bf D90}
  (2014)  013010},
\href{http://arxiv.org/abs/1405.7651}{{\tt arXiv:1405.7651 [hep-ph]}}.

\bibitem{Ellis:2014dva}
J.~Ellis, V.~Sanz, and T.~You, {\em {Complete Higgs Sector Constraints on
  Dimension-6 Operators}\/},
  \href{http://dx.doi.org/10.1007/JHEP07(2014)036}{JHEP {\bf 1407} (2014)
  036},
\href{http://arxiv.org/abs/1404.3667}{{\tt arXiv:1404.3667 [hep-ph]}}.

\bibitem{Klute:2013cx}
M.~Klute, R.~Lafaye, T.~Plehn, M.~Rauch, and D.~Zerwas, {\em {Measuring Higgs
  Couplings at a Linear Collider}\/},
  \href{http://dx.doi.org/10.1209/0295-5075/101/51001}{Europhys. Lett. {\bf
  101} (2013)  51001},
\href{http://arxiv.org/abs/1301.1322}{{\tt arXiv:1301.1322 [hep-ph]}}.

\bibitem{Fujii:2017vwa}
K.~Fujii et al., {\em {Physics Case for the 250 GeV Stage of the International
  Linear Collider}\/},
\href{http://arxiv.org/abs/1710.07621}{{\tt arXiv:1710.07621 [hep-ex]}}.

\bibitem{Sun:2016bel}
Q.-F. Sun, F.~Feng, Y.~Jia, and W.-L. Sang, {\em {Mixed electroweak-QCD
  corrections to e+e-->HZ at Higgs factories}\/},
  \href{http://dx.doi.org/10.1103/PhysRevD.96.051301}{Phys. Rev. {\bf D96}
  (2017) no.~5, 051301},
\href{http://arxiv.org/abs/1609.03995}{{\tt arXiv:1609.03995 [hep-ph]}}.

\bibitem{Gong:2016jys}
Y.~Gong, Z.~Li, X.~Xu, L.~L. Yang, and X.~Zhao, {\em {Mixed QCD-EW corrections
  for Higgs boson production at $e^+e^-$ colliders}\/},
  \href{http://dx.doi.org/10.1103/PhysRevD.95.093003}{Phys. Rev. {\bf D95}
  (2017) no.~9, 093003},
\href{http://arxiv.org/abs/1609.03955}{{\tt arXiv:1609.03955 [hep-ph]}}.

\bibitem{Lepage:2014fla}
G.~P. Lepage, P.~B. Mackenzie, and M.~E. Peskin, {\em {Expected Precision of
  Higgs Boson Partial Widths within the Standard Model}\/},
\href{http://arxiv.org/abs/1404.0319}{{\tt arXiv:1404.0319 [hep-ph]}}.

\bibitem{Denner:2011mq}
A.~Denner, S.~Heinemeyer, I.~Puljak, D.~Rebuzzi, and M.~Spira, {\em {Standard
  Model Higgs-Boson Branching Ratios with Uncertainties}\/},
  \href{http://dx.doi.org/10.1140/epjc/s10052-011-1753-8}{Eur. Phys. J. {\bf
  C71} (2011)  1753},
\href{http://arxiv.org/abs/1107.5909}{{\tt arXiv:1107.5909 [hep-ph]}}.

\bibitem{Almeida:2013jfa}
L.~G. Almeida, S.~J. Lee, S.~Pokorski, and J.~D. Wells, {\em {Study of the
  standard model Higgs boson partial widths and branching fractions}\/},
  \href{http://dx.doi.org/10.1103/PhysRevD.89.033006}{Phys. Rev. {\bf D89}
  (2014)  033006},
\href{http://arxiv.org/abs/1311.6721}{{\tt arXiv:1311.6721 [hep-ph]}}.

\bibitem{Bernaciak:2014pna}
C.~Bernaciak, T.~Plehn, P.~Schichtel, and J.~Tattersall, {\em {Spying an
  invisible Higgs boson}\/},
  \href{http://dx.doi.org/10.1103/PhysRevD.91.035024}{Phys. Rev. {\bf D91}
  (2015)  035024},
\href{http://arxiv.org/abs/1411.7699}{{\tt arXiv:1411.7699 [hep-ph]}}.

\bibitem{Barr:1990vd}
S.~M. Barr and A.~Zee, {\em {Electric Dipole Moment of the Electron and of the
  Neutron}\/},  \href{http://dx.doi.org/10.1103/PhysRevLett.65.21}{Phys. Rev.
  Lett. {\bf 65} (1990)  21--24}.
[Erratum: Phys. Rev. Lett.65,2920(1990)].

\bibitem{Fan:2013qn}
J.~Fan and M.~Reece, {\em {Probing Charged Matter Through Higgs Diphoton Decay,
  Gamma Ray Lines, and EDMs}\/},
  \href{http://dx.doi.org/10.1007/JHEP06(2013)004}{JHEP {\bf 06} (2013)  004},
\href{http://arxiv.org/abs/1301.2597}{{\tt arXiv:1301.2597 [hep-ph]}}.

\bibitem{Baron:2013eja}
{ACME} Collaboration, {\em {Order of Magnitude Smaller Limit on the Electric
  Dipole Moment of the Electron}\/},
  \href{http://dx.doi.org/10.1126/science.1248213}{Science {\bf 343} (2014)
  269--272},
\href{http://arxiv.org/abs/1310.7534}{{\tt arXiv:1310.7534 [physics.atom-ph]}}.

\bibitem{Chien:2015xha}
Y.~T. Chien, V.~Cirigliano, W.~Dekens, J.~de~Vries, and E.~Mereghetti, {\em
  {Direct and indirect constraints on CP-violating Higgs-quark and Higgs-gluon
  interactions}\/},  \href{http://dx.doi.org/10.1007/JHEP02(2016)011}{JHEP {\bf
  02} (2016)  011},
\href{http://arxiv.org/abs/1510.00725}{{\tt arXiv:1510.00725 [hep-ph]}}.

\bibitem{Harnik:2013aja}
R.~Harnik, A.~Martin, T.~Okui, R.~Primulando, and F.~Yu, {\em {Measuring CP
  violation in $h \to \tau^+ \tau^-$ at colliders}\/},
  \href{http://dx.doi.org/10.1103/PhysRevD.88.076009}{Phys. Rev. {\bf D88}
  (2013)  076009},
\href{http://arxiv.org/abs/1308.1094}{{\tt arXiv:1308.1094 [hep-ph]}}.

\bibitem{Hagiwara:1993ck}
K.~Hagiwara, S.~Ishihara, R.~Szalapski, and D.~Zeppenfeld, {\em {Low-energy
  effects of new interactions in the electroweak boson sector}\/},
\href{http://dx.doi.org/10.1103/PhysRevD.48.2182}{Phys. Rev. {\bf D48} (1993)
  2182--2203}.

\bibitem{Gounaris:1996rz}
G.~Gounaris et al., {\em {Triple gauge boson couplings}\/},
\href{http://arxiv.org/abs/hep-ph/9601233}{{\tt arXiv:hep-ph/9601233
  [hep-ph]}}.

\bibitem{Bian:2015zha}
L.~Bian, J.~Shu, and Y.~Zhang, {\em {Prospects for Triple Gauge Coupling
  Measurements at Future Lepton Colliders and the 14 TeV LHC}\/},
  \href{http://dx.doi.org/10.1007/JHEP09(2015)206}{JHEP {\bf 09} (2015)  206},
\href{http://arxiv.org/abs/1507.02238}{{\tt arXiv:1507.02238 [hep-ph]}}.

\bibitem{Falkowski:2001958}
A.~Falkowski, {\em {Higgs Basis: Proposal for an EFT basis choice for LHC
  HXSWG}\/}, LHCHXSWG-INT-2015-001 (March, 2015)  .
  \url{https://cds.cern.ch/record/2001958}.

\bibitem{ATL-PHYS-PUB-2013-014}
{ATLAS} Collaboration, {\em {Projections for measurements of Higgs boson cross
  sections, branching ratios and coupling parameters with the ATLAS detector at
  a HL-LHC}\/},  ATL-PHYS-PUB-2013-014 (2013)  .
  \url{https://cds.cern.ch/record/1611186}.

\bibitem{ATL-PHYS-PUB-2014-012}
{ATLAS} Collaboration, {\em {HL-LHC projections for signal and background yield
  measurements of the $H\to\gamma\gamma$ when the Higgs boson is produced in
  association with $t$ quarks, $W$ or $Z$ bosons}\/},  ATL-PHYS-PUB-2014-012
  (2014)  . \url{https://cds.cern.ch/record/1741011}.

\bibitem{ATL-PHYS-PUB-2014-006}
{ATLAS} Collaboration, {\em {Update of the prospects for the $H\to Z\gamma$
  search at the High-Luminosity LHC}\/},  ATL-PHYS-PUB-2014-006 (2014)  .
  \url{https://cds.cern.ch/record/1703276}.

\bibitem{ATL-PHYS-PUB-2014-011}
{ATLAS} Collaboration, {\em {Prospects for the study of the Higgs boson in the
  VH(bb) channel at HL-LHC}\/},  ATL-PHYS-PUB-2014-011 (2014)  .
  \url{https://cds.cern.ch/record/1740962}.

\bibitem{ATL-PHYS-PUB-2014-018}
{ATLAS} Collaboration, {\em {Studies of the VBF $H\rightarrow\tau_l\tau_{had}$
  analysis at High Luminosity LHC conditions}\/},  ATL-PHYS-PUB-2014-018 (2014)
   . \url{https://cds.cern.ch/record/1956732}.

\bibitem{Falkowski:2015jaa}
A.~Falkowski, M.~Gonzalez-Alonso, A.~Greljo, and D.~Marzocca, {\em {Global
  constraints on anomalous triple gauge couplings in effective field theory
  approach}\/},  \href{http://dx.doi.org/10.1103/PhysRevLett.116.011801}{Phys.
  Rev. Lett. {\bf 116} (2016)  011801},
\href{http://arxiv.org/abs/1508.00581}{{\tt arXiv:1508.00581 [hep-ph]}}.

\bibitem{Contino:2016jqw}
R.~Contino, A.~Falkowski, F.~Goertz, C.~Grojean, and F.~Riva, {\em {On the
  Validity of the Effective Field Theory Approach to SM Precision Tests}\/},
  \href{http://dx.doi.org/10.1007/JHEP07(2016)144}{JHEP {\bf 07} (2016)  144},
\href{http://arxiv.org/abs/1604.06444}{{\tt arXiv:1604.06444 [hep-ph]}}.

\bibitem{Falkowski:2016cxu}
A.~Falkowski, M.~Gonzalez-Alonso, A.~Greljo, D.~Marzocca, and M.~Son, {\em
  {Anomalous Triple Gauge Couplings in the Effective Field Theory Approach at
  the LHC}\/},  \href{http://dx.doi.org/10.1007/JHEP02(2017)115}{JHEP {\bf 02}
  (2017)  115},
\href{http://arxiv.org/abs/1609.06312}{{\tt arXiv:1609.06312 [hep-ph]}}.

\bibitem{Zhang:2016zsp}
Z.~Zhang, {\em {Time to Go Beyond Triple-Gauge-Boson-Coupling Interpretation of
  $W$ Pair Production}\/},
  \href{http://dx.doi.org/10.1103/PhysRevLett.118.011803}{Phys. Rev. Lett. {\bf
  118} (2017)  011803},
\href{http://arxiv.org/abs/1610.01618}{{\tt arXiv:1610.01618 [hep-ph]}}.

\bibitem{ATL-PHYS-PUB-2015-043}
{ATLAS} Collaboration, {\em {Search for the Standard Model Higgs and $Z$ Boson
  decays to $J/\psi\,\gamma$: HL-LHC projections}\/},  ATL-PHYS-PUB-2015-043
  (2015)  . \url{http://cds.cern.ch/record/2054550}.

\bibitem{Bodwin:2013gca}
G.~T. Bodwin, F.~Petriello, S.~Stoynev, and M.~Velasco, {\em {Higgs boson
  decays to quarkonia and the $H\bar{c}c$ coupling}\/},
  \href{http://dx.doi.org/10.1103/PhysRevD.88.053003}{Phys. Rev. {\bf D88}
  (2013)  053003},
\href{http://arxiv.org/abs/1306.5770}{{\tt arXiv:1306.5770 [hep-ph]}}.

\bibitem{Perez:2015aoa}
G.~Perez, Y.~Soreq, E.~Stamou, and K.~Tobioka, {\em {Constraining the charm
  Yukawa and Higgs-quark coupling universality}\/},
  \href{http://dx.doi.org/10.1103/PhysRevD.92.033016}{Phys. Rev. {\bf D92}
  (2015)  033016},
\href{http://arxiv.org/abs/1503.00290}{{\tt arXiv:1503.00290 [hep-ph]}}.

\bibitem{Brivio:2015fxa}
I.~Brivio, F.~Goertz, and G.~Isidori, {\em {Probing the Charm Quark Yukawa
  Coupling in Higgs+Charm Production}\/},
  \href{http://dx.doi.org/10.1103/PhysRevLett.115.211801}{Phys. Rev. Lett. {\bf
  115} (2015)  211801},
\href{http://arxiv.org/abs/1507.02916}{{\tt arXiv:1507.02916 [hep-ph]}}.

\bibitem{Bishara:2016jga}
F.~Bishara, U.~Haisch, P.~F. Monni, and E.~Re, {\em {Constraining Light-Quark
  Yukawa Couplings from Higgs Distributions}\/},
\href{http://arxiv.org/abs/1606.09253}{{\tt arXiv:1606.09253 [hep-ph]}}.

\bibitem{Carpenter:2016mwd}
L.~M. Carpenter, T.~Han, K.~Hendricks, Z.~Qian, and N.~Zhou, {\em {Higgs Boson
  Decay to Light Jets at the LHC}\/},
  \href{http://dx.doi.org/10.1103/PhysRevD.95.053003}{Phys. Rev. {\bf D95}
  (2017)  053003},
\href{http://arxiv.org/abs/1611.05463}{{\tt arXiv:1611.05463 [hep-ph]}}.

\bibitem{ATL-PHYS-PUB-2017-001}
{ATLAS} Collaboration, {\em {Study of the double Higgs production channel
  $H(\rightarrow b\bar{b})H(\rightarrow \gamma\gamma)$ with the ATLAS
  experiment at the HL-LHC}\/},   ATL-PHYS-PUB-2017-001, CERN, Geneva, Jan,
  2017.
\newblock \url{https://cds.cern.ch/record/2243387}.

\bibitem{Contino:2016spe}
R.~Contino et al., {\em {Physics at a 100 TeV pp collider: Higgs and EW
  symmetry breaking studies}\/},
  \href{http://dx.doi.org/10.23731/CYRM-2017-003.255}{CERN Yellow Report (2017)
   255--440},
\href{http://arxiv.org/abs/1606.09408}{{\tt arXiv:1606.09408 [hep-ph]}}.

\bibitem{McCullough:2013rea}
M.~McCullough, {\em {An Indirect Model-Dependent Probe of the Higgs
  Self-Coupling}\/},  \href{http://dx.doi.org/10.1103/PhysRevD.90.015001}{Phys.
  Rev. {\bf D90} (2014)  015001},
\href{http://arxiv.org/abs/1312.3322}{{\tt arXiv:1312.3322 [hep-ph]}}.

\bibitem{DiVita:2017eyz}
S.~Di~Vita, C.~Grojean, G.~Panico, M.~Riembau, and T.~Vantalon, {\em {A global
  view on the Higgs self-coupling}\/},
  \href{http://dx.doi.org/10.1007/JHEP09(2017)069}{JHEP {\bf 09} (2017)  069},
\href{http://arxiv.org/abs/1704.01953}{{\tt arXiv:1704.01953 [hep-ph]}}.

\bibitem{Azatov:2015oxa}
A.~Azatov, R.~Contino, G.~Panico, and M.~Son, {\em {Effective field theory
  analysis of double Higgs boson production via gluon fusion}\/},
  \href{http://dx.doi.org/10.1103/PhysRevD.92.035001}{Phys. Rev. {\bf D92}
  (2015)  035001},
\href{http://arxiv.org/abs/1502.00539}{{\tt arXiv:1502.00539 [hep-ph]}}.

\bibitem{HiggsTop}
Z.~Liu, I.~Low, and L.-T. Wang, {\em Higgs-Top Interactions at Future Circular
  $e^+e^-$ Colliders\/},  \href{http://arxiv.org/abs/2018.nnnn}{{\tt
  arXiv:2018.nnnn}}.

\bibitem{Li:2015kxc}
G.~Li, H.-R. Wang, and S.-h. Zhu, {\em {Probing CP-violating $h\bar{t}t$
  coupling in $e^{+}e^{-}\rightarrow h \gamma$}\/},
\href{http://arxiv.org/abs/1506.06453}{{\tt arXiv:1506.06453 [hep-ph]}}.

\bibitem{Vryonidou:2018eyv}
E.~Vryonidou and C.~Zhang, {\em {Dimension-six electroweak top-loop effects in
  Higgs production and decay}\/},
\href{http://arxiv.org/abs/1804.09766}{{\tt arXiv:1804.09766 [hep-ph]}}.

\bibitem{Durieux:2018tev}
G.~Durieux, M.~Perello, M.~Vos, and C.~Zhang, {\em {Global and optimal probes
  for the top-quark effective field theory at future lepton colliders}\/},
  \href{http://dx.doi.org/10.1007/JHEP10(2018)168}{JHEP {\bf 10} (2018)  168},
\href{http://arxiv.org/abs/1807.02121}{{\tt arXiv:1807.02121 [hep-ph]}}.

\bibitem{Boselli:2018zxr}
S.~Boselli, R.~Hunter, and A.~Mitov, {\em {Prospects for the determination of
  the top-quark Yukawa coupling at future $e^+e^-$ colliders}\/},
\href{http://arxiv.org/abs/1805.12027}{{\tt arXiv:1805.12027 [hep-ph]}}.

\bibitem{AguilarSaavedra:2008zc}
J.~A. Aguilar-Saavedra, {\em {A Minimal set of top anomalous couplings}\/},
  \href{http://dx.doi.org/10.1016/j.nuclphysb.2008.12.012}{Nucl. Phys. {\bf
  B812} (2009)  181--204},
\href{http://arxiv.org/abs/0811.3842}{{\tt arXiv:0811.3842 [hep-ph]}}.

\bibitem{AguilarSaavedra:2009mx}
J.~A. Aguilar-Saavedra, {\em {A Minimal set of top-Higgs anomalous
  couplings}\/},
  \href{http://dx.doi.org/10.1016/j.nuclphysb.2009.06.022}{Nucl. Phys. {\bf
  B821} (2009)  215--227},
\href{http://arxiv.org/abs/0904.2387}{{\tt arXiv:0904.2387 [hep-ph]}}.

\bibitem{Gao:2010qx}
Y.~Gao, A.~V. Gritsan, Z.~Guo, K.~Melnikov, M.~Schulze, and N.~V. Tran, {\em
  {Spin determination of single-produced resonances at hadron colliders}\/},
  \href{http://dx.doi.org/10.1103/PhysRevD.81.075022}{Phys. Rev. {\bf D81}
  (2010)  075022},
\href{http://arxiv.org/abs/1001.3396}{{\tt arXiv:1001.3396 [hep-ph]}}.

\bibitem{Bolognesi:2012mm}
S.~Bolognesi, Y.~Gao, A.~V. Gritsan, K.~Melnikov, M.~Schulze, N.~V. Tran, and
  A.~Whitbeck, {\em {On the spin and parity of a single-produced resonance at
  the LHC}\/},  \href{http://dx.doi.org/10.1103/PhysRevD.86.095031}{Phys. Rev.
  {\bf D86} (2012)  095031},
\href{http://arxiv.org/abs/1208.4018}{{\tt arXiv:1208.4018 [hep-ph]}}.

\bibitem{Anderson:2013afp}
I.~Anderson et al., {\em {Constraining anomalous HVV interactions at proton and
  lepton colliders}\/},
  \href{http://dx.doi.org/10.1103/PhysRevD.89.035007}{Phys. Rev. {\bf D89}
  (2014)  035007},
\href{http://arxiv.org/abs/1309.4819}{{\tt arXiv:1309.4819 [hep-ph]}}.

\bibitem{Alwall:2007st}
J.~Alwall, P.~Demin, S.~de~Visscher, R.~Frederix, M.~Herquet, F.~Maltoni,
  T.~Plehn, D.~L. Rainwater, and T.~Stelzer, {\em {MadGraph/MadEvent v4: The
  New Web Generation}\/},
  \href{http://dx.doi.org/10.1088/1126-6708/2007/09/028}{JHEP {\bf 09} (2007)
  028},
\href{http://arxiv.org/abs/0706.2334}{{\tt arXiv:0706.2334 [hep-ph]}}.

\bibitem{Curtin:2013fra}
D.~Curtin, R.~Essig, S.~Gori, P.~Jaiswal, A.~Katz, et al., {\em {Exotic decays
  of the 125 GeV Higgs boson}\/},
  \href{http://dx.doi.org/10.1103/PhysRevD.90.075004}{Phys.Rev. {\bf D90}
  (2014)  075004},
\href{http://arxiv.org/abs/1312.4992}{{\tt arXiv:1312.4992 [hep-ph]}}.

\bibitem{Liu:2016zki}
Z.~Liu, L.-T. Wang, and H.~Zhang, {\em {Exotic decays of the 125 GeV Higgs
  boson at future $e^+e^-$ lepton colliders}\/},
\href{http://arxiv.org/abs/1612.09284}{{\tt arXiv:1612.09284 [hep-ph]}}.

\bibitem{Liu:2018wte}
J.~Liu, Z.~Liu, and L.-T. Wang, {\em {Long-lived particles at the LHC: catching
  them in time}\/},
\href{http://arxiv.org/abs/1805.05957}{{\tt arXiv:1805.05957 [hep-ph]}}.

\bibitem{Thamm:2015zwa}
A.~Thamm, R.~Torre, and A.~Wulzer, {\em {Future tests of Higgs compositeness:
  direct vs indirect}\/},
  \href{http://dx.doi.org/10.1007/JHEP07(2015)100}{JHEP {\bf 07} (2015)  100},
\href{http://arxiv.org/abs/1502.01701}{{\tt arXiv:1502.01701 [hep-ph]}}.

\bibitem{Gu:2017ckc}
J.~Gu, H.~Li, Z.~Liu, S.~Su, and W.~Su, {\em {Learning from Higgs Physics at
  Future Higgs Factories}\/},
  \href{http://dx.doi.org/10.1007/JHEP12(2017)153}{JHEP {\bf 12} (2017)  153},
\href{http://arxiv.org/abs/1709.06103}{{\tt arXiv:1709.06103 [hep-ph]}}.

\bibitem{Fan:2014axa}
J.~Fan, M.~Reece, and L.-T. Wang, {\em {Precision Natural SUSY at CEPC, FCC-ee,
  and ILC}\/},
\href{http://arxiv.org/abs/1412.3107}{{\tt arXiv:1412.3107 [hep-ph]}}.

\bibitem{Essig:2017zwe}
R.~Essig, P.~Meade, H.~Ramani, and Y.-M. Zhong, {\em {Higgs-Precision
  Constraints on Colored Naturalness}\/},
  \href{http://dx.doi.org/10.1007/JHEP09(2017)085}{JHEP {\bf 09} (2017)  085},
\href{http://arxiv.org/abs/1707.03399}{{\tt arXiv:1707.03399 [hep-ph]}}.

\bibitem{Chacko:2005pe}
Z.~Chacko, H.-S. Goh, and R.~Harnik, {\em {The Twin Higgs: Natural electroweak
  breaking from mirror symmetry}\/},
  \href{http://dx.doi.org/10.1103/PhysRevLett.96.231802}{Phys. Rev. Lett. {\bf
  96} (2006)  231802},
\href{http://arxiv.org/abs/hep-ph/0506256}{{\tt arXiv:hep-ph/0506256
  [hep-ph]}}.

\bibitem{Burdman:2006tz}
G.~Burdman, Z.~Chacko, H.-S. Goh, and R.~Harnik, {\em {Folded supersymmetry and
  the LEP paradox}\/},
  \href{http://dx.doi.org/10.1088/1126-6708/2007/02/009}{JHEP {\bf 02} (2007)
  009},
\href{http://arxiv.org/abs/hep-ph/0609152}{{\tt arXiv:hep-ph/0609152
  [hep-ph]}}.

\bibitem{Craig:2013xia}
N.~Craig, C.~Englert, and M.~McCullough, {\em {New Probe of Naturalness}\/},
  \href{http://dx.doi.org/10.1103/PhysRevLett.111.121803}{Phys. Rev. Lett. {\bf
  111} (2013)  121803},
\href{http://arxiv.org/abs/1305.5251}{{\tt arXiv:1305.5251 [hep-ph]}}.

\bibitem{Kajantie:1996mn}
K.~Kajantie, M.~Laine, K.~Rummukainen, and M.~E. Shaposhnikov, {\em {Is there a
  hot electroweak phase transition at m(H) larger or equal to m(W)?}\/},
  \href{http://dx.doi.org/10.1103/PhysRevLett.77.2887}{Phys. Rev. Lett. {\bf
  77} (1996)  2887--2890},
\href{http://arxiv.org/abs/hep-ph/9605288}{{\tt arXiv:hep-ph/9605288
  [hep-ph]}}.

\bibitem{Profumo:2014opa}
S.~Profumo, M.~J. Ramsey-Musolf, C.~L. Wainwright, and P.~Winslow, {\em
  {Singlet-catalyzed electroweak phase transitions and precision Higgs boson
  studies}\/},  \href{http://dx.doi.org/10.1103/PhysRevD.91.035018}{Phys. Rev.
  {\bf D91} (2015)  035018},
\href{http://arxiv.org/abs/1407.5342}{{\tt arXiv:1407.5342 [hep-ph]}}.

\bibitem{Katz:2014bha}
A.~Katz and M.~Perelstein, {\em {Higgs Couplings and Electroweak Phase
  Transition}\/},  \href{http://dx.doi.org/10.1007/JHEP07(2014)108}{JHEP {\bf
  07} (2014)  108},
\href{http://arxiv.org/abs/1401.1827}{{\tt arXiv:1401.1827 [hep-ph]}}.

\bibitem{Noble:2007kk}
A.~Noble and M.~Perelstein, {\em {Higgs self-coupling as a probe of electroweak
  phase transition}\/},
  \href{http://dx.doi.org/10.1103/PhysRevD.78.063518}{Phys. Rev. {\bf D78}
  (2008)  063518},
\href{http://arxiv.org/abs/0711.3018}{{\tt arXiv:0711.3018 [hep-ph]}}.

\bibitem{Henning:2014gca}
B.~Henning, X.~Lu, and H.~Murayama, {\em {What do precision Higgs measurements
  buy us?}\/},
\href{http://arxiv.org/abs/1404.1058}{{\tt arXiv:1404.1058 [hep-ph]}}.

\bibitem{Profumo:2007wc}
S.~Profumo, M.~J. Ramsey-Musolf, and G.~Shaughnessy, {\em {Singlet Higgs
  phenomenology and the electroweak phase transition}\/},
  \href{http://dx.doi.org/10.1088/1126-6708/2007/08/010}{JHEP {\bf 08} (2007)
  010},
\href{http://arxiv.org/abs/0705.2425}{{\tt arXiv:0705.2425 [hep-ph]}}.

\bibitem{Curtin:2014jma}
D.~Curtin, P.~Meade, and C.-T. Yu, {\em {Testing Electroweak Baryogenesis with
  Future Colliders}\/},  \href{http://dx.doi.org/10.1007/JHEP11(2014)127}{JHEP
  {\bf 11} (2014)  127},
\href{http://arxiv.org/abs/1409.0005}{{\tt arXiv:1409.0005 [hep-ph]}}.

\bibitem{Cao:2017oez}
Q.-H. Cao, F.~P. Huang, K.-P. Xie, and X.~Zhang, {\em {Testing the electroweak
  phase transition in scalar extension models at lepton colliders}\/},
  \href{http://dx.doi.org/10.1088/1674-1137/42/2/023103}{Chin. Phys. {\bf C42}
  (2018)  023103},
\href{http://arxiv.org/abs/1708.04737}{{\tt arXiv:1708.04737 [hep-ph]}}.

\bibitem{Huang:2016cjm}
P.~Huang, A.~J. Long, and L.-T. Wang, {\em {Probing the Electroweak Phase
  Transition with Higgs Factories and Gravitational Waves}\/},
  \href{http://dx.doi.org/10.1103/PhysRevD.94.075008}{Phys. Rev. {\bf D94}
  (2016)  075008},
\href{http://arxiv.org/abs/1608.06619}{{\tt arXiv:1608.06619 [hep-ph]}}.

\end{thebibliography}\endgroup
}
\end{multicols}   

\newpage



\end{document}